\documentclass[twocolumn]{aastex63}
\submitjournal{ApJ}
\shorttitle{A VLBA Study of DPAGN}
\shortauthors{Kharb et al.}
\begin{document}
\title{The Nature of Jets in Double-Peaked Emission Line AGN in the KISSR Sample}
\correspondingauthor{P. Kharb}\email{kharb@ncra.tifr.res.in}
\author[0000-0003-3203-1613]{P. Kharb}
\affil{National Centre for Radio Astrophysics (NCRA) - Tata Institute of Fundamental Research (TIFR),
S. P. Pune University Campus, Post Bag 3, Ganeshkhind, Pune 411007, India}
\author[0000-0002-5331-6098]{S. Subramanian}
\affil{Indian Institute of Astrophysics, II Block, Koramangala, Bangalore 560034, India}
\author[0000-0001-8996-6474]{M. Das}
\affil{Indian Institute of Astrophysics, II Block, Koramangala, Bangalore 560034, India}
\author[0000-0003-3295-6595]{S. Vaddi}
\affil{Arecibo Observatory, Route 625 Bo. Esperanza Arecibo, Puerto Rico, PR 00612, USA}
\author[0000-0002-5195-335X]{Z. Paragi}
\affil{Joint Institute for VLBI ERIC, Postbus 2, 7990 AA Dwingeloo, The Netherlands}

\begin{abstract}
Double-peaked emission line AGN (DPAGN) have been regarded as binary black hole candidates. We present here results from parsec-scale radio observations with the Very Long Baseline Array (VLBA) of five DPAGN belonging to the KISSR sample of emission-line galaxies. This work concludes our pilot study of nine type 2 Seyfert and LINER DPAGN from the KISSR sample. In the nine sources, dual compact cores are only detected in the ``offset AGN'', KISSR\,102. The overall incidence of jets however, in the eight sources detected with the VLBA, is $\ge$60\%. We find a difference in the ``missing flux density'' going from the Very Large Array (VLA) to VLBA scales between Seyferts and LINERs, with LINERs showing less missing flux density on parsec-scales. Using the emission-line modeling code, MAPPINGS III, we find that the emission lines are likely to be influenced by jets in {5/9} sources. Jet-medium interaction is the likely cause of the emission-line splitting observed in the SDSS spectra of these sources. Jets in radio-quiet AGN are therefore energetically capable of influencing their parsec- and kpc-scale environments, making them agents of ``radio AGN feedback'', similar to radio-loud AGN.
\end{abstract}

\section{Introduction}
The presence of prominent emission lines in the optical/UV spectra is one of the hallmarks of an active galactic nucleus (AGN). AGN are believed to be the energetic centres of galaxies that derive their enormous luminosities from the release of gravitational potential energy as matter accretes onto supermassive black holes (M$_\mathrm{BH} \sim10^6-10^9$~M$_\sun$) via accretion disks \citep{Rees84}. The conservation of angular momentum ensures the launch of bipolar outflows in the form of collimated jets or broad winds from the black hole-accretion-disk systems \citep[see][]{Peterson88}. Fast and slower-moving line-emitting gas clouds form respectively, the broad and narrow line regions (BLR and NLR). A dusty obscuring torus hides the BLR from certain lines of sight, resulting in type 2 AGN. Certain other lines of sight allow an unobstructed view of the BLR and NLR, resulting in type 1 AGN \citep{Antonucci93}. 

\begin{table*}[t]
\begin{center}
\caption{Source and Calibrator Properties}
\begin{tabular}{lcccccccc}
\hline\hline
{Source} & {RA, Dec} &$z$& {Type} & {S$^\mathrm{peak}_\mathrm{FIRST}$} & {S$^\mathrm{total}_\mathrm{FIRST}$} & {PhaseCal} & {X, Y Error, Separ.}\\ 
{Name}   & {h m s, $\degr$~$\arcmin$~$\arcsec$}     & {}             & {}    & {mJy/beam} & {mJy}    &    & {mas, mas, $\degr$}\\ \hline
KISSR\,618 & 15~02~28.776, +28~58~15.60 &0.072940 & Sy 2    & 1.69$\pm$0.14 & 2.47$\pm$0.33 & J1454+2955 & 0.21, 0.60, 1.98 \\
KISSR\,872 & 15~50~09.805, +29~11~07.23 &0.083064 & LINER & 4.80$\pm$0.15 & 5.25$\pm$0.27 & J1539+2744 & 0.35, 0.81, 2.72 \\
KISSR\,967 & 16~06~31.785, +29~27~56.99 &0.092067 & LINER & 2.36$\pm$0.15 & 2.36$\pm$0.25 & J1605+3001 & 0.36, 0.54, 0.60 \\
KISSR\,1154 & 11~56~32.871, +42~59~39.27 &0.072032 & Sy 2   & 0.66$\pm$0.12 & 3.52$\pm$0.78 & J1150+4332 & 0.88, 1.40, 1.26 \\
KISSR\,1321 & 12~30~31.596, +42~58~22.18 &0.063856& Sy 2   & 1.43$\pm$0.14 & 1.60$\pm$0.25 & J1226+4340 & 2.74, 7.17, 0.96 \\
\hline
\end{tabular}
\end{center}
{\small Columns~1, 2: Source name, right ascension and declination. {Column~3: Source redshift.} Column~4: {Spectral type from NED\footnote{NASA/IPAC Extragalactic Database. URL: https://ned.ipac.caltech.edu}.} Columns~5, 6: Peak intensity and total flux density from VLA FIRST {1.4~GHz} image (beam $=5.4\arcsec \times 5.4\arcsec$) {with errors obtained from the {\tt AIPS} task {\tt JMFIT}}. Columns~7, 8: Phase reference calibrator with positional errors in X and Y directions in milli-arcseconds and separation in degrees from targets.}
\label{tab1}
\end{table*}

A small fraction of AGN \citep[$\sim1\%$ in the SDSS\footnote{Sloan Digital Sky Survey \citep{York00}} survey;][]{Wang09} show double-peaked emission lines in their optical spectra; these are double-peaked AGN (DPAGN). The presence of double-peaked `narrow' emission lines has been suggested to arise either due to the presence of binary supermassive black holes (predicted in hierarchical galaxy evolution models) with individual associated NLRs, disk-like NLRs, NLR kinematics or jet-NLR interaction \citep{Smith10,Liu10a,Smith12,Fu12}.

While Seyfert and low-ionization nuclear emission-line region (LINER) galaxies \citep{Heckman80} are classed together as ``radio-quiet'' AGN, they do possess jets/outflows ranging in extents from tens to hundreds of parsecs on one end, and a few kpc to $\sim$10~kpc on the other \citep{Wilson80,UlvestadWilson84,HoUlvestad01,Nagar02,Gallimore06,Kharb16}. The origin of these radio outflows has been debated; both AGN-related and starburst-related processes have been invoked to explain the observed radio emission. Most likely, AGN-related and stellar-related outflows coexist in these low luminosity AGN \citep[LLAGN;][]{Ulvestad81,Baum93,Colbert96,Hota06,Panessa19,Sebastian19,Sebastian20,Silpa20}. However, milli-arcsecond (mas)-scale radio imaging using the technique of Very Long Baseline Interferometry (VLBI) has provided greater support to the AGN-related outflow model.

\begin{figure*}[t]
\centerline{
\includegraphics[width=6cm,trim=30 160 0 150]{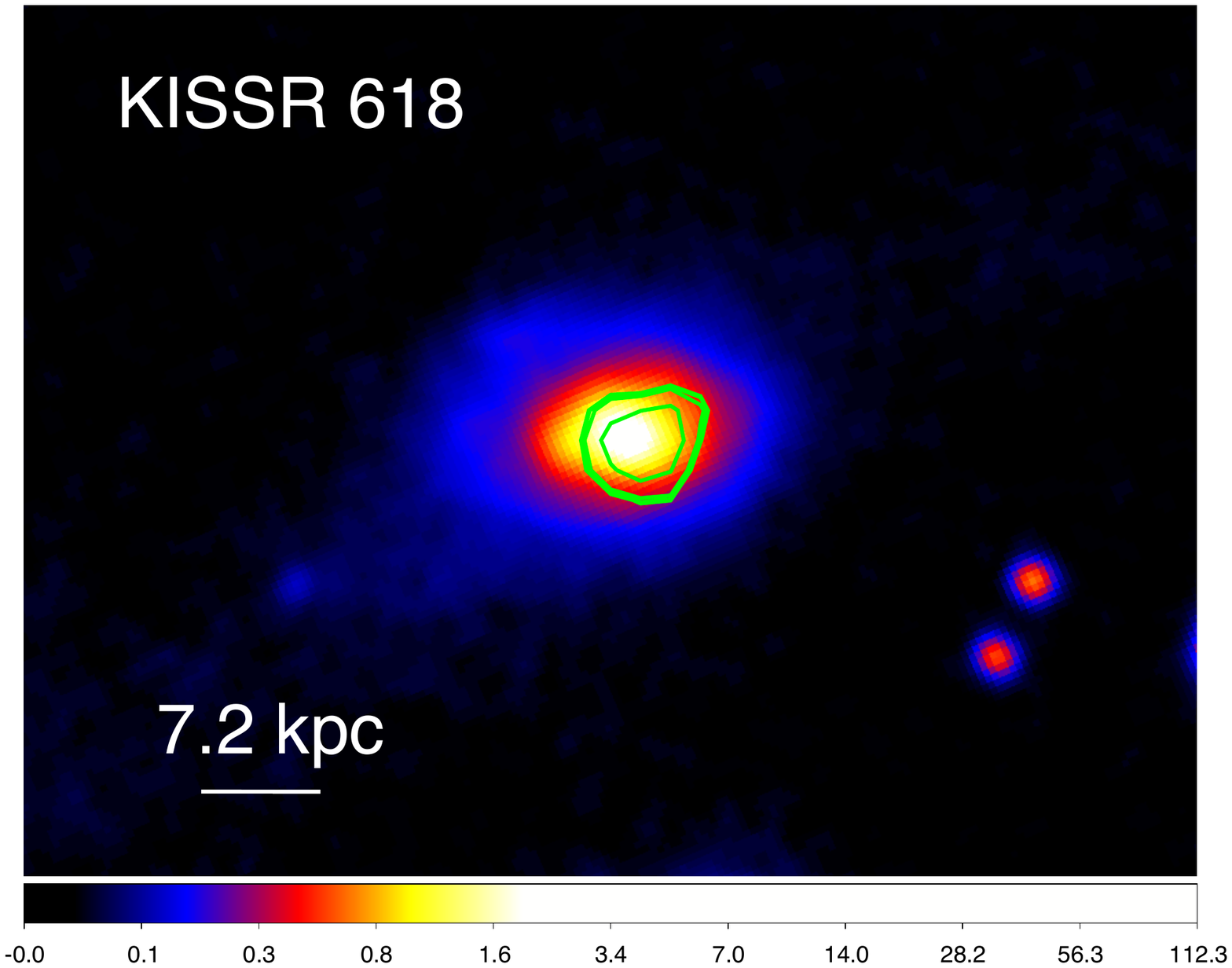}
\includegraphics[width=6cm,trim=30 160 0 150]{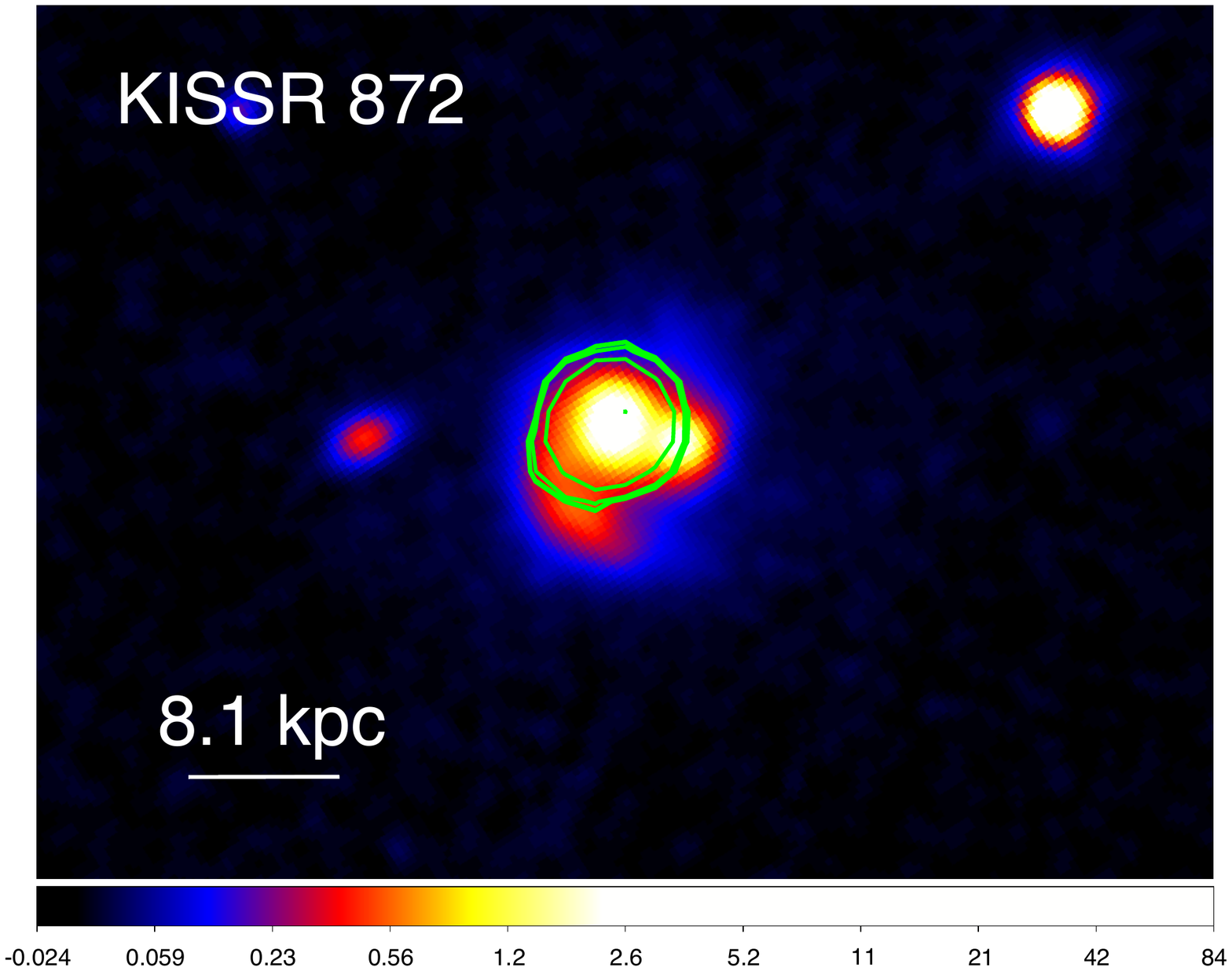}
\includegraphics[width=5.5cm, trim=0 -50 0 150]{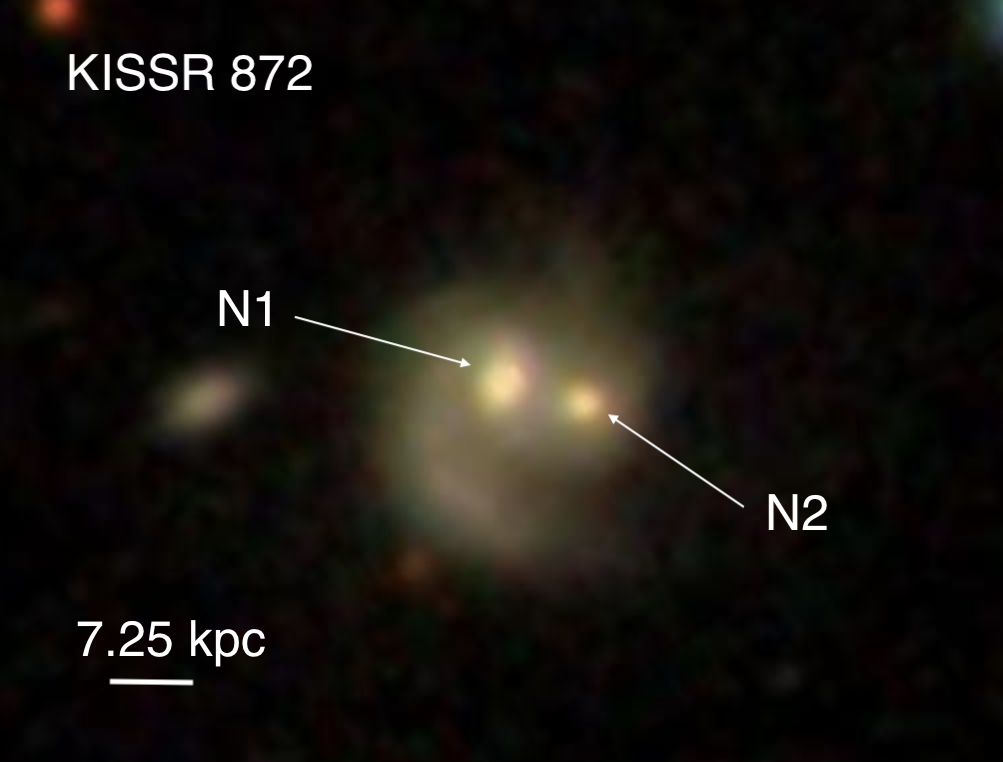}}
\centerline{
\includegraphics[width=6cm,trim=0 160 30 180]{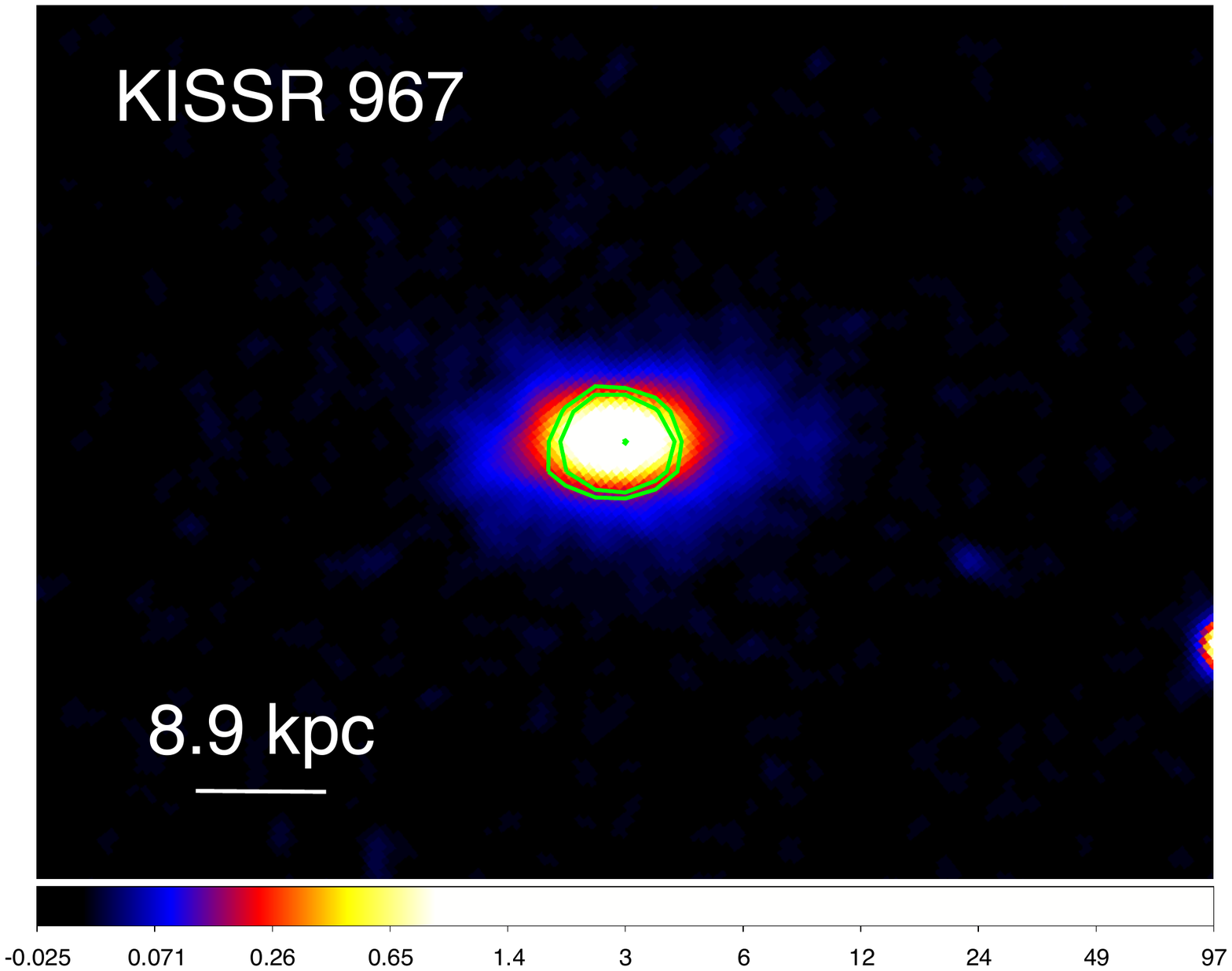}
\includegraphics[width=6cm,trim=5 160 30 180]{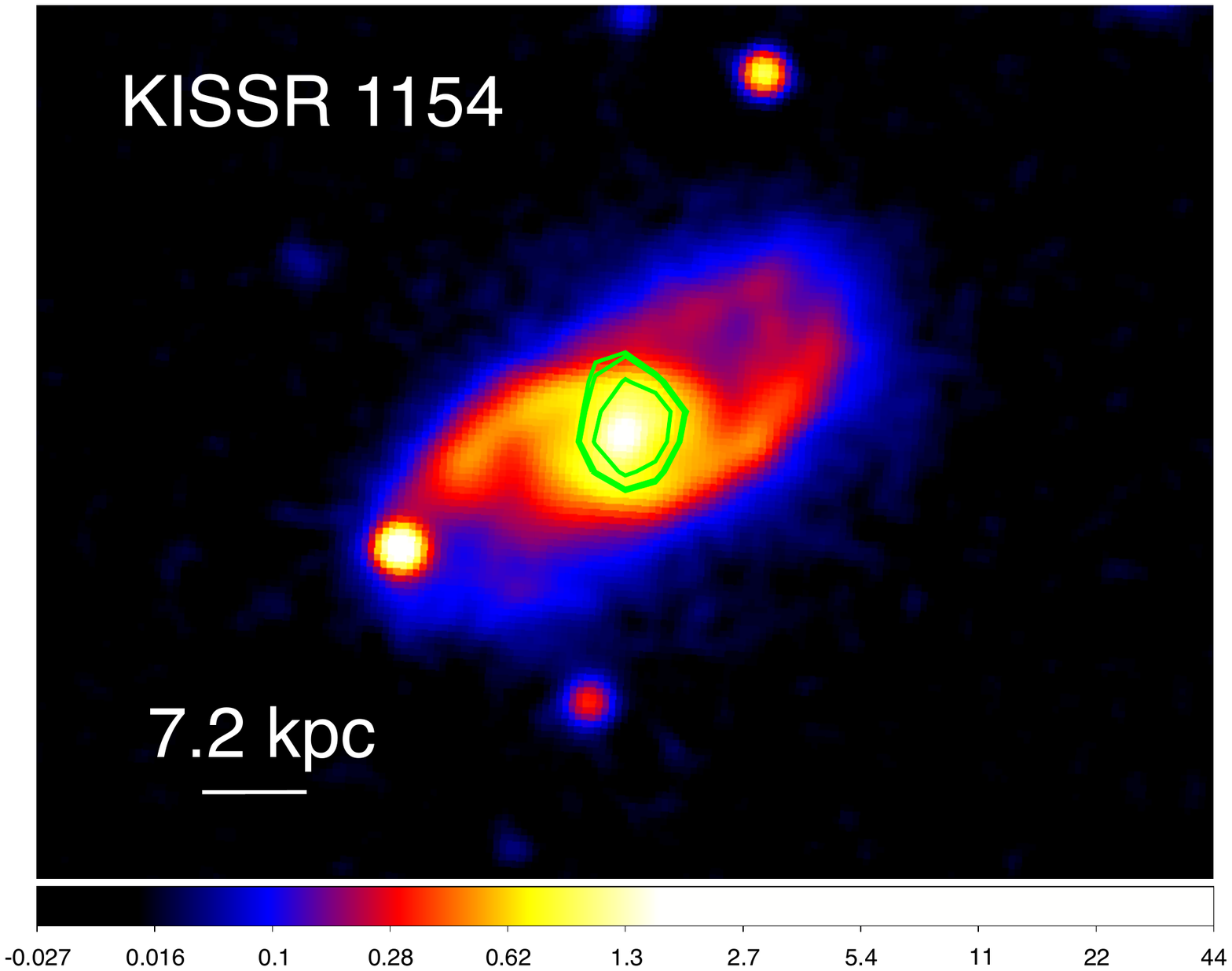}
\includegraphics[width=6cm,trim=5 160 30 180]{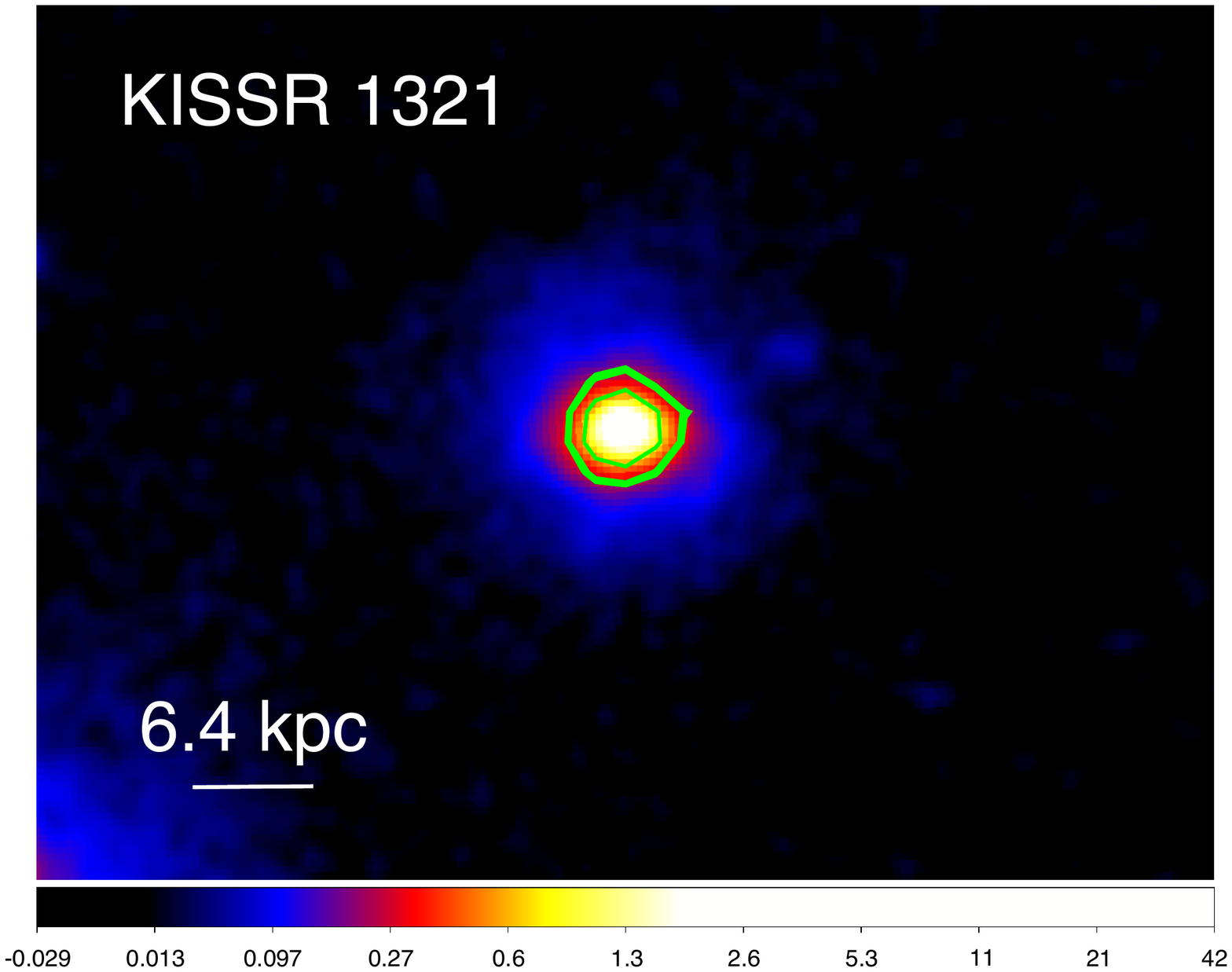}}
\caption{\small Radio-optical overlays with green contours from VLA FIRST {at 1.4~GHz} and color from green ($g$, $\lambda 4770 \AA$) band SDSS images. (Top row) Left: KISSR\,618. Centre: KISSR\,872. Right: {SDSS {\it gri} color composite image} of KISSR\,872 showing two optical nuclei, N1, N2. The SDSS colour scale is in units of 3.631~$\mu$Jy. (Bottom row) Left: KISSR\,967. Centre: KISSR\,1154. Right: KISSR\,1321. The radio contours in units of mJy~beam$^{-1}$ are KISSR\,618: 0.61, 0.62, 0.69, 1.11, 
KISSR\,872: 0.62, 0.64, 0.74, 1.30, 4.44,
KISSR\,967: 0.96, 1.13, 1.78,
KISSR\,1154: 0.60, 0.61, 0.67, and
KISSR\,1321: 0.60, 0.60, 0.62, 0.74, 1.37.}
\label{fig1}
\end{figure*}

VLBI observations of Seyfert and LINER galaxies have revealed the presence of weak radio cores and one or two sided radio jets in them \citep{Falcke00,Middelberg04,Nagar05,Orienti10,Kharb10a,Baldi18}. The brightness temperatures (T$_\mathrm{B}$) of the radio cores are typically of the order of $10^6-10^{11}$~K, supporting an AGN origin \citep{Nagar05,Doi13}. While the radio cores typically exhibit flat or inverted spectral indices \citep{Wilson98,Mundell00,Ho08}, steep spectrum radio cores have also been reported in several LLAGN in the literature \citep{Roy00,Bontempi12,Kharb19,Silpa20}. These cores could be contaminated by the presence of steep spectrum jet/lobe emission, or could be coronal synchrotron emission \citep[e.g.,][]{Kharb15b}. Alternatively, the {\it actual} radio cores may have failed to get detected at the observing frequency and may show up at higher frequencies, due to inverted spectra \citep[e.g.,][]{Kharb17b,Rubinur20}. 

We identified an initial sample of nine sources belonging to the KPNO Internal Spectroscopic Survey Red (KISSR) of spiral/disk emission line galaxies \citep{Wegner03}, that showed double peaked or asymmetric lines in their SDSS spectra, and which had been detected in the Very Large Array (VLA) FIRST\footnote{Faint Images of the Radio Sky at Twenty-Centimeters \citep{Becker95}} survey ({1.4~GHz}, resolution $\theta\sim5\arcsec$). {The sources were selected based on a visual inspection of the SDSS spectra; sources that either showed double peaks, or asymmetries in one or more emission lines like [S {\sc ii}], H$\alpha$, [N {\sc ii}], H$\beta$ were chosen. Sources which showed asymmetry only in the [O {\sc iii}] lines were not considered for selection, as such asymmetries could be signatures of emission-line gas outflows.}
Dual frequency VLBA observations were carried out on these nine DPAGN. Results from observations of KISSR\,1494, KISSR\,1219, KISSR\,434, and KISSR\,102, have been presented in \citet{Kharb15b,Kharb17b,Kharb19,Kharb20}, respectively. 

In this paper, we present the results for the five remaining DPAGN, viz., KISSR\,618, KISSR\,872,  KISSR\,967, KISSR\,1154 and KISSR\,1321. Radio-optical overlays of the five galaxies with SDSS optical images and FIRST radio contours are presented in Figure~\ref{fig1}. In this paper, we adopt the cosmology with H$_0$ = 73~km~s$^{-1}$~Mpc$^{-1}$, $\Omega_{mat}$ = 0.27, $\Omega_{vac}$ =  0.73. Spectral index $\alpha$ is defined such that flux density at frequency $\nu$ is $S_\nu \propto \nu^{\alpha}$.

\begin{figure*}[t]
\centering{
\includegraphics[width=8cm,trim=0 200 30 150]{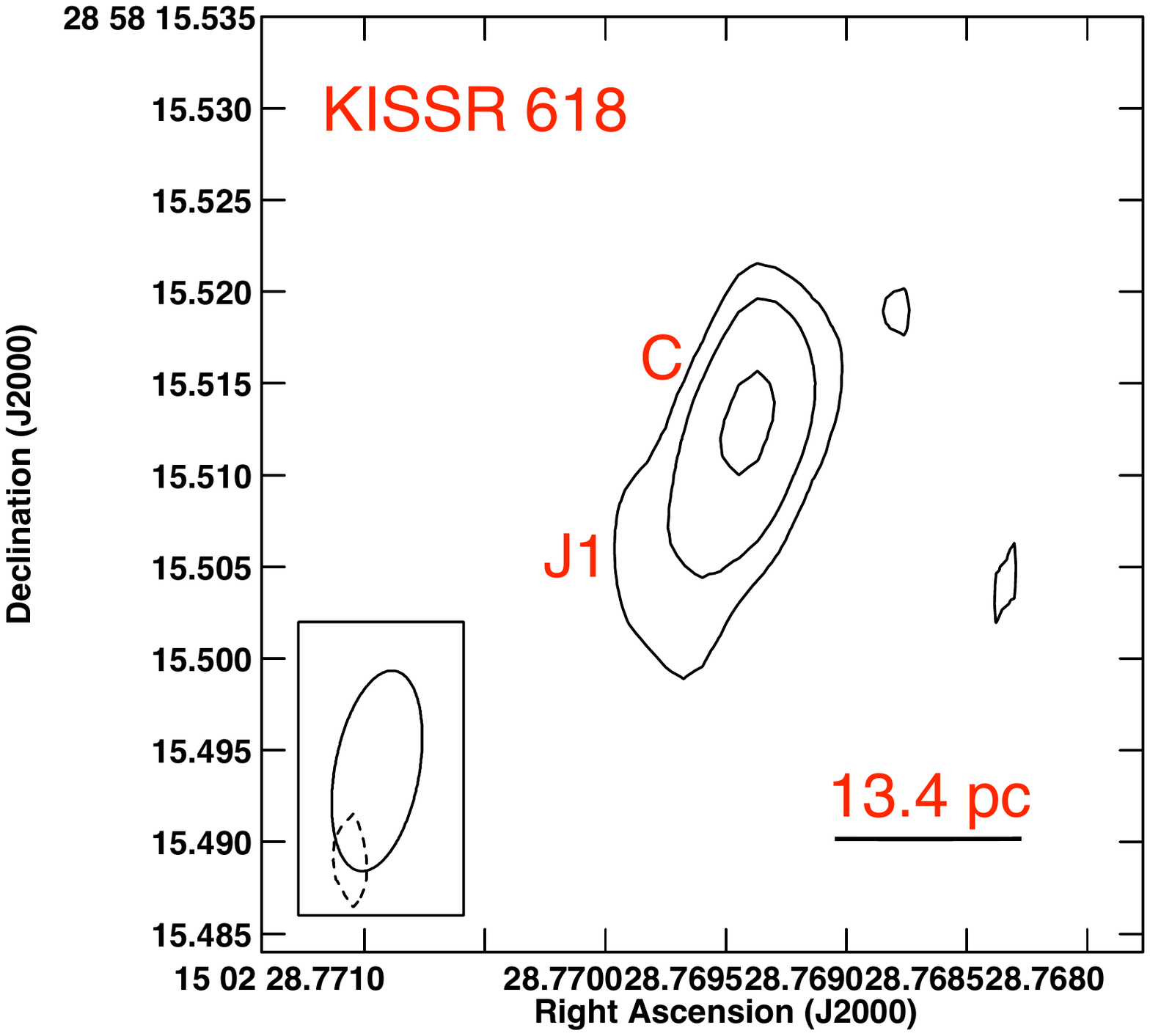}
\includegraphics[width=7.5cm,trim=0 185 30 150]{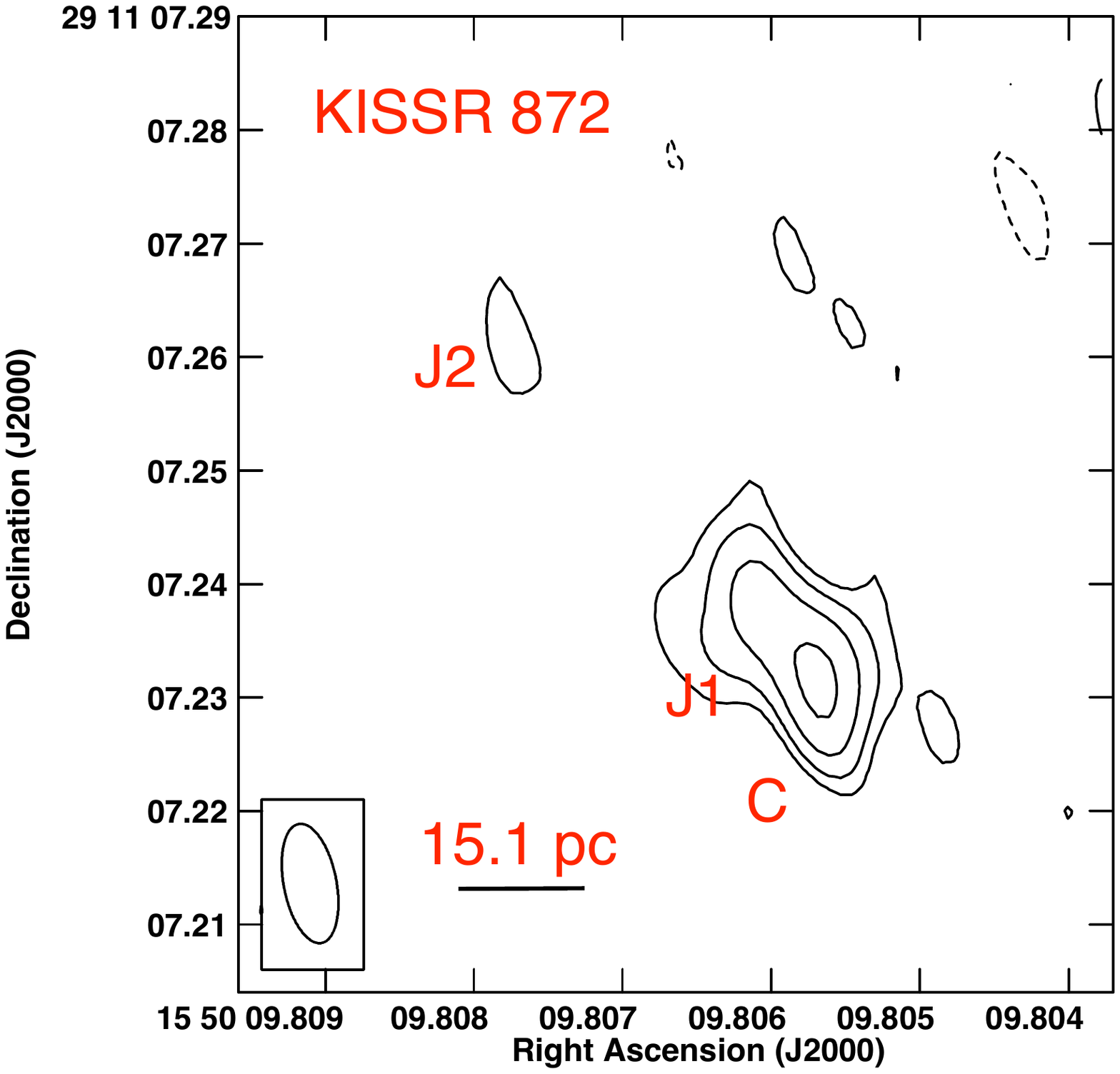}
\includegraphics[width=8cm,trim=-35 200 70 170]{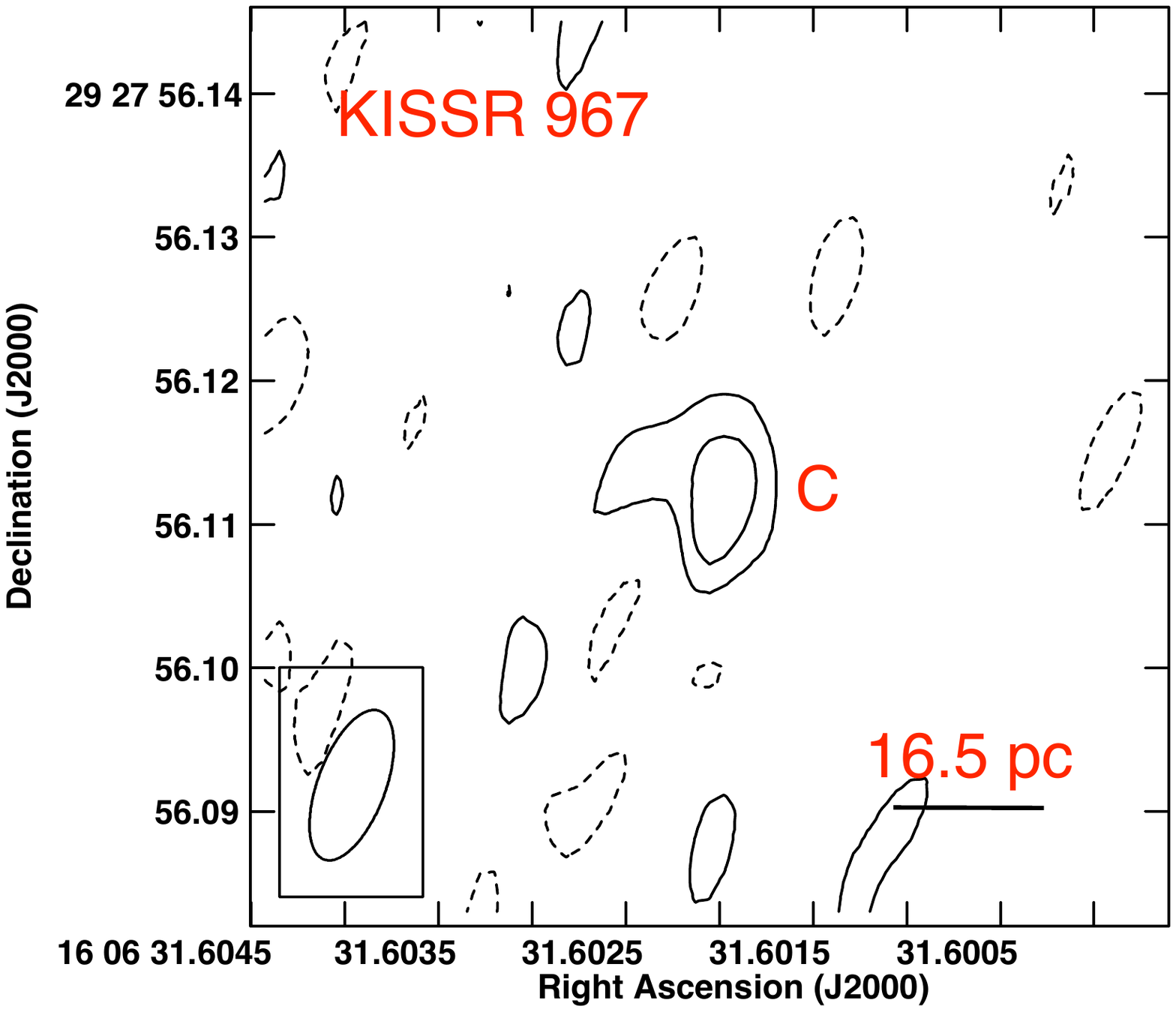}
\includegraphics[width=8.4cm,trim=-10 200 10 170]{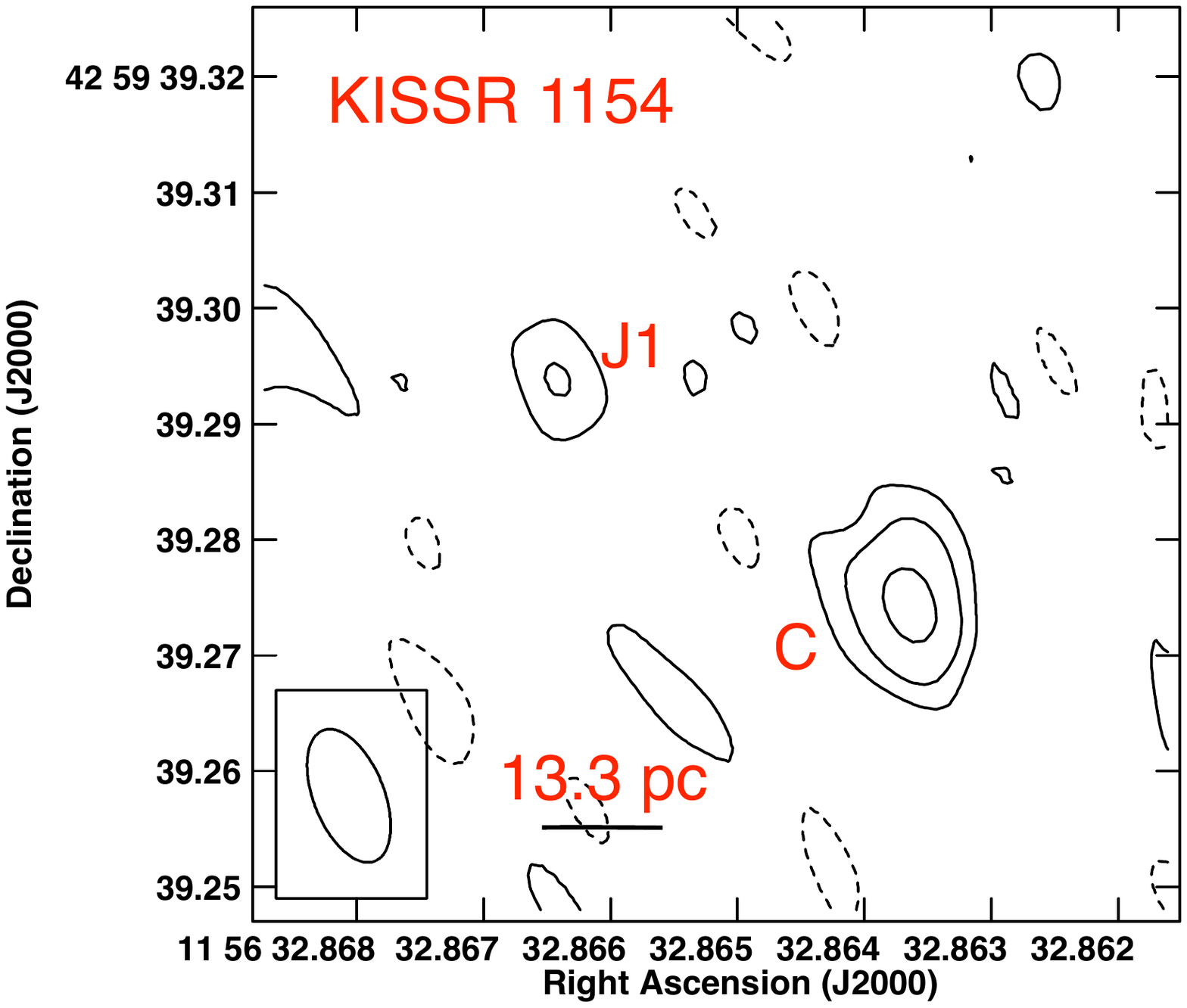}}
\caption{\small 1.5~GHz VLBA contour images of (top left) KISSR\,618, (top right) KISSR\,872, (bottom left) KISSR\,967, and (bottom right) KISSR\,1154. The contour levels are in percentage of peak intensity (=0.31, 0.84, 0.16, and 0.29 mJy~beam$^{-1}$, respectively) and increase in steps of 2 with the lowest contour level being at $\pm$ 22.5, 10.0, 32.0, and 20.0\%, respectively. The respective synthesized beam-sizes are listed in Table~\ref{tab2}. C, J1, J2 indicate the core and jet components. }
\label{fig2}
\end{figure*}

\section{Radio Data}
We observed KISSR\,618, KISSR\,872,  KISSR\,967, KISSR\,1154 and KISSR\,1321, with eight to ten antennas of the VLBA in a phase-referencing experiment at 1.39 and 4.85~GHz on December 11, 26, 28 and 31, 2018, as well as January 1, 6, and 14, 2019 (Project IDs BK219A, BK219B). We observed all sources for $\approx$120~min each at each of the two frequencies; the weakest source in FIRST data, viz., KISSR\,1321, was observed for $\approx$180~min at each frequency. Nearby compact calibrators were used as phase reference calibrators (see Table~\ref{tab1}). The targets and the phase reference calibrators were observed in a ``nodding'' mode in a 5~min cycle (2~min on calibrator and 3~min on source), for good phase calibration. 

The data were processed using the VLBA data calibration pipeline procedure {\tt VLBARUN\footnote{URL: http://www.aips.nrao.edu/vlbarun.shtml}} in the Astronomical Image Processing System \citep[{\tt AIPS};][]{Greisen03}. {\tt VLBARUN} is based on the {\tt VLBAUTIL} procedures; their step-by-step usage is described in Appendix C of the {\tt AIPS} Cookbook\footnote{URL: http://www.aips.nrao.edu/cook.html}. Los Alamos (LA) was used as the reference antenna for the entire experiment. The source images were not self-calibrated. All images were created using uniform weighting with {\tt ROBUST +1}. The resultant {\it rms} noise in the images was $\sim30~\mu$Jy~beam$^{-1}$ at 1.5 GHz and $\sim20~\mu$Jy~beam$^{-1}$ at 4.9 GHz. {\tt AIPS} verbs {\tt TVMAXF}, {\tt TVSTAT} and {\tt TVDIST} were used respectively to obtain the peak intensity, total flux density and component distances. 

As missing short spacings at the higher frequency of the VLBA may cause diffuse flux density to be missed \citep{Wrobel95}, we created the $1.5-4.9$~GHz spectral index image for KISSR\,872 (the only source detected at both frequencies in the current sample), by choosing the appropriate ($u, v$) plane weighting functions during imaging at each frequency. The two frequency images were then created with the beam of the lower frequency image (9.0~{mas}~$\times$~2.8~{mas}, PA=13$\degr$) and the core positions were made coincident using the AIPS task {\tt OGEOM}, before using the AIPS task {\tt COMB} to create the spectral index image. Pixels with flux density values below three times the {\it rms} noise were blanked in {\tt COMB}. A spectral index noise image was created as well. The spectral index and noise values are reported in Section~\ref{sec872} ahead.

\begin{table*}[t]
\caption{Source Properties on VLBA Scales}
\begin{center}
\begin{tabular}{lccccccccc}
\hline\hline
{Source} & {}& {S$^\mathrm{peak}_\mathrm{VLBA}$}& {S$^\mathrm{total}_\mathrm{VLBA}$} & {{\it rms} Noise} & {Beam}& {Scale} &{Jet} &{$\mathrm{R_J}$} & {$\beta\equiv v/c$} \\ 
{Name}   & {} & {mJy/beam} & {mJy} & {$\mu$Jy/beam}& {mas~$\times$~mas, PA$\degr$} &{kpc/$\arcsec$} & {parsec} &{}&{}\\ \hline
KISSR\,618   &C                & 0.32$\pm$0.03 &0.52 & 26.8 & 11.1$\times$4.6, $-10.3$ & 1.342 & 23 & 1.20 & 0.05 \\
                      &J1               & ...               & 0.09& ...                                                     & ... & ... & ... &... & ...\\
KISSR\,872~$L-band$ &C & 0.84$\pm$0.03 & 1.85 & 27.9& 10.7$\times$4.6, $+10.9$ & 1.506 & 200 & 5.40 & 0.40 \\
                      &J1               & ...               & 0.46 & ...                                                     & ... & ... & 26$^\star$ &... & ...\\
                      &J2               & 0.14$\pm$0.03 & ... & ...                                                     & ... & ... & 63$^\star$ &... & ...\\
                      &J3               & 0.12$\pm$0.03 & ... & ...                                                     & ... & ... & ...& ... & ...\\
KISSR\,872~$C-band$ &C& 0.29$\pm$0.02 & 0.35 & 18.3 & 3.6$\times$1.7, $+20.4$ & 1.506 & 17 & 2.70 & 0.30\\
                      &J0               & 0.15$\pm$0.02 & 0.18 & ...                                                & ... & ... & ... & ... & ...\\
KISSR\,967 &C                  & 0.16$\pm$0.03 & 2.17 & 28.0 & 11.1$\times$4.7, $-20.6$ & 1.650 & ...  & ... & ... \\
KISSR\,1154 &C                & 0.29$\pm$0.03 & 0.36 & 27.6 & 12.2$\times$6.1, $+21.6$  &  1.331 & 49 & 1.01 & 0.003 \\
                      &J1              & 0.12$\pm$0.03 & 0.12 & ...                                        & ...  &  ...           &... & ... & ...\\
KISSR\,1321 &...&... &...  &32.6                                               & ...  & 1.192 & ...  & ... & ... \\
\hline
\end{tabular}
\end{center}
{Column~1: Source name. Column~2: Radio components with C denoting the core and J the jet components. Columns~3, 4: Peak intensity {with error from {\tt AIPS} task {\tt JMFIT}}, and total flux density from {\it L$-$band} (1.5~GHz) VLBA image, unless otherwise stated. {\it C$-$band} data is at 4.9~GHz. {Total flux density was estimated using the {\tt AIPS} verb {\tt TVSTAT} for TV selected regions and have typical systematic errors of less than 5\%.} Column~5: {\it rms} noise in final image. Column~6: Synthesized beam in milli-arcsec and beam position angle measured from North through East. Column~7: Scale in kpc/arcsec at source. Column~8: Total core-jet extent in parsecs. $^\star$ Jet extents for components C-J1 and C-J2 in KISSR\,0872. Column~9: Jet-to-counterjet intensity ratio using 3 times {\it rms} noise for counterjet emission. Column~10: Estimated jet speed for an assumed jet inclination of $50\degr$ for these type 2 sources.}
\label{tab2}
\end{table*}

\begin{figure*}[t]
\centering{
\includegraphics[width=9.85cm,trim=100 210 0 250]{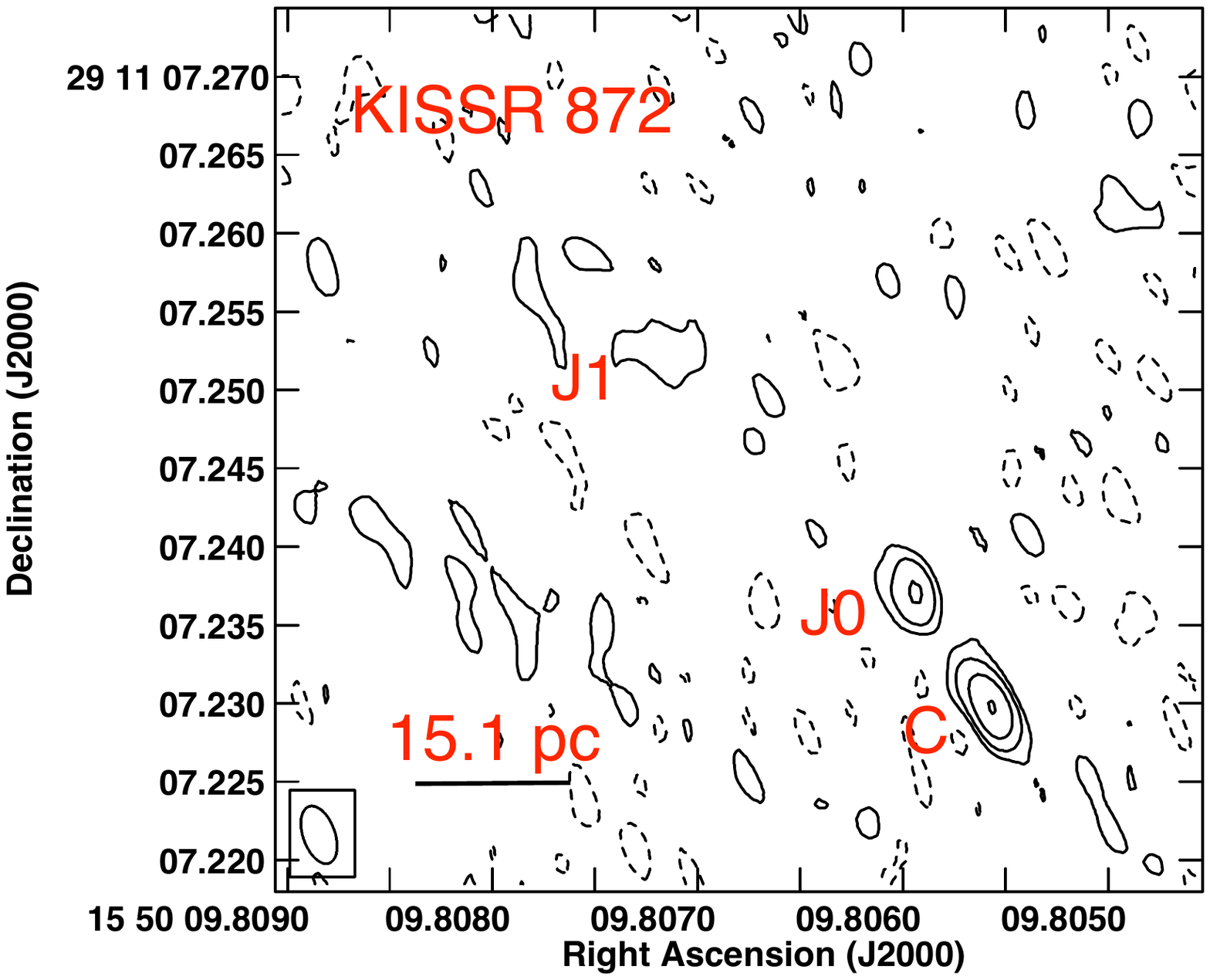}
\includegraphics[height=7.7cm,trim=80 120 20 80]{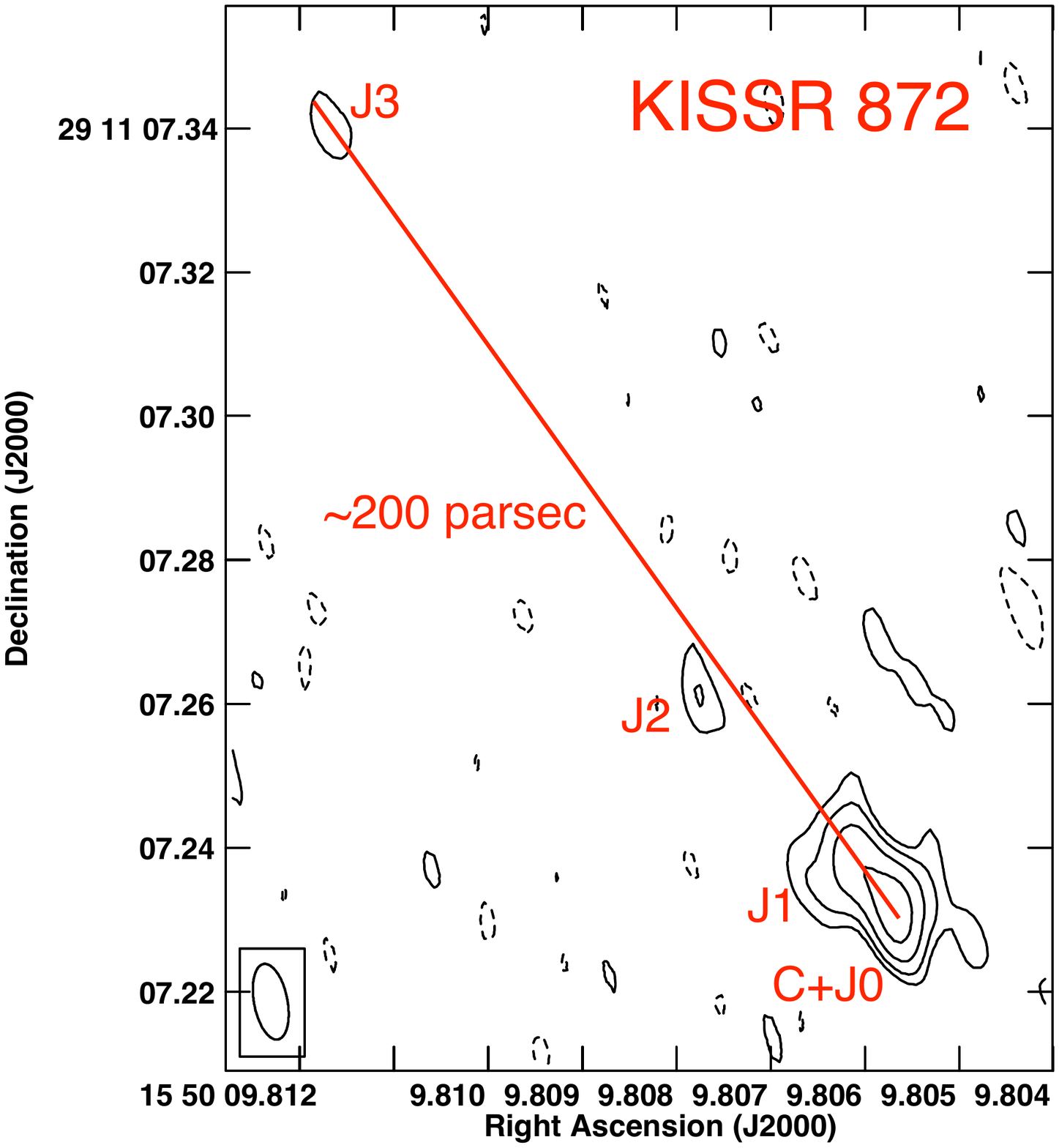}}
\caption{\small (Left) 4.9~GHz and (Right) 1.5~GHz VLBA image of KISSR\,872. The full extent of the $\sim$200~parsec jet is seen in the 1.5~GHz image. The contour levels are in percentage of peak intensity (top: 0.30~mJy~beam$^{-1}$, bottom: 0.84~mJy~beam$^{-1}$) and increase in steps of 2 with the lowest contour level being at $\pm$12\% (top) and $\pm$8\% (bottom). C, J0, J1, J2, J3 indicate the core and jet components. }
\label{fig3}
\end{figure*}

\section{Results}
\subsection{Radio Properties}
Parsec-scale radio emission is detected at one or both radio frequencies in four out of five DPAGN (Figure~\ref{fig2}). Only the Seyfert galaxy with the faintest radio emission in the VLA FIRST image, viz., KISSR\,1321 (S$^\mathrm{total}_\mathrm{FIRST}=1.6$~mJy), is not detected with the VLBA at either radio frequency. The detection rate of parsec-scale emission for KISSR sources with VLA FIRST core flux density $\ge2.4$~mJy is 100\%. Clear jet-like features are detected in KISSR\,618, KISSR\,872, and a possible jet component (J1) is detected at the 5$\sigma$ level in KISSR\,1154 about $\sim50$~parsec from the radio core. The double optical nuclei galaxy, KISSR\,872, shows the longest parsec-scale radio jet ($\sim$200~parsec), which is detected at both 1.5 and 4.9~GHz (the inner 10~parsec). It is unclear if the VLBA image of KISSR\,967 shows the beginning of a curved jet in Figure~\ref{fig2}.

The VLBA observations fail to detect any emission at 4.9~GHz in KISSR\,618, KISSR\,967 and KISSR\,1154. This could imply that the core-jet emission has a steep radio spectrum. In the only source detected at both frequencies (KISSR\,872), the measured average spectral indices are indeed steep ($\alpha\sim-1$, see Section~\ref{sec872}). We estimate the brightness temperature of the unresolved cores in KISSR\,967 and KISSR\,1154 to be $6\times10^6$~K and $7\times10^6$~K, respectively, using the relation by \citet{Ulvestad05}. The peak flux densities and beam-deconvolved sizes could not be uniquely obtained using the Gaussian-fitting task {\tt JMFIT} for KISSR\,618 and KISSR\,872. A resolved component C and an unresolved component J0 are  observed in the 4.9~GHz image of KISSR\,872 (Figure~\ref{fig3}). We obtained T$_\mathrm{B}=1.2\times10^7$~K and 1.6$\times10^7$~K, for components C and J0, respectively, attesting to their AGN origin. 

\subsection{Prevalence of Jets}
Parsec-scale jets are clearly observed in three of the four sources detected with the VLBA in the current study. Even KISSR\,967 may be showing the beginning of a curved jet in Figure~\ref{fig2}. For the eight sources detected with the VLBA, the overall incidence of jets (ignoring KISSR\,967) is $\ge$60\% (5/8). 
{A high incidence of jets has been reported for statistically significant samples of LLAGN as well. Sub-parsec-scale (or larger) radio jets have been reported in 45\% of the Palomar sample comprising of 44 LLAGN in VLBA (mas-scales) observations by \citet{Nagar05}. Jets were detected in nearly 50\% of the LeMMINGs sample with MERLIN (100-mas-scales) by \citet{Baldi18}. The LeMMINGs sample comprises 103 nearby galaxies from the Palomar sample that are categorised as Seyferts, LINERs and HII galaxies. $\sim$40\% of the 7 sources from the NUGA sample reveal extended jet-like features in either 100-mas-scale MERLIN images or mas-scale EVN images, as presented by \citet{Krips07}. It is worth noting that most of these sources, except for a subset of the LeMMINGs sample, have total mas-scale flux densities of a few mJy to few tens of mJy, whereas the KISSR sources typically have sub-mJy flux densities. The relatively high incidence of jets in the KISSR sources could be a result of the selection criterion, which was the presence of double-peaked/asymmetric line spectra. As discussed ahead in Section~\ref{secdiscuss}, that in turn could support the idea that double-peaked lines are a consequence of jet-medium interaction. }

\subsection{Jet One-Sidedness}
The jet-like features detected in all the four sources appear to be one-sided (even though jet detection in KISSR\,967 is not as robust). The counter-jet emission could either be Doppler-dimmed or free-free absorbed. We have estimated limits on the jet-to-counterjet surface brightness ratios ($\mathrm{R_J}$) for all sources using their VLBA images. These are tabulated in Table~\ref{tab2}. We find that Doppler boosting/dimming effects with jet speeds varying between $0.003c$ and $0.4c$ for all four sources showing jets, can explain the jet-sidedness ratio limits for jet inclinations of $\sim50\degr$ \citep[using relations in][]{Urry95}. These jet inclinations are consistent with the fact that these are type 2 AGN and $50\degr$ is the typical torus half opening angle \citep[e.g.,][]{Simpson96}. 

KISSR\,872 has the highest implied jet speed of $\sim0.3c - 0.4c$ in the Doppler boosting/dimming scenario. Such jet speeds have been observed in several Seyfert galaxies through multi-epoch VLBI observations \citep[e.g.,][]{Roy01,Peck03}, making this a plausible scenario. If the missing counterjet emission is a result of free-free absorption on the other hand, the required electron densities of the ionized gas can be estimated. The optical depth at frequency $\nu$ is $\tau_\mathrm{ff} (\nu)\approx8.235\times10^{-2}~T^{-1.35}~(\nu/\mathrm{GHz})^{-2.1}~EM$ \citep{Mezger67}, where $T$ is the gas temperature in units of $10^4$~K and $EM$ is the emission measure in units of cm$^{-6}$~pc. The electron density, $n_e$ can be estimated for a given path length $l$ in parsec as $n_e=\sqrt{EM/l}$. An optical depth of $\tau_\mathrm{ff} (1.5~GHz)\gtrsim2$ would be needed to account for the observed jet-to-counterjet brightness ratio  in KISSR\,872, using $\exp(-\tau)=1/R_J$. Assuming a path length of $\sim$0.1~pc through ionized gas with temperature $10^4$~K, ionized gas densities with $n_e\gtrsim1\times10^4$~cm$^{-3}$ would be needed in the inner parsec of the galaxy. The ionized gas densities must fall to $700$~cm$^{-3}$ to explain the missing counterjet emission $\sim$200 parsec away from the nucleus in KISSR\,872. 

Such ionized gas densities can be found in NLR gas clouds. However the volume filling factor of NLR clouds is of the order of $10^{-4}$ \citep[e.g.,][]{Alexander99}. This makes them unlikely candidates for absorbers for, say, a $\sim$200 parsec long counterjet. However, intercloud NLR gas, which would have lower gas densities but higher temperatures \citep[see Equation~9 in][]{Filippenko84} would have a much larger covering factor and could in principle act as the free-free absorbing medium. 

If the parsec-scale core and jets are indeed Doppler-boosted, it is worth noting that these features are not as bright or compact as those observed in radio-loud AGN, even the lower luminosity radio-loud AGN like Fanaroff-Riley type I (FRI) radio galaxies \citep[e.g.,][]{Kharb08,Kharb12a}. Whether this difference is due to a difference in plasma outflow speeds or outflow composition, is something to be examined in the future. 

\subsection{Jet Kinetic Power}
Estimates of jet kinetic power ($Q_\mathrm{jet}$) have typically been obtained from empirical correlations between radio luminosity, narrow emission-line luminosity or $p$d$V$ work from X-ray cavities, for radio powerful AGN \citep{Willott99,Merloni07}. If we assume that the kpc-scale emission is from an AGN outflow in the KISSR sources, we can estimate the time-averaged kinetic jet power of these outflows following the relations for radio-powerful galaxies listed by \citet{Punsly11} \citep[see also][]{Willott99}. We derive $F_{151}$, the flux density at 151~MHz that is needed for the calculation, using the 1.4~GHz flux density from the kpc-scale FIRST images and a jet/lobe spectral index of $-0.7$. 

The kinetic jet powers obtained this way are tabulated in Table~\ref{tab4}. $Q_\mathrm{jet}$ estimates range from $1.6\times10^{41}$~erg~s$^{-1}$ to $1.3\times10^{42}$~erg~s$^{-1}$ for the nine KISSR sources. These lie in the range of $Q_\mathrm{jet}$ values derived in lower luminosity radio galaxies of the PR2010 sample of \citet{Mezcua14}, which have in turn been estimated using the empirical relation between $Q_\mathrm{jet}$ and jet radio luminosity (${L_R}$) obtained by \citet{Merloni07}, that is $\mathrm{log}~Q_\mathrm{jet} = 0.81$~log~$L_R + 11.9$. 
\citet{Merloni07} had obtained jet kinetic powers using the $p$d$V$ work done to inflate cavities and bubbles in the hot X-ray emitting atmosphere in these galaxies. 

\citet{Merloni07} used 5~GHz jet luminosity and corrected their relation for relativistic beaming. For the one source detected at 5~GHz in the current VLBA study, viz.,  KISSR\,872, we therefore obtain $L_R = 4.23\times10^{38}$~erg~s$^{-1}$. This translates to a $Q_\mathrm{jet}$ of $1.54\times10^{43}$~erg~s$^{-1}$ in KISSR\,872. Interestingly, this value lies in the $Q_\mathrm{jet}$ value range of FRI radio galaxies \citep[][]{Rawlings91}. This underscores the fact that the jets of LINER and Seyfert galaxies are not distinctly different from FRI radio galaxies in terms of radio or kinetic power, undermining the existence of a clear radio-loud/radio-quiet dichotomy in LLAGN (see Section~4).

We however, do not find a link between $Q_\mathrm{jet}$ and jet extent. Such a connection would normally be expected and has indeed been found by \citet{Mezcua14} in their LLAGN sample with sources having large-scale radio jets ($>$100~parsec) when their $Q_\mathrm{jet}>10^{42}$~erg~s$^{-1}$. We find instead that the source with the highest $Q_\mathrm{jet}$ value ($=1.3\times10^{42}$~erg~s$^{-1}$), viz., KISSR\,1494, did not exhibit a jet in its VLBA image and was a point source in its $\sim5\arcsec$ VLA FIRST image as well. The source with the second highest $Q_\mathrm{jet}$ value ($\sim10^{42}$~erg~s$^{-1}$), viz., KISSR\,102, also did not exhibit a jet in its VLBA or VLA FIRST image. The possibility that higher resolution images compared to FIRST, or higher sensitivity images compared to the VLBA, may pick up jet emission, however remains, and must be tested. 

\begin{table*}[t]
\caption{Global Source Properties}
\begin{center}
\begin{tabular}{lccccccccc}
\hline\hline
{Source} &  {L$_\mathrm{H\alpha}$}& {SFR}               & {L$_\mathrm{Edd}$} & {L$_\mathrm{bol}$}\\ 
{Name}   &  {erg~s$^{-1}$} & {M$_\sun$~yr$^{-1}$} & {erg~s$^{-1}$} & {erg~s$^{-1}$}         \\  \hline
KISSR\,618 &  $4.5\pm0.3\times10^{40}$& $0.36\pm0.03$ & $6.20\times10^{45}$& $1.09\times10^{44}$  \\
KISSR\,872 &  $1.2\pm0.1\times10^{41}$&$0.93\pm0.08$&$5.60\times10^{45}$&$2.26\times10^{44}$     \\
KISSR\,967 &  $1.7\pm0.2\times10^{41}$ & $1.36\pm0.20$ & $1.78\times10^{46}$ & $2.30\times10^{43}$  \\
KISSR\,1154 &  $1.7\pm0.5\times10^{40}$ & $0.13\pm0.04$ & $4.50\times10^{45}$ & $8.70\times10^{43}$   \\
KISSR\,1321 &  $4.5\pm0.7\times10^{40}$ & $0.36\pm0.06$ & $3.82\times10^{45}$ & $9.89\times10^{43}$   \\
\hline
\end{tabular}
\end{center}
{\small Column~1: Source name. 3: H$\alpha$ line luminosity. 4: Star-formation rate using the H$\alpha$ line luminosity. 5: Eddington luminosity $\equiv1.25\times10^{38}~\mathrm{M_{BH}/M}_\sun$. 6: Bolometric luminosity using the [O III] $\lambda5007$ line luminosity.}
\label{tab3}
\end{table*}

\begin{figure}[b]
\includegraphics[width=8.4cm,trim=45 190 40 175]{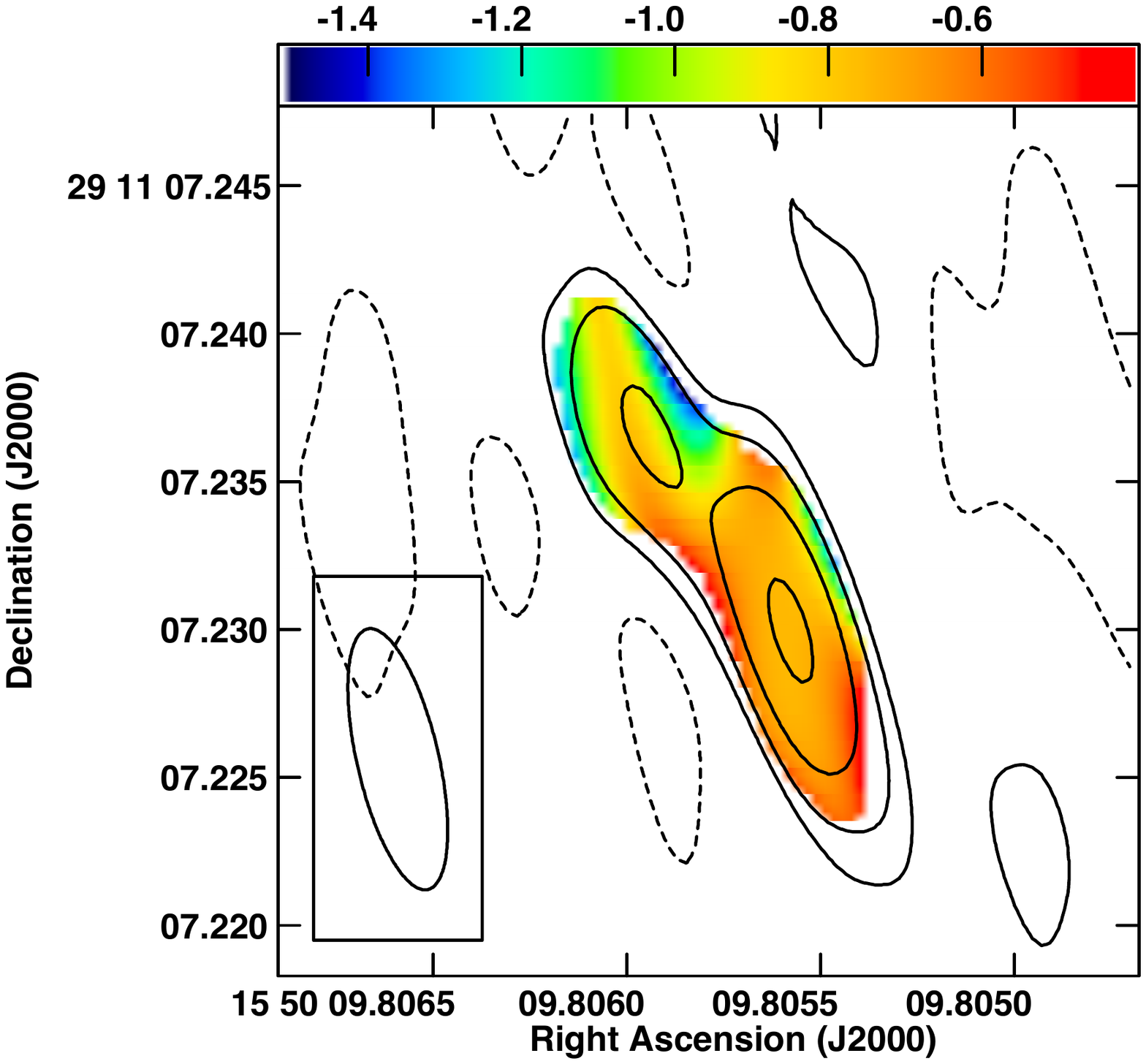}
\caption{\small $1.5-4.9$~GHz spectral index image of KISSR\,872 with 4.9~GHz radio contours overlaid. The contour levels are in percentage of peak intensity (=0.33~mJy~beam$^{-1}$) and increase in steps of 2 with the lowest level being at $\pm$11.3\%.}
\label{fig4}
\end{figure} 

\subsection{Missing Flux Density from VLA to VLBA}
We find a suggestion that going from kpc-scale radio emission in VLA FIRST images ($\theta\sim5\arcsec$) to parsec-scale emission from the VLBA, the ``recovered flux density'' in type 2 Seyfert galaxies is on average close to $\sim20$\%, similar to that reported by \citet{Orienti10}. However, it is $\sim$65\% for the three LINER galaxies in our sample. There is a suggestion of an inverse correlation between Eddington ratios and the ``recovered flux density''; i.e., sources with lower Eddington ratios have greater recovered flux density on parsec-scales. This anti-correlation becomes significant at the 97\% level when KISSR\,872 (an ongoing galactic merger) is excluded (using the Kendall tau test). This implies that LINERs have lower Eddington ratios and greater recovered flux density on parsec-scales. In other words, their radio outflows are mostly concentrated on parsec-scales compared to the Seyfert galaxies. 
{While this result appears intriguing at first, it is only so when the fact of the small sample size is taken into consideration. In their VLBI study, \citet{Nagar05} found Seyfert nuclei did not have detected parsec-scale jets, but had larger (100 pc- or kpc-scale) jets, while LINER nuclei had sub-parsec jets but weak or no known larger scale jets. Similarly, \citet{Baldi18} found that LINERs showed more ``core-brightened'' structures in $\le$200~mas resolution MERLIN images, compared to Seyfert galaxies, suggesting that their flux density was concentrated on parsec-scales. \citet{Nagar05} suggested that LINER jets get decollimated on parsec-scales, possibly due to jet-medium interaction. We suggest instead that the shorter jet extents in some LINERs could be indicative of shorter lived jets (see Section~\ref{secdiscuss}).}

We now present results from the emission line modeling of the SDSS spectra of these DPAGN. To put these results into perspective, we present them for the entire sample of nine sources. 

\begin{table*}[t]
\caption{Entire Sample of 9 KISSR Sources}
\begin{center}
\begin{tabular}{lccccccllcc}
\hline\hline
{Source} & {$z$}    & {Type}& {S$^\mathrm{total}_\mathrm{FIRST}$} & {S$^\mathrm{total}_\mathrm{VLBA}$} & {Recovered} & {Structure} & {Black hole} & {Eddington} & {$Q_\mathrm{jet}$} &{Ref}\\ 
{Name}   & {}                 & {}        & {mJy}     & {mJy}   & {flux \%}   & {VLBA}                & {mass (M$_\sun$)}        & {ratio}  &{erg~s$^{-1}$}   & {} \\ 
\hline
KISSR 1219 & 0.037580  & Sy 2 & 5.6$\pm$0.3  & 1.7  & 31 & J & $2.1\times10^7$ & 0.020  &$1.9\times10^{41}$&1\\ 
KISSR 618 & 0.072940 & Sy 2    & 2.5$\pm$0.3     &  0.6 & 24 & J & $4.6\times10^7$ & 0.018  &$2.9\times10^{41}$&2 \\
KISSR 434 & 0.064128 & Sy 2    & 6.0$\pm$0.3    &  0.9 & 15  & J & $1.3\times10^8$ & 0.012  &$4.9\times10^{41}$&3 \\
KISSR 1154 & 0.072032 & Sy 2  & 3.5$\pm$0.8    & 0.5 & 14 & C-J& $3.6\times10^7$ & 0.019  &$3.8\times10^{41}$&2\\
KISSR 1494 & 0.057446  & Sy 2  & 23.0$\pm$0.3 & 0.65 & 3 & C  & $1.4\times10^8$ & 0.020   &$1.3\times10^{42}$&4\\
KISSR 1321 & 0.063856 & Sy 2  & 1.6$\pm$0.2    & $<$0.03 & $<$6 &\nodata &$3.0\times10^7$&0.026&$1.6\times10^{41}$&2\\
\hline
KISSR 872 & 0.083064 & LINER & 5.2$\pm$0.3   & 2.3 & 44 &J & $4.4\times10^7$  & 0.040      &$6.7\times10^{41}$&2 \\
KISSR 102 & 0.066320 & LINER & 11.3$\pm$0.3 &  7.0 & 62 &C-J/C-C& $1.7\times10^9$ & $3.0\times10^{-4}$ &$8.9\times10^{41}$&5 \\
KISSR 967 & 0.092067 & LINER & 2.4$\pm$0.2   & 2.2 & 90 &C  & $1.4\times10^8$ & 0.0013 &$4.1\times10^{41}$&2  \\
\hline
\end{tabular}
\end{center}
{\small Column~1: Source name. Column~2: Redshift. Column~4: Total flux density in VLA FIRST image {with errors obtained from {\tt AIPS} task {\tt JMFIT}. Column~5: Total flux density in the 1.5~GHz VLBA image obtained using component entries in Table~\ref{tab2}, with typical errors being less than 5\%. Column~6: Total flux density recovered in VLBA images compared to the VLA FIRST core flux density, with typical errors being less than 10\%.} Column~7: Radio structure observed in VLBA images. J = Jet, C = Core, C-J = Core-Jet, C-C = Core-Core. Column~10: Jet kinetic power. Column~11: References for radio data: 1: \citet{Kharb17a}, 2: This paper, 3: \citet{Kharb19}, 4: \citet{Kharb15b}, 5: \citet{Kharb20}.}
\label{tab4}
\end{table*}

\subsection{Emission Line Properties of Entire Sample}
\label{sec:line}
{The observed SDSS DR12 spectra of all galaxies are corrected for reddening using the E(B-V) value from \citet{Schlegel98}. The best fit model for the underlying stellar population is obtained using the penalized pixel-fitting stellar kinematics extraction code, pPXF \citep{Cappellari04,Cappellari17} and subtracted from the observed reddening corrected spectra of these galaxies. During this procedure the stellar velocity dispersions of these galaxies are also obtained. The pure emission line spectra are analyzed to obtain the parameters of the emission lines. Details of the line-fitting procedure have been provided earlier in \citet{Kharb17a, Kharb19}. For all the sample galaxies, each component of the [S~{\sc ii}] doublet required two components for a satisfactory fit (i.e., the reduced $\chi^2$ improved by $\ge$20\% compared to a single component fit). The [S~{\sc ii}] model was used as a template to fit the narrow H$\alpha$, H$\beta$, H$\gamma$ and [N~{\sc ii}] doublet lines. As the [O~{\sc iii}] profile does not typically match with that of [S~{\sc ii}] \citep{Greene05}, the [O~{\sc iii}] and [O~{\sc i}] doublets are modeled independently using two Gaussian component models. {\tt IDL's MPFIT} program for non-linear least-square optimization was used to fit the emission line profiles with Gaussian components and to obtain the best fit parameters and associated errors. These fits are reported in Tables~\ref{tabline1} to \ref{tabline5}. Figures~\ref{fig5} to \ref{fig9} show the pure emission line spectra, different emission line regions and the best fit for the emission lines.}

Global source properties like star-formation rates (SFR) were estimated using all components of the H$\alpha$ line and the relation of \citet{Kennicutt98}, bolometric luminositites (L$_\mathrm{bol}$) were estimated using the [O~{\sc iii}] $\lambda5007$ line components and the relation from \citet{Heckman04} and black hole masses were estimated using the stellar velocity dispersion ($\sigma_\star$) from spectral line fitting and the M$_\mathrm{BH}-\sigma_\star$ relation from \citet{McConnell13}. These properties are presented in Tables~\ref{tab3} and \ref{tab4}.

We explored the physical conditions of the NLR by comparing the observed line ratios with those predicted by the plasma modeling code, MAPPINGS III \citep{Sutherland93}. This is a shock and photoionization modeling code that can predict emission line spectra of low and medium density emission line nebulae when exposed to an external source of radiation. 
{The code assumes a plane-parallel isochoric model for the gas nebula; the plane-parallel model assumes that the nebula is an infinitely long and wide slab of gas. }
The gas is specified with an initial constant hydrogen gas density. The single ionizing source models that we use are AGN photoionization (dusty or dust-free) and shock heating (shock-only or shock+precursor). {We note that between the two fitted line components, we have chosen the stronger line component with the higher peak flux for the MAPPINGS III analysis.}

The IDL Tool for Emission-line Ratio Analysis \citep[ITERA;][]{Groves2010} was used for generating the line ratio diagrams used for MAPPINGS III. {These line ratios have the benefit of small separation in wavelengths, thereby reducing the effects of reddening. The predicted line ratios are then compared with the observed line ratios to understand the source of ionization of the NLR gas.} The source of ionization in AGN follows a non-thermal power-law; the strength of the photon distribution is determined by the ionization parameter (U) and the shape of the distribution is set by the power-law index, $\alpha$. These two variables are used on the grid run for the AGN model. U is varied from $10^{-4}$ to 1.0 and $\alpha$ is varied from $-2.0$ to $-1.2$. 

{When ionizing radiation (extreme UV and soft X-ray photons) generated by the cooling of hot gas behind a shock front, creates a strong radiation field leading to significant photoionization, the shock+precursor model comes into play \citep{Allen2008}. This model takes as input the abundance and pre-shock gas densities (n) and generates a grid of emission line ratios for a range of shock velocities and magnetic parameters defined as $B/\sqrt{n}$. For the shock heating models, shock velocities range from 100 km~s$^{-1}$ to 1000 km~s$^{-1}$ while the magnetic parameter varies from $10^{-4}$ to 10 $\mu$G~cm$^{3/2}$.} The shock velocities determine the shape of the ionizing spectrum, while the magnetic field controls the compression of the gas, effectively controlling the ionization parameter of the post-shock recombination zone. 

As the estimated gas density using the [S {\sc ii}] line typically lies in the range $\approx100-200$~cm$^{-3}$ for all the sources, a gas density of 100~cm$^{-3}$ and solar metallicity have been used for all sources to generate the line ratio plots. The model generated output spectrum are displayed in the form of line ratio diagrams using [O {\sc iii}]/H$\beta$, [S {\sc ii}]/H$\beta$, [N {\sc ii}]/H$\alpha$ in Figures~\ref{fig10} to \ref{fig14}. As seen in Figure~\ref{fig12}, meaningful conclusions cannot be drawn about the best-fit model in KISSR\,967, because the errors on the line (e.g., [O {\sc iii}]) flux measurements are too high. %The estimated gas density using the [S II] line is $\sim$180~cm$^{-3}$ in KISSR\,967.

\subsubsection{AGN Photoionization Model}
The AGN photoionisation model, either dusty or dust-free, works for all the sources whose spectra could be analysed successfully with MAPPINGS III, viz., KISSR\,618, KISSR\,872, KISSR\,1154 and KISSR\,1321. {The grid is generated for solar metallicity. Super solar metallicities do not reproduce most of the line ratios. To explain the [N II]/H$\alpha$ misfit, a higher nitrogen abundance is traditionally invoked in the literature; this may hold true for our sample sources as well. Other solutions involve internal micro-turbulent clouds that are dissipative in nature. These dissipative turbulent clouds convert the turbulent motion into heat thereby increasing the electron temperature and the nitrogen line emissivity 
\citep{Bottorff00,Kraemer07,Mignoli19}.} 
However, as described ahead, models other than AGN photoionisation also fit some of the sources. As presented in \citet{Kharb19}, the dusty AGN photoionization model fits the data well for the Seyfert galaxy, KISSR\,434, while the shock+precursor model or the shock-only model is not a good fit. This was surprising because KISSR\,434 exhibits a $\sim$150~parsec long curved jet in its 1.5 GHz VLBA image. A possibly precessing jet could be stirring up NLR clouds, causing the emission line peak splits; it is however not shock-ionising the NLR gas. The dusty AGN photoionization model fits the data well also for the Seyfert galaxy, KISSR\,1494, while the shock+precursor model does not \citep{Kharb15b}. This source does not reveal a jet in VLBA images; rather a weak steep-spectrum radio component detected here is suggested to be the base of a coronal wind.

\subsubsection{Shock-only Model}
A dust-free AGN photoionization and shock-only model with velocities between 100$-$300~km~s$^{-1}$ fit the observed emission-line properties of the Seyfert galaxy KISSR\,1154. The gas density estimated from the Sulphur lines is $\sim$400~cm$^{-3}$ and the temperature is 10$^4$ K \citep[using the PyNeb software;][]{Luridiana15}. The shock+precursor model requires a higher value of magnetic fields in the medium, making it unrealistic. Its VLBA image shows a ``core'' and a possible jet knot $\sim$50~parsec away. Similarly, for KISSR\,1321, a dusty AGN photoionization model as well as the shock-only model work. The estimated gas density using the [S II] line is $\sim$110~cm$^{-3}$. It is interesting though that this source is not detected with the VLBA. It remains unclear if there is a parsec-scale radio outflow in the source. For the LINER and dual-optical-nuclei galaxy, KISSR\,102, the shock-only model works best, while the AGN photoionization model underpredicts the observed line ratios \citep{Kharb20}. Its VLBA images reveal three compact radio cores, while its host galaxy appears to be in late-stage merger, consistent with a turbulent galactic environment.

\subsubsection{Shock + Precursor Model}
The dusty AGN photoionization model as well as a shock+precursor model with 300$-$500 km~s$^{-1}$ shock velocities fits the line data well in the Seyfert galaxy, KISSR\,618. The gas density is $\sim$150~cm$^{-3}$ and the temperature is 10$^4$ K. The 1.5~GHz VLBA image reveals a $\sim$30~parsec long radio jet. Similarly, the dust-free AGN photoionization and shock+precursor model with shock velocities between 200$-$300 km~s$^{-1}$ is able to reproduce the lines (within error bars) for the LINER and merging galaxy with two optical nuclei, KISSR\,872. This source exhibits a $\sim$200~parsec long jet in its VLBA image. While the dusty AGN photoionization model fits the data well (although not all lines), the shock+precursor model fits the lines better in the Seyfert galaxy, KISSR\,1219 \citep{Kharb17a}. Its VLBA image reveals a $\sim$70~parsec long radio jet.

\subsection{The Double Optical Nuclei Galaxy KISSR\,872}\label{sec872}
The double nuclei associated with the LINER galaxy KISSR\,872 (N1 and N2 in Figure~\ref{fig1}) turn out to be two individual galaxies at around the same redshift, undergoing a merger. The projected separation of optical centres N1 and N2 is $\sim$7.25~kpc; a tidal tail is observed to the north. While KISSR\,872 is a $\sim10^{11}$ solar mass galaxy, the second galaxy is a $\sim10^{10.5}$ solar mass galaxy that falls in the composite region of the BPT diagram \citep{Baldwin81} \citep[stellar masses are from the SDSS database which are calculated using the method of][]{Chen12}. The black hole mass in this galaxy is $9.3\times10^6$~M$_\sun$ from the M$_\mathrm{BH}-\sigma_\star$ relation using $\sigma_\star=$ 121.24~km~s$^{-1}$ \citep{Thomas13}. No radio emission is detected from the optical nucleus N2 in the FIRST image.

In the 1.5~GHz VLBA image, the core-jet (C to J3, Figure~\ref{fig3}) extent is nearly 200 parsec (135~mas) at a PA of $-145\degr$. The jet appears to be remarkably straight along its length. The distance between components C to J2 (see Figure~\ref{fig2}) is $\sim60$~parsec. At 4.9~GHz, only the inner $\sim$10 parsec jet is detected (C and J0 in the left panel of Figure~\ref{fig3}). The $1.5 - 4.9$~GHz spectral index image is presented in Figure~\ref{fig4}. The average spectral index is around $-0.71\pm0.26$ in the core-inner-jet region (C$-$J0 in the right panel of Figure~\ref{fig3}) and  $-0.95\pm0.33$ in the outer jet region (J1 in the right panel of Figure~\ref{fig3}).
  
The width of one of the components of the $[\mathrm {O~III}]$ lines is relatively broad in the SDSS spectrum of KISSR\,872 (see Table~\ref{tabline2}). Also, the peak of the broad component is redshifted with respect to the other component. This could be a signature of inflowing gas. The significance of KISSR\,872 being in the middle of a galactic merger with a second supermassive black hole only $\sim$7 kpc away from the primary, and the fact that it possesses one of the longest parsec-scale radio jets in our sample of nine sources, needs to be explored further. The ongoing galaxy merger may be driving gas inwards into KISSR\,872, influencing the accretion rate on to its primary black hole and consequently its radio outflow.

\section{Discussion}\label{secdiscuss}
We have looked for radio emission on parsec-scales in nine type 2 Seyfert and LINER galaxies that show double-peaked emission lines in their optical spectra. While we detect parsec-scale radio emission in eight of the sources, in only one (KISSR\,102) have we found indications of dual accreting black holes \citep[see][]{Kharb20}. The detection rate of parsec-scale emission for KISSR sources with FIRST core flux density $\ge2.4$~mJy is 100\%. We have detected jet-like features in the majority (5/8 = 63\%) of the sources. This is consistent with results in the literature for larger samples of LLAGN \citep[e.g.,][]{Baldi18}. 

It is probable that jet-NLR interaction is the primary reason for the splitting of the emission line peaks in these sources. Indeed, from the MAPPINGS III emission-line modeling, it appears that in sources possessing  parsec-scale radio jets, the shock+precursor model can explain the observed line ratios, consistent with the idea of jet-medium interaction. The only exception is KISSR\,434 where a curved $\ge$150 parsec long radio jet is observed with VLBI but the shock+precursor model does not fit the emission line data. Sources whose emission line ratios are consistent with the shock-only model, either show compact radio emission in VLBA images or reside in a host galaxy that is undergoing a galactic merger. The exception in this case is KISSR\,872, where a $\ge$200 parsec long jet is observed and the shock+precursor model fits the emission line ratios best. 

{We find that the double peaks of the emission lines are typically separated by velocities of $\sim100-300$~km~s$^{-1}$ for most sources (Tables~\ref{tabline1} to \ref{tabline5}). Widths of the lines correspond to velocities of $\sim100-200$~km~s$^{-1}$ except for the [O {\sc iii}] lines of KISSR\,872 which is around $\sim600$~km~s$^{-1}$. Typical velocity differences between the lines due to ``outflows'' have been found to be $\sim300 - 600$~km~s$^{-1}$ with widths of the outflow components being $\sim200 - 800$~km~s$^{-1}$ \citep{Karouzos16}. Therefore, except for the [O {\sc iii}] lines of KISSR\,872, none of the lines in the other sources show typical signatures of outflowing emission line gas. In the case of KISSR\,872, the wide component of [O {\sc iii}] is in fact redshifted, suggesting an inflow (see Section~\ref{sec872}). The observed velocities in the sample sources are much smaller, by factors of several hundred, than the expected jet speeds. This therefore suggests that the emission line gas could be pushed in a direction lateral to the jet \citep[e.g.,][]{Kharb17a} or could arise in wider, slower-moving winds around the jets. Indeed, nested biconical outflows have been invoked to explain the origin of double-peaked emission lines in low luminosity AGN by \citet{Nevin16,Nevin18}. Importantly, wide wind-like outflows can be efficient agents of AGN feedback \citep[see the review by][]{Harrison18}.}

We note that the spatial scales sampled by the SDSS fiber with a diameter of 3$\arcsec$, corresponding to $\sim3 - 6$~kpc at the distance of the KISSR sources, are much larger than the $\sim$100 parsec-scale VLBA jets. However, as \citet{Schmitt03a,Schmitt03b} have noted, the NLR ranges from a few 100 parsecs to a few kpc in Seyfert galaxies. The emission-line modeling is consistent with the idea that radio-quiet AGN are energetically capable of influencing their parsec- and kpc-scale environments, making them agents of ``radio AGN feedback'', similar to radio-loud AGN like FRI radio galaxies \citep[e.g.,][]{DeYoung10}.

When jet-like features are observed in our sample LINERs and Seyferts, they always appear to be one-sided. One-sided radio jets are typically understood to arise due to relativistic motion of jet plasma and subsequent Doppler boosting/dimming effects in the approaching/receding jets. If this remains true in our sample sources, then jet speeds ranging from $0.003c$ to $0.4c$ are implied for these sources. Here we have assumed the jet inclinations to be $\ge50\degr$, commensurate with their type 2 classification, and expected torus half-opening angles \citep[e.g.,][]{Simpson96}. 
Alternately, free-free absorption of the counterjet radio emission from intercloud NLR gas, rather than the NLR clouds themselves which have small covering factors, could be instrumental. 
Multi-epoch VLBI observations which can detect proper motions in jet components, will be essential to confirm the mechanism for the jet one-sidedness.

We find that the LINER galaxies in our KISSR sample have greater recovered flux densities on parsec-scales. That is, their radio outflows are mostly concentrated on parsec-scales compared to the Seyfert galaxies in our sample. Interestingly, \citet{Baldi18} have found that the LINERs show radio structures that are more ``core-brightened'' than Seyferts in their LeMMINGs sample, broadly consistent with our results. As their study was carried out using the eMERLIN at 1.5~GHz with a resolution of $\le$200 mas, their ``cores'' would in principle have included the milli-arcsecond-scale jets that we observe with the VLBA.

\citet{Nagar05} have suggested that LINER galaxies may be in the ``low/hard'' state (low Eddington ratios and a greater ability to launch collimated jets), in analogy to Galactic black hole candidates, compared to Seyfert galaxies which were in a ``high'' state (higher Eddington ratios and less likely to launch collimated jets). Our results may be suggesting that the size of the launched jets in sources with lower Eddington ratios (i.e., LINERs) are smaller than the jets launched in sources with higher Eddington ratios (i.e., Seyferts), or that the latter may have less collimated or more diffuse jets (like in FRI radio galaxies on kpc-scales), so that VLBI observations pick up only a small fraction of the flux density. We note that \citet{Orienti10} have suggested that the ``missing flux'' in Seyfert galaxies could be emission from jets that lost collimation due to interaction with the surrounding medium, broadly consistent with the latter suggestion. Therefore, while LINERs may possess a greater ability to launch collimated jets, these jets may not be long lived to extend to 10s of kpc-scales. {This idea is similar to the one proposed by \citet{Sanders84} for explaining the shorter jets in Seyfert 1s compared to Seyfert 2s; they suggested that shorter jets were a result of smaller accreted masses in the former, leading to shorter-lived accretion disks in them \citep[although we now agree that unification schemes can more easily explain the jet extent differences between Seyfert 1s and 2s as jet fore-shortening due to smaller inclinations in Seyfert 1s;][]{Antonucci93}. }

Using empirical relations from \citet{Punsly11} and \citet{Merloni07}, we have estimated the jet kinetic power, $Q_\mathrm{jet}$, for the KISSR sources. We find that their $Q_\mathrm{jet}$ lies in the same range as other lower luminosity radio galaxies in the literature \cite{Mezcua14}; in KISSR\,872, the $Q_\mathrm{jet}$ lies in the range observed for FRI radio galaxies \citep{Rawlings91}. \citet{Baldi20} have also found indications that LINERs show FRI-like core-brightened radio structures in their LeMMINGs-II sample. Similarly, \citet{Kharb14a} find an overlap in the radio loudness parameters of Seyferts and FRIs. These results underscore the idea that jets in LINER and Seyfert galaxies are not distinctly different from FRI radio galaxies in terms of radio or kinetic power, undermining the existence of a clear radio-loud/radio-quiet divide. Recently, \citet{Kunert20} have reported a rapid transition from the radio-quiet to the radio-loud mode in the quasar 013815+00, in keeping with this idea. More examples of the same have recently been presented by \citet{Nyland20} {and \citep{Wolowska21}.}

However, we do not find a clear correlation between $Q_\mathrm{jet}$ and jet extent. This correlation has been reported for a small sample of LLAGN by \citet{Mezcua14}. Barring the possibility of the efficiency factor ($\eta$), that allows for work done on the external medium and enters into the calculation of $Q_\mathrm{jet}$, being different in radio-quiet AGN compared to radio-loud AGN, we may simply be suffering from inadequate sensitivity in our VLBA images. 

\section{Summary}
This paper completes our dual-frequency (1.5 and 4.9~GHz) phase-referenced VLBA study of nine double-peaked emission-line AGN, five of which are Seyfert galaxies and three of which are LINERs. Here, we have presented the radio and emission-line data of five of these nine sources, viz., KISSR\,618, KISSR\,872, KISSR\,967, KISSR\,1154 and KISSR\,1321. The primary results of this study are:

\begin{enumerate}
\item Of the five DPAGN all but one source, viz., KISSR\,1321, are detected at one or both of the frequencies with the VLBA. One-sided core-jet features are detected in all four sources, although the jet identification is not robust in the case of KISSR\,967.

\item Doppler boosting/dimming effects with parsec-scale jet speeds varying between $0.003c$ and $0.4c$ can explain the jet-sidedness ratios for jet inclinations of $\ge50\degr$ in these five DPAGN. If the counterjet emission is free-free absorbed however, the medium could be the intercloud NLR gas, rather than the NLR clouds themselves, which have low covering factors. 

\item For the nine sources, we detect dual parsec-scale cores which could be the signature of binary supermassive black holes, only in the case of KISSR\,102 \citep{Kharb20}. Jet-NLR interaction instead, appears to be the primary reason for the splitting of the emission line peaks in these sources. 

\item Comparing the total flux density recovered on parsec-scales with the VLBA, to the kpc-scale flux density in the FIRST images, we find a weak inverse correlation between Eddington ratios and ``recovered flux density'' on parsec-scales; the sizes of the launched jets in sources with lower Eddington ratios (i.e., LINERs) are smaller than the jets launched in sources with higher Eddington ratios (i.e., Seyferts). {This could be indicative of shorter lived jets in some LINERs compared to Seyferts.} Alternately, the latter have less collimated jets, so that VLBA picks only a small fraction of their flux density. 

\item We find that the jet kinetic powers, $Q_\mathrm{jet}$, in some KISSR sources overlap with those observed in radio-loud FRI radio galaxies, undermining a clear radio-loud/radio-quiet divide. We do not however find that sources with larger $Q_\mathrm{jet}$ values have longer jets. Future higher sensitivity VLBI observations or high resolution ($\theta\sim0.2\arcsec$) VLA observations will be able to re-look for this correlation in the KISSR sample.

\item Emission-line modeling using the MAPPINGS-III code indicates that for the Seyferts and LINERs exhibiting parsec-scale radio jets, the shock+precursor model can explain the observed line ratios, consistent with the idea of jet-medium interaction in them. Jets in radio-quiet AGN are therefore energetically capable of influencing their parsec- and kpc-scale environments, making them agents of ``radio AGN feedback'', similar to radio-loud AGN like FRI radio galaxies.
\end{enumerate} 

\acknowledgments
{{We thank the referee for their insightful suggestions which have significantly  improved this paper.} SS acknowledges support from the Science and Engineering Research Board, India through the Ramanujan Fellowship. The National Radio Astronomy Observatory is a facility of the National Science Foundation operated under cooperative agreement by Associated Universities, Inc. We acknowledge the support of the Department of Atomic Energy, Government of India, under the project 12-R\&D-TFR-5.02-0700. This research has made use of the NASA/IPAC Extragalactic Database (NED), which is funded by the National Aeronautics and Space Administration and operated by the California Institute of Technology.
Funding for the SDSS and SDSS-II has been provided by the Alfred P. Sloan Foundation, the Participating Institutions, the National Science Foundation, the U.S. Department of Energy, the National Aeronautics and Space Administration, the Japanese Monbukagakusho, the Max Planck Society, and the Higher Education Funding Council for England. The SDSS Web Site is http://www.sdss.org/.}

\appendix
\section{Appendix information}
The SDSS emission line fitting and modeling results from MAPPINGS III.

\begin{figure*}
\centering{
\includegraphics[width=10cm,trim=100 280 100 100]{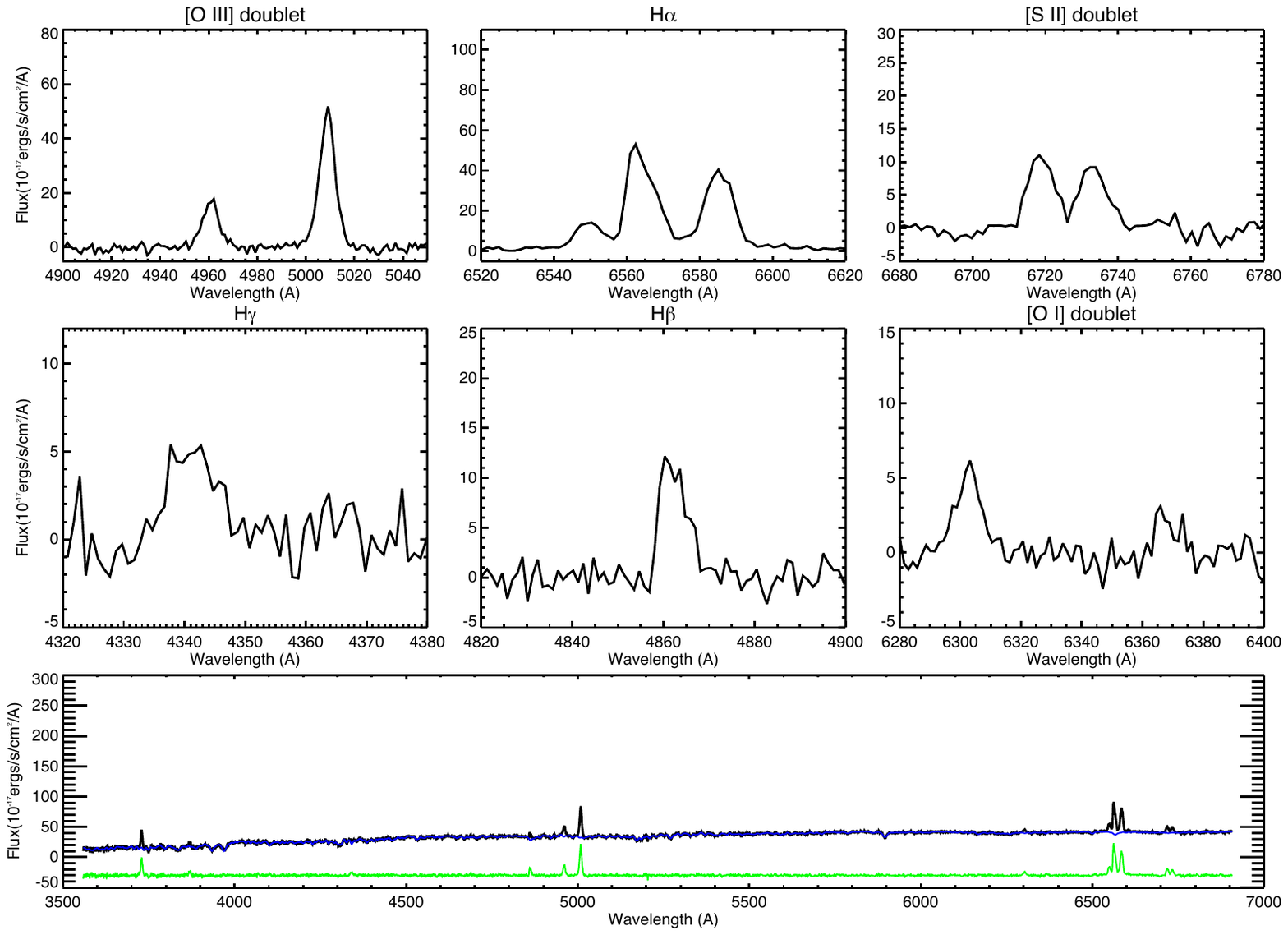}
\includegraphics[width=10.5cm,trim=100 380 100 220]{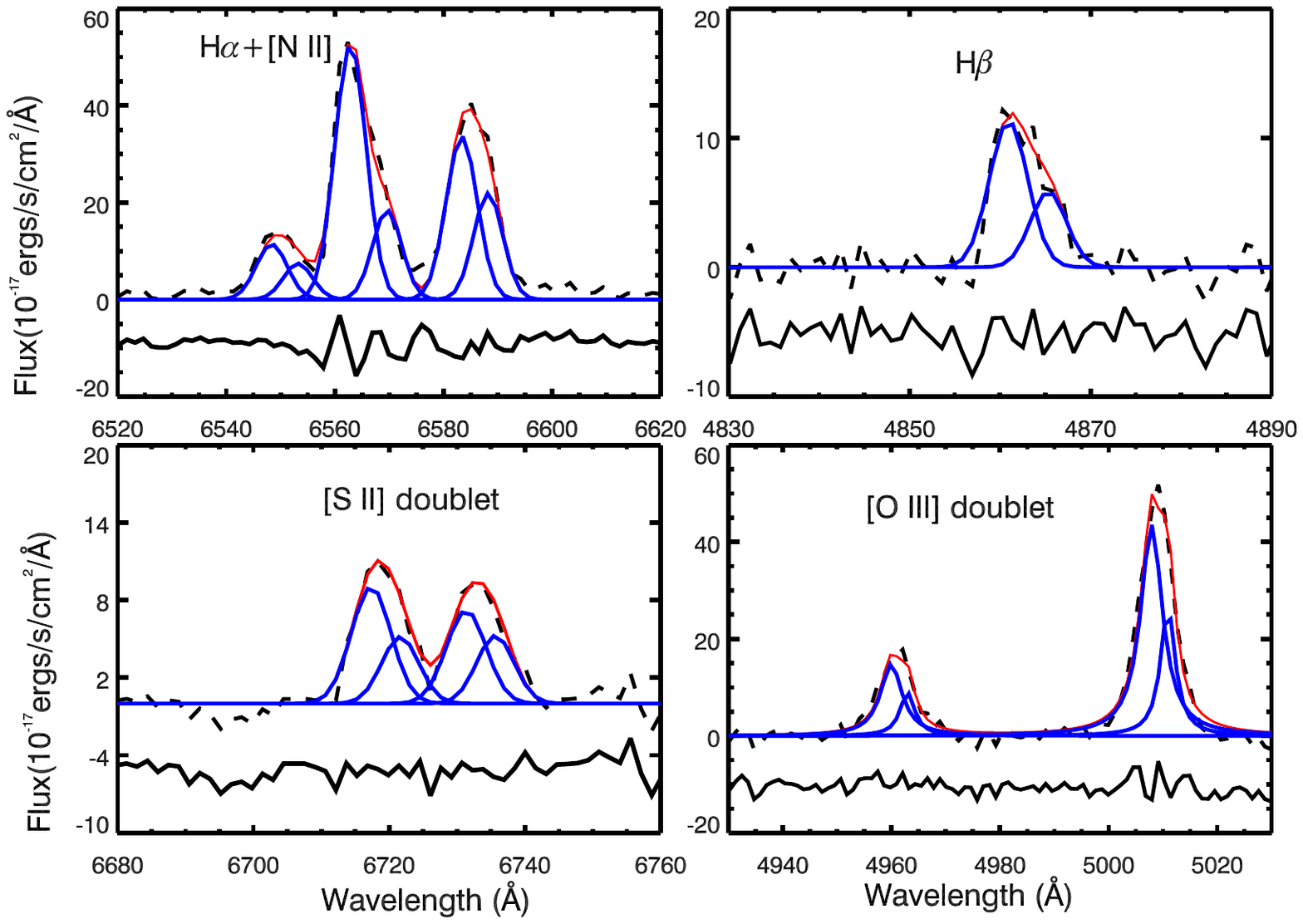}}
\caption{\small SDSS spectrum of KISSR\,618 showing double peaks in the de-reddened H$\alpha$, H$\beta$, H$\gamma$ and other emission lines is presented in the top two panels. The middle panel shows the entire spectrum along with the fitted stellar continuum in blue and the pure emission-line spectrum in green, shifted down for better visualization. The bottom two panels show the individual fitted Gaussian components in blue to the actual lines in black (dashed lines) and the total fit in red; residuals are shown at the bottom of the panels in black solid line. See Section~\ref{sec:line} for details.}
\label{fig5}
\end{figure*}

\begin{figure*}
\centering{
\includegraphics[width=12.4cm,trim=50 370 50 50]{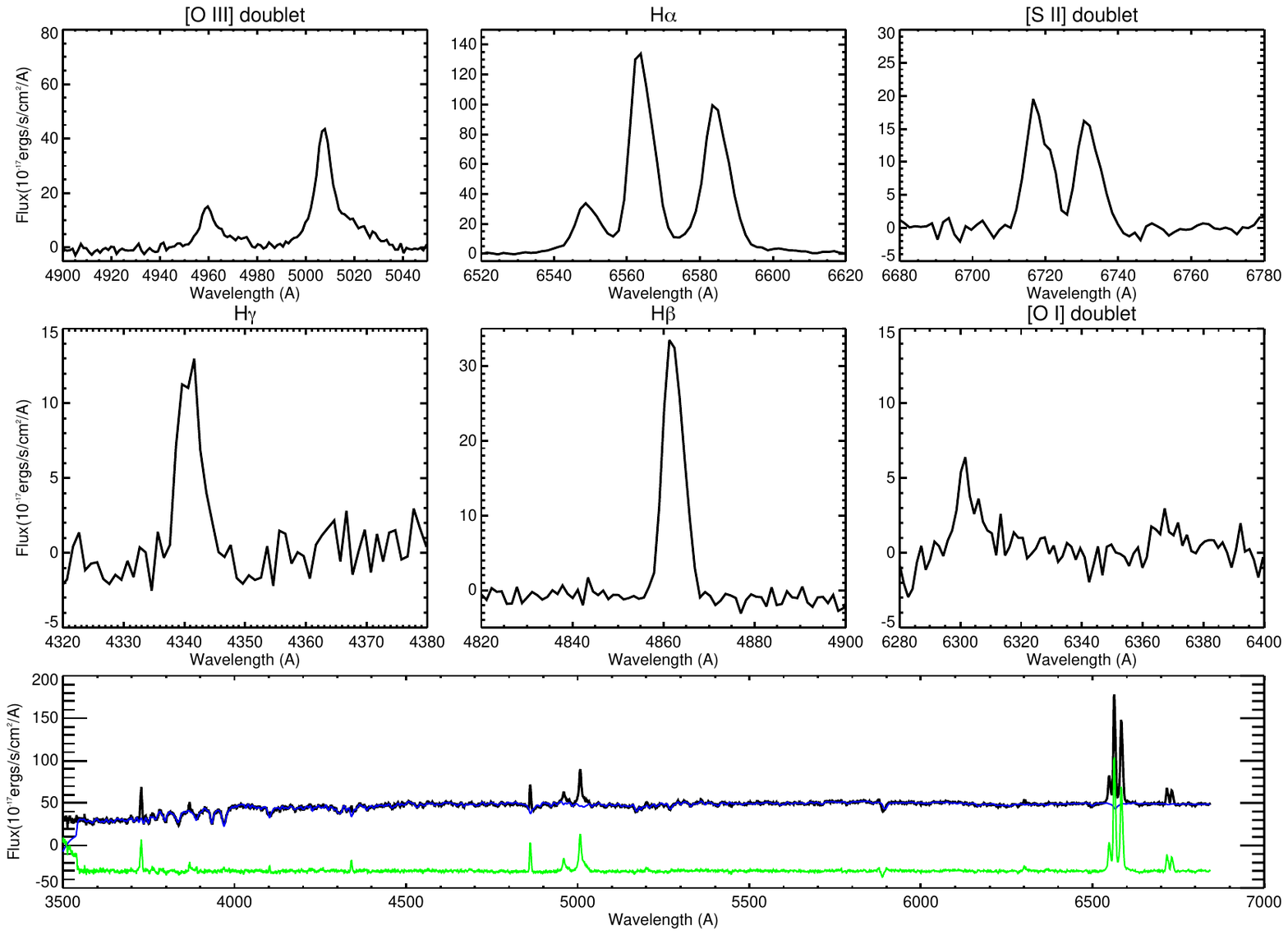}
\includegraphics[width=10.5cm,trim=100 340 100 70]{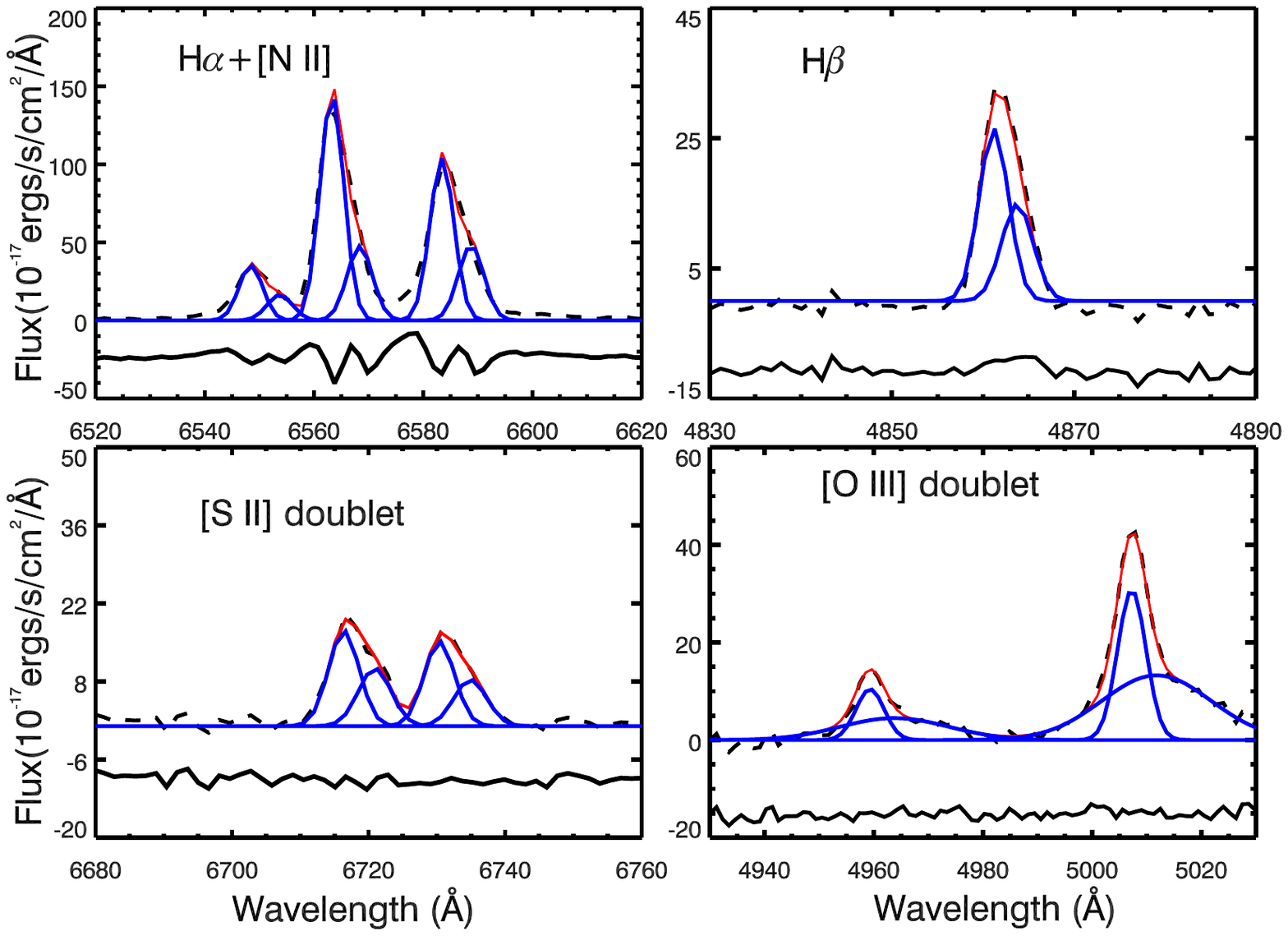}}
\caption{\small SDSS spectrum of KISSR\,872. Details same as in Figure~\ref{fig5}.}
\label{fig6}
\end{figure*}

\begin{figure*}
\centering{
\includegraphics[width=10cm,trim=100 280 100 100]{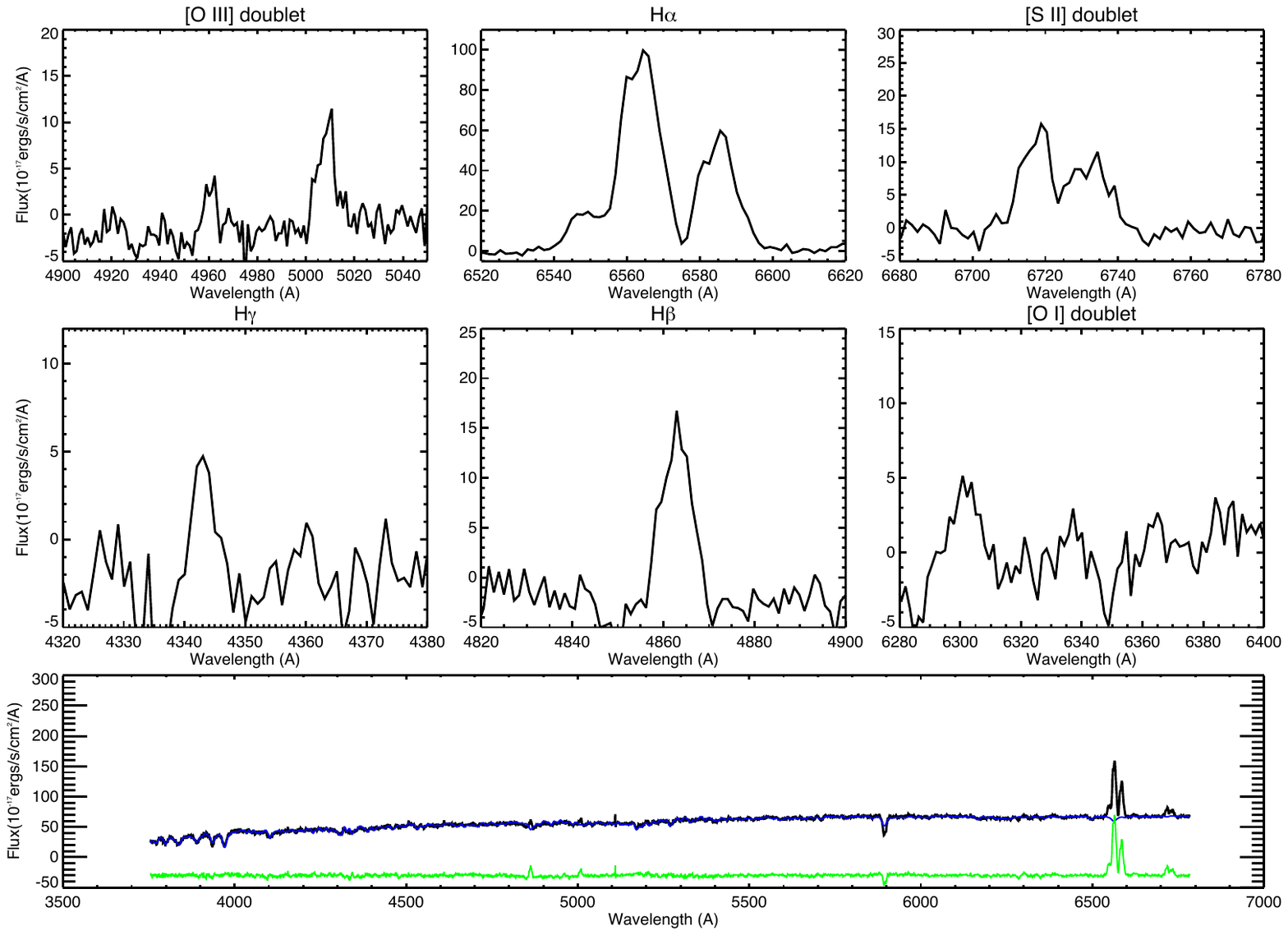}
\includegraphics[width=10.5cm,trim=100 380 100 220]{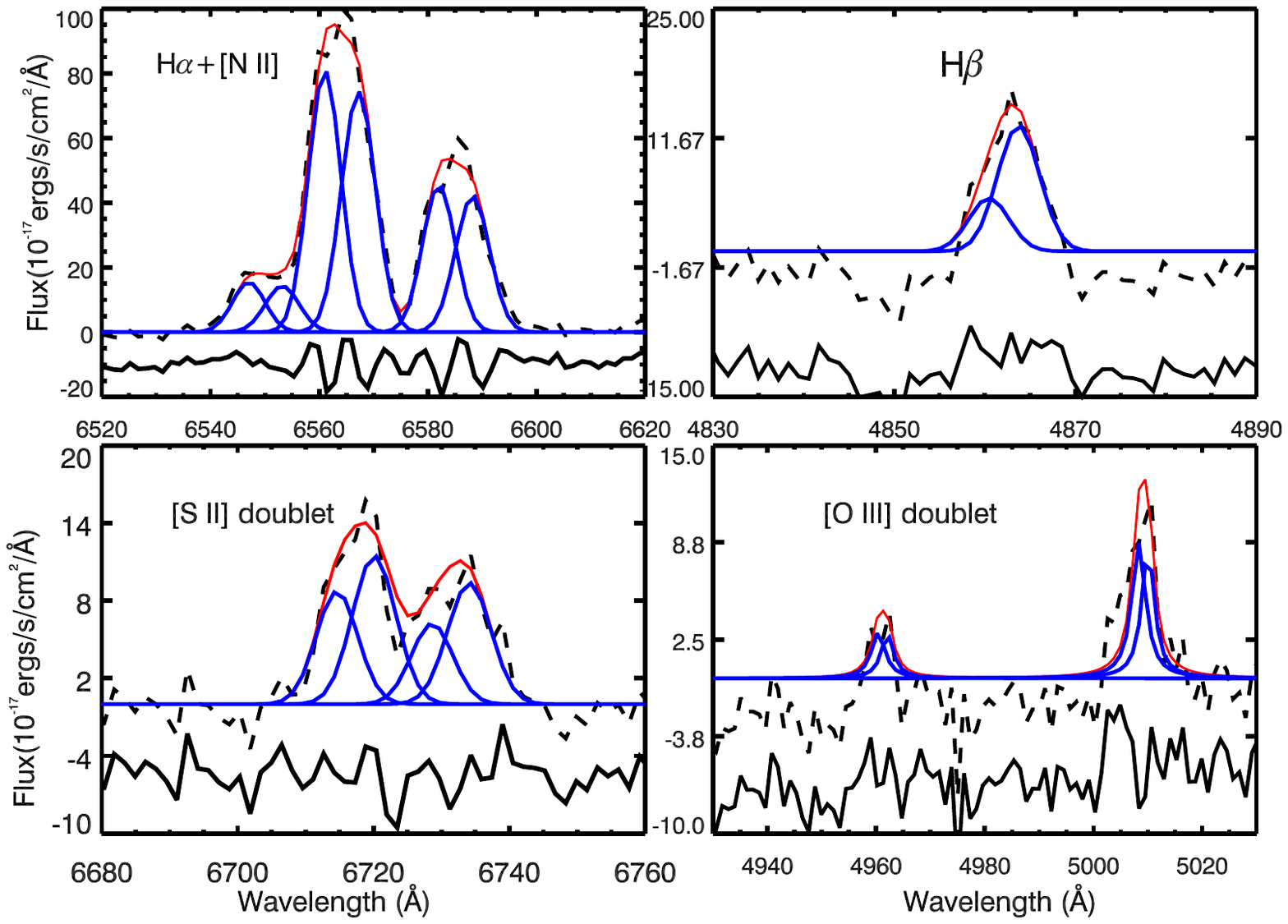}}
\caption{\small SDSS spectrum of KISSR\,967. Details same as in Figure~\ref{fig5}.}
\label{fig7}
\end{figure*}

\begin{figure*}
\centering{
\includegraphics[width=10cm,trim=100 280 100 100]{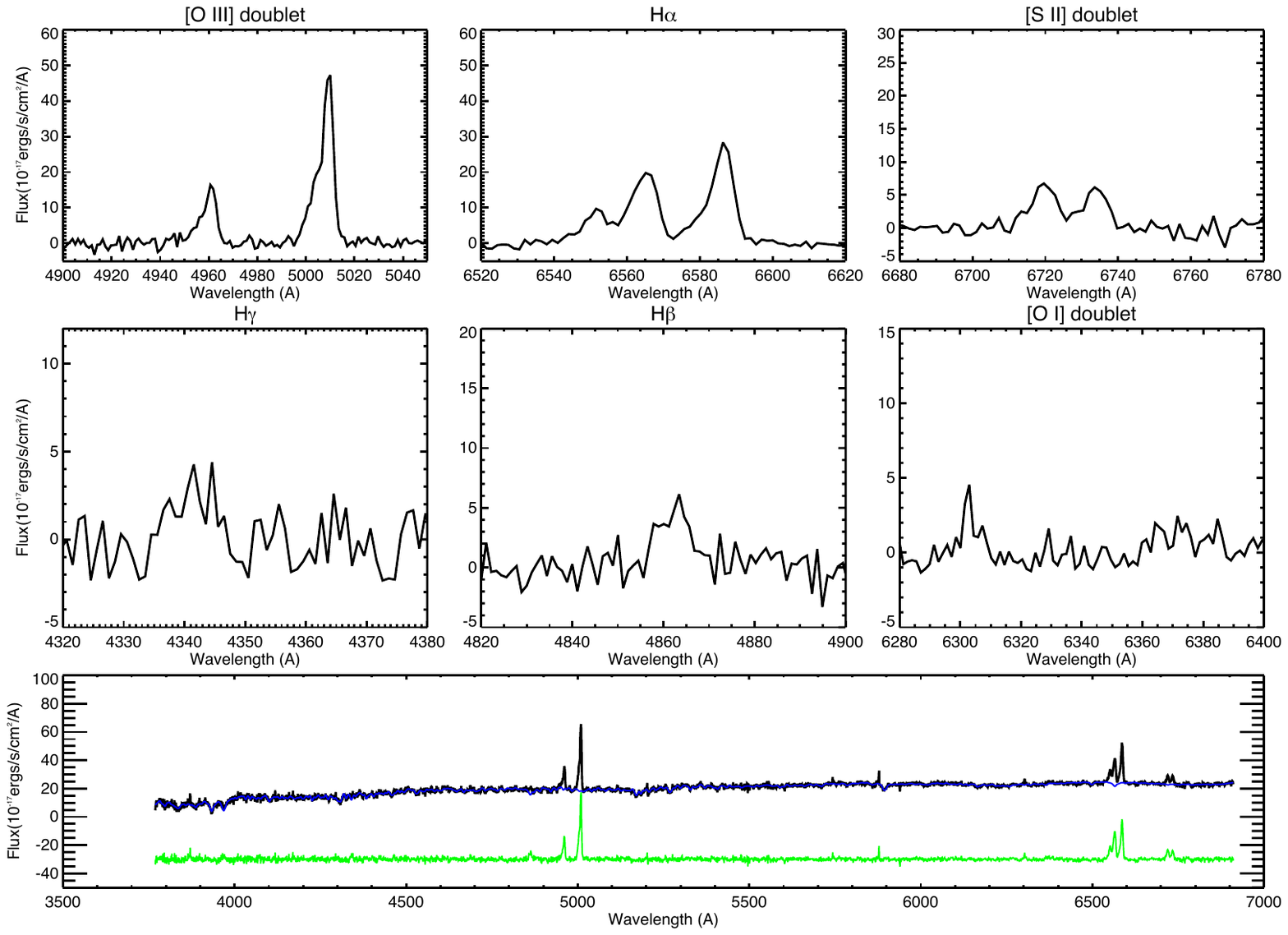}
\includegraphics[width=10.5cm,trim=100 380 100 220]{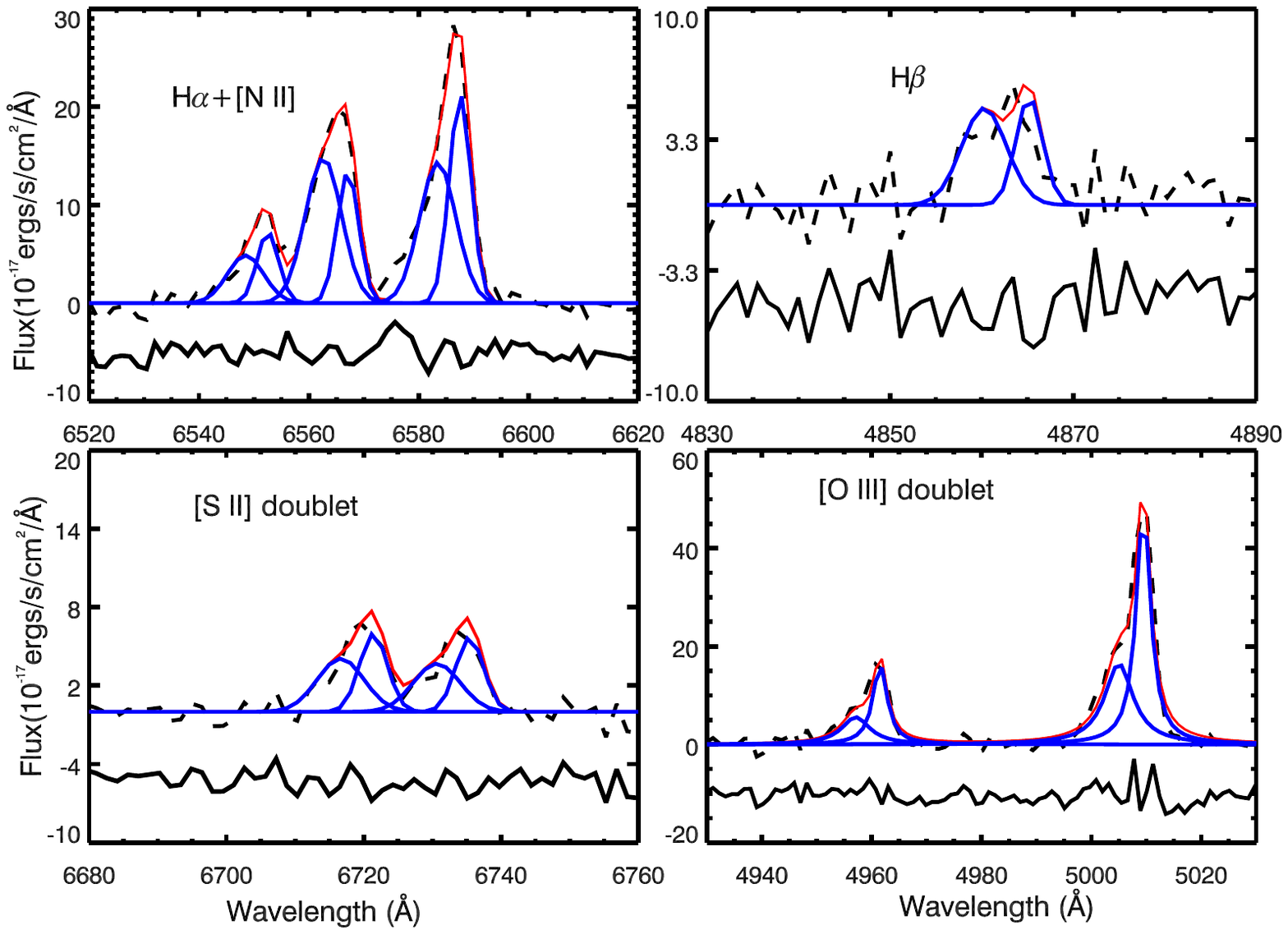}}
\caption{\small SDSS spectrum of KISSR\,1154. Details same as in Figure~\ref{fig5}.}
\label{fig8}
\end{figure*}

\begin{figure*}
\centering{
\includegraphics[width=10cm,trim=100 280 100 100]{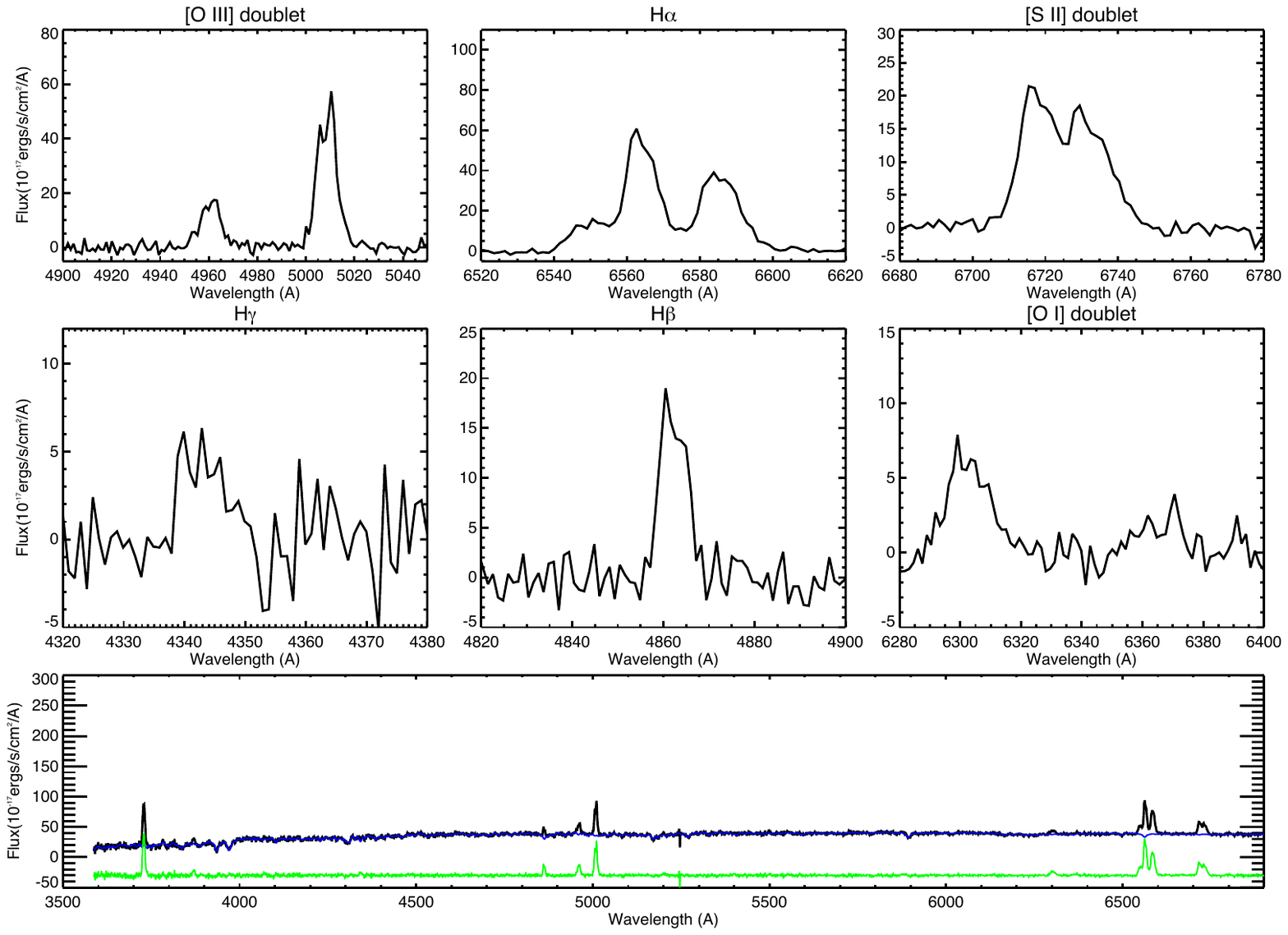}
\includegraphics[width=10.5cm,trim=100 380 100 220]{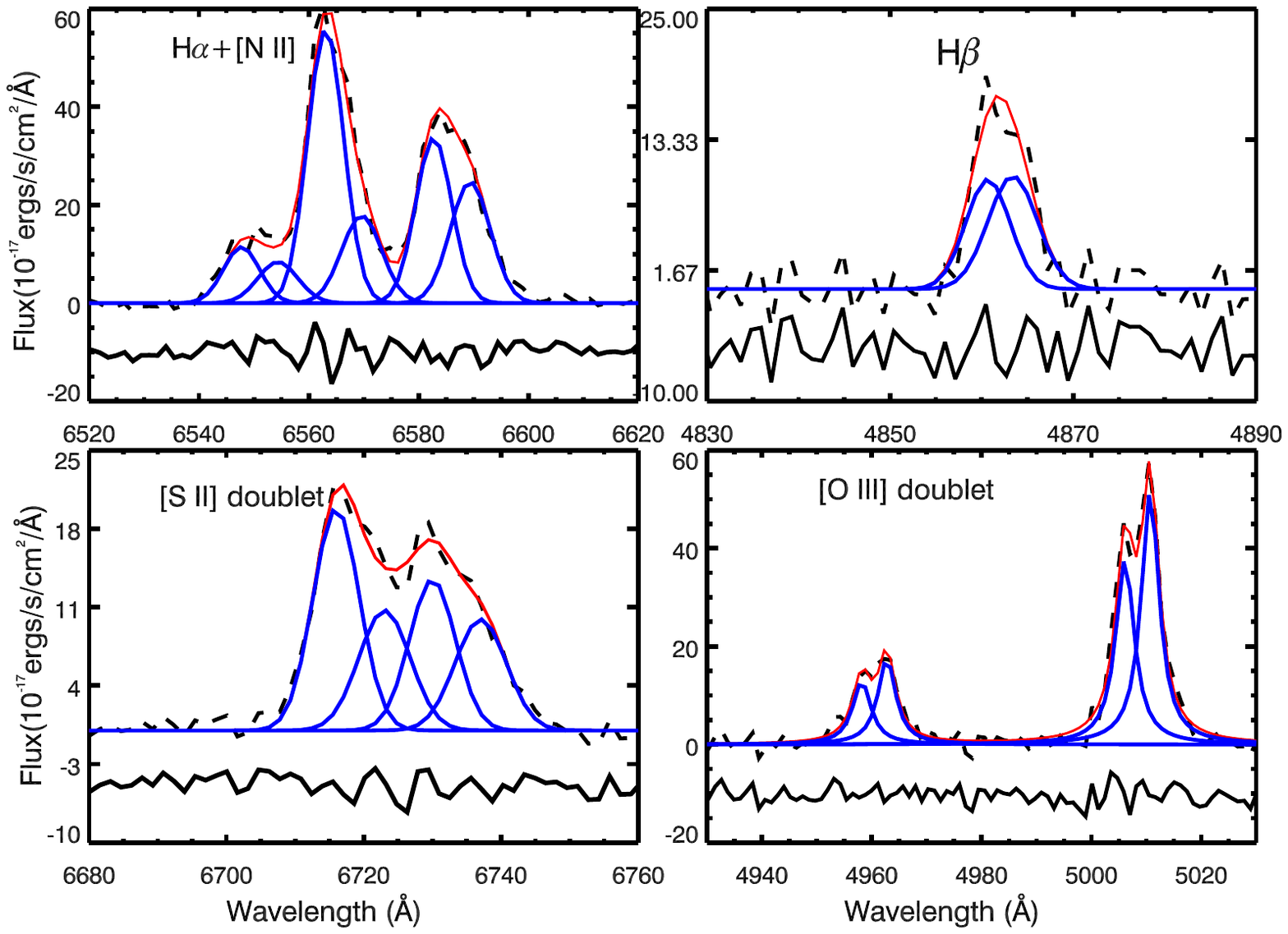}}
\caption{\small SDSS spectrum of KISSR\,1321. Details same as in Figure~\ref{fig5}.}
\label{fig9}
\end{figure*}

\begin{figure*}
\centering{
\includegraphics[width=7cm,trim=0 310 0 200]{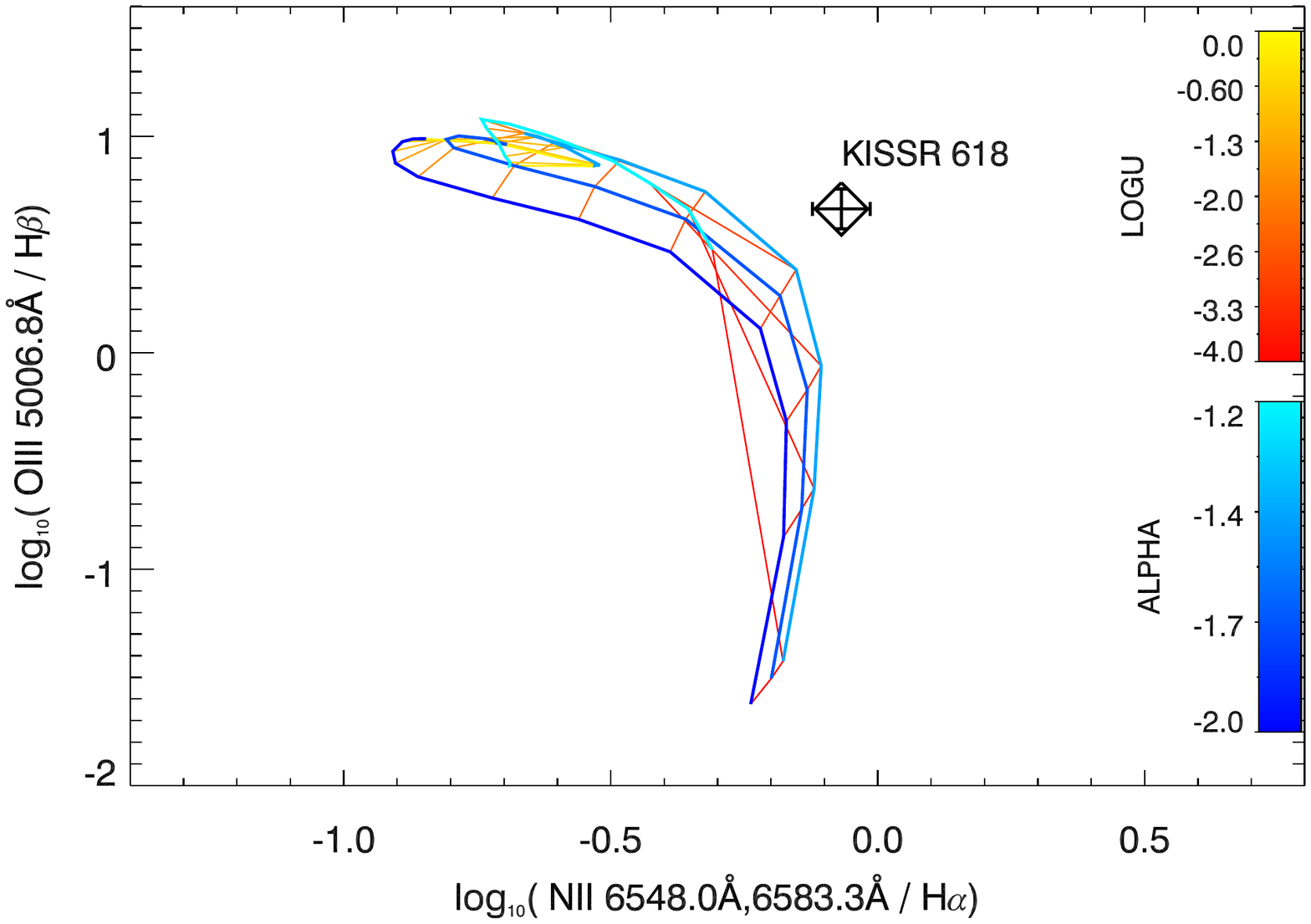}
\includegraphics[width=7cm,trim=0 310 0 200]{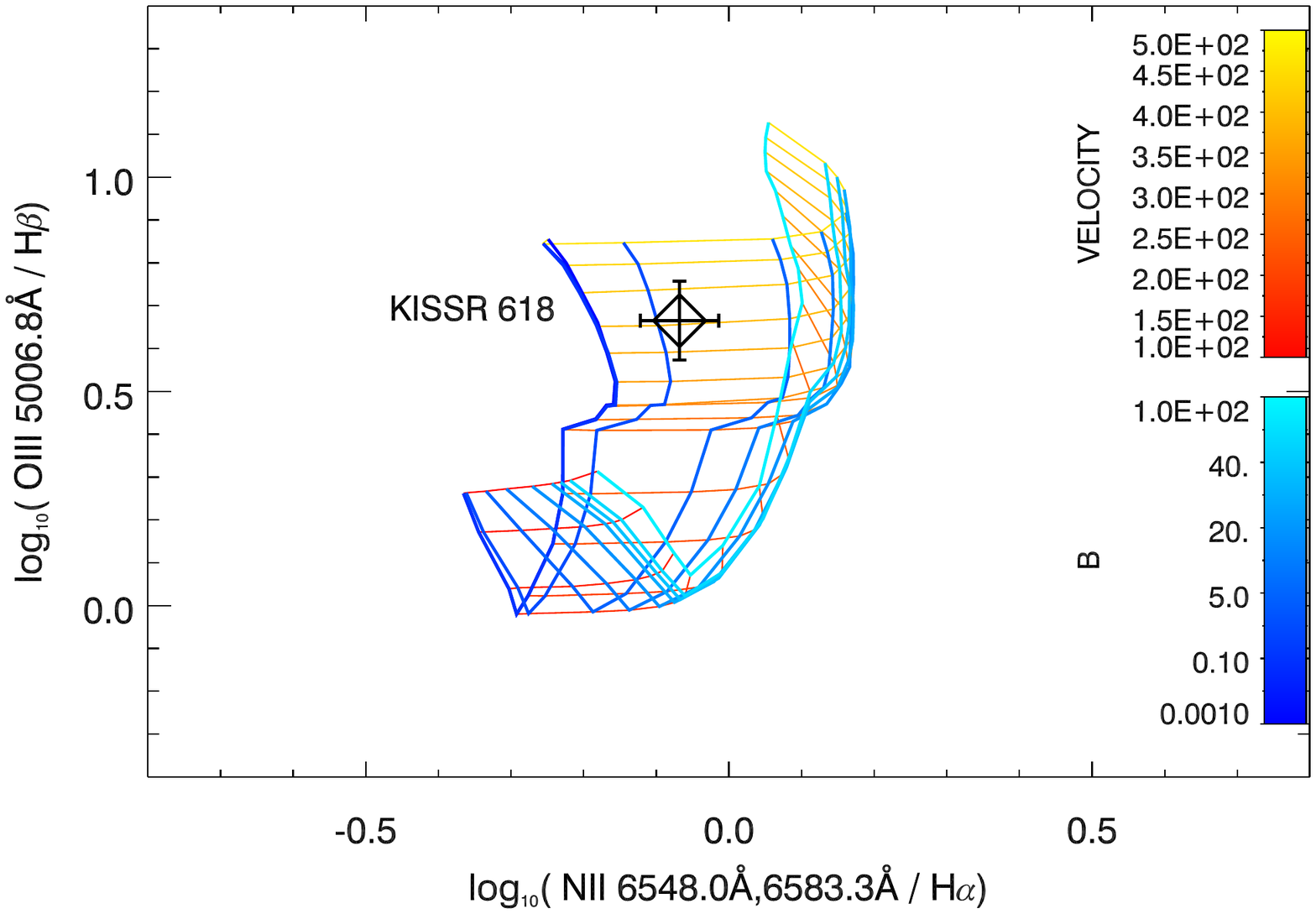}
\includegraphics[width=7cm,trim=0 220 0 100]{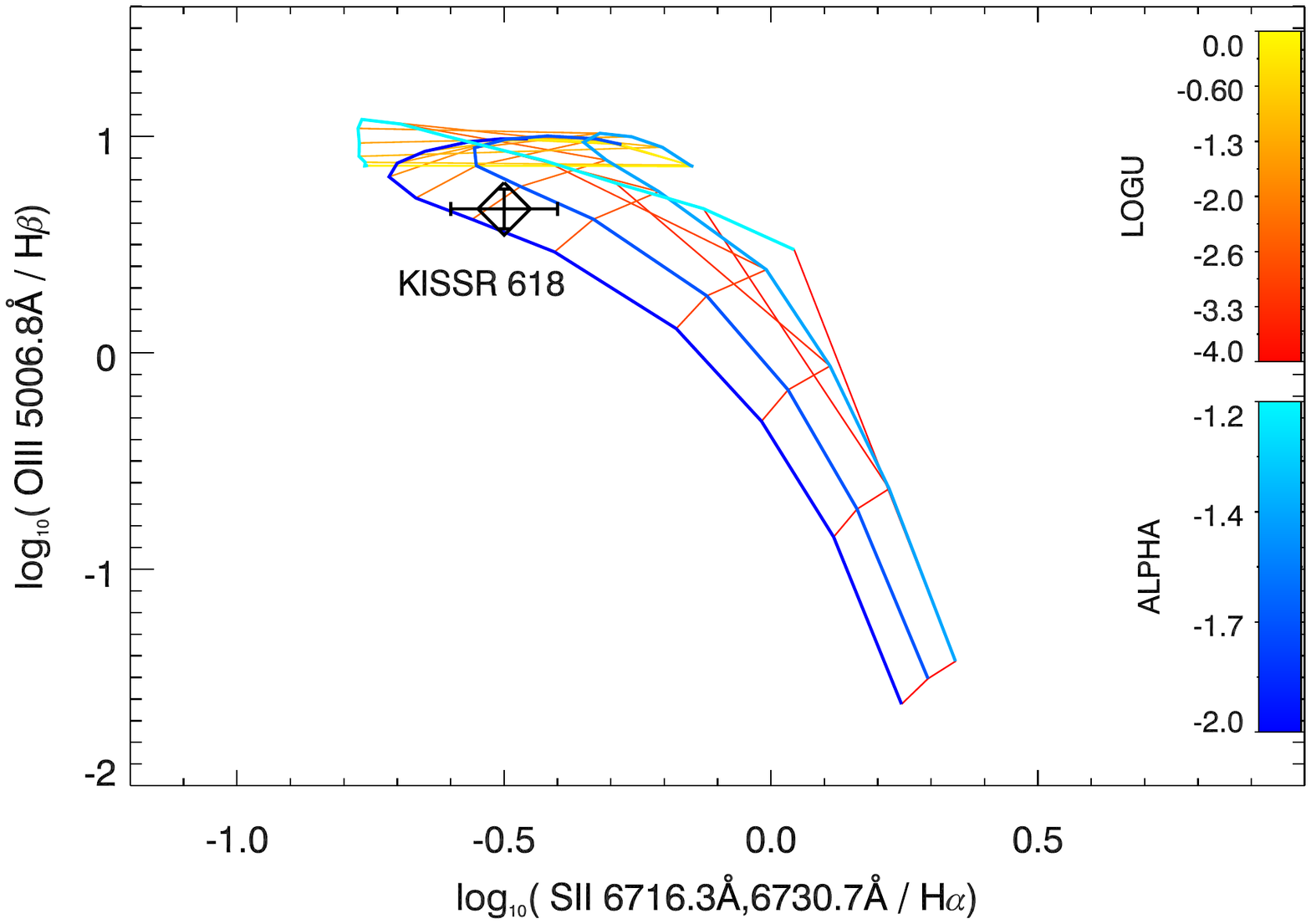}
\includegraphics[width=7cm,trim=0 220 0 100]{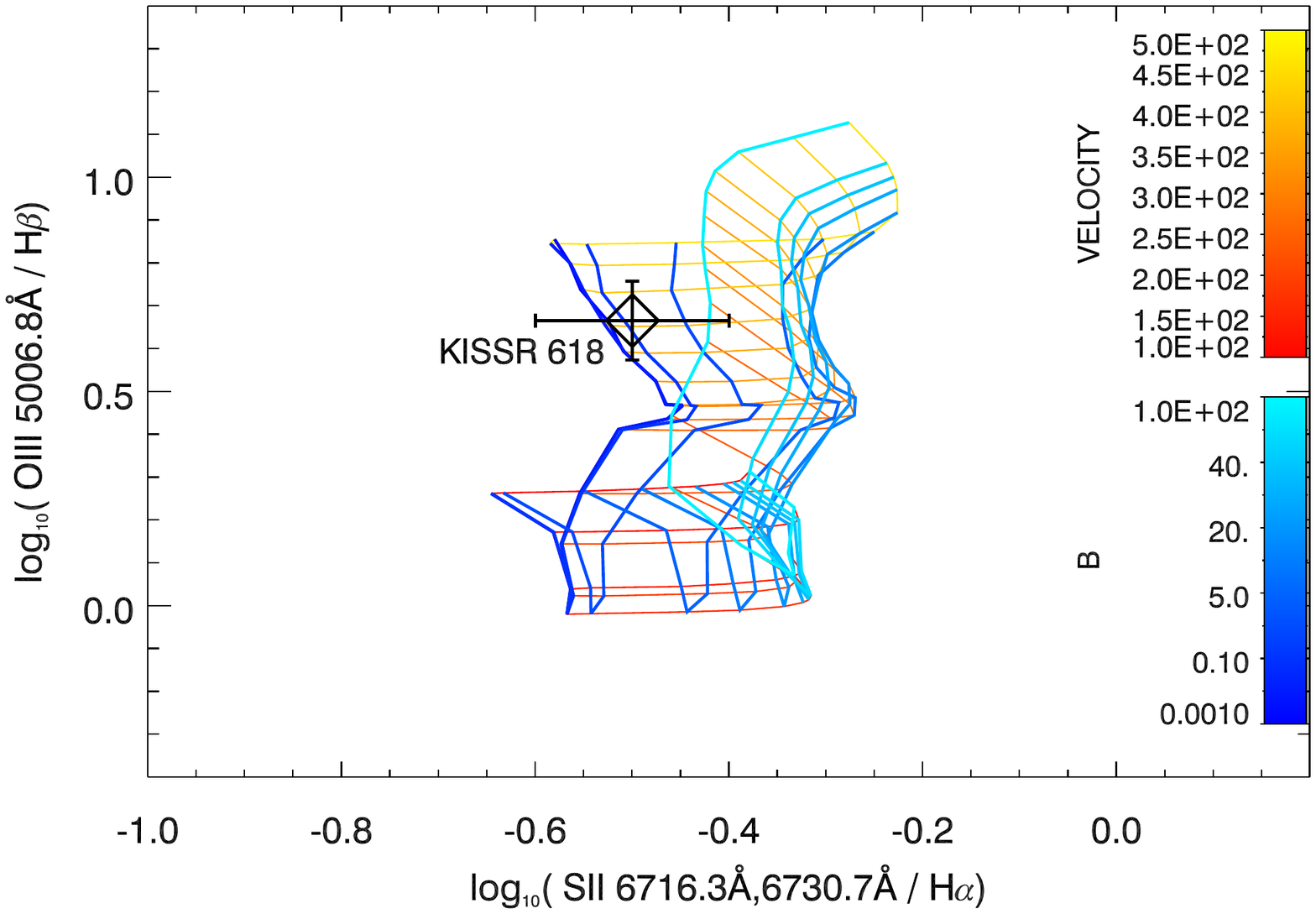}
}
\caption{\small 
(Top) [N {\sc ii}] $\lambda$6583/H$\alpha$ vs [O {\sc iii}] $\lambda$5007/H$\beta$ and 
(Bottom) [S {\sc ii}] $\lambda$6716, 6731/H$\alpha$ vs [O {\sc iii}] $\lambda$5007/H$\beta$ diagnostic diagram for a density of 100~cm$^{-3}$ and solar abundance, for KISSR\,618. Of the double line components, the line with the greater peak flux was used in the analysis. (Left) AGN dusty photoionization model grids for varying ionization parameters ($-4<$ Log U $<0$) and power-law indices ($-2< \alpha <-1.2$). Lines of constant $\alpha$ are shaded in blue and lines of constant U are red-yellow shaded. (Right) Shock+precursor model grids for  shock velocities ($100 < \,v \, < 500$~km~s$^{-1}$) and magnetic fields ($0.001 < B < 100~\mu$G). Lines of constant magnetic field are blue shaded and lines of constant shock velocities are red-yellow shaded. See Section~\ref{sec:line} for details.}
\label{fig10}
\end{figure*}

\begin{figure*}
\centering{
\includegraphics[width=7cm,trim=0 310 0 200]{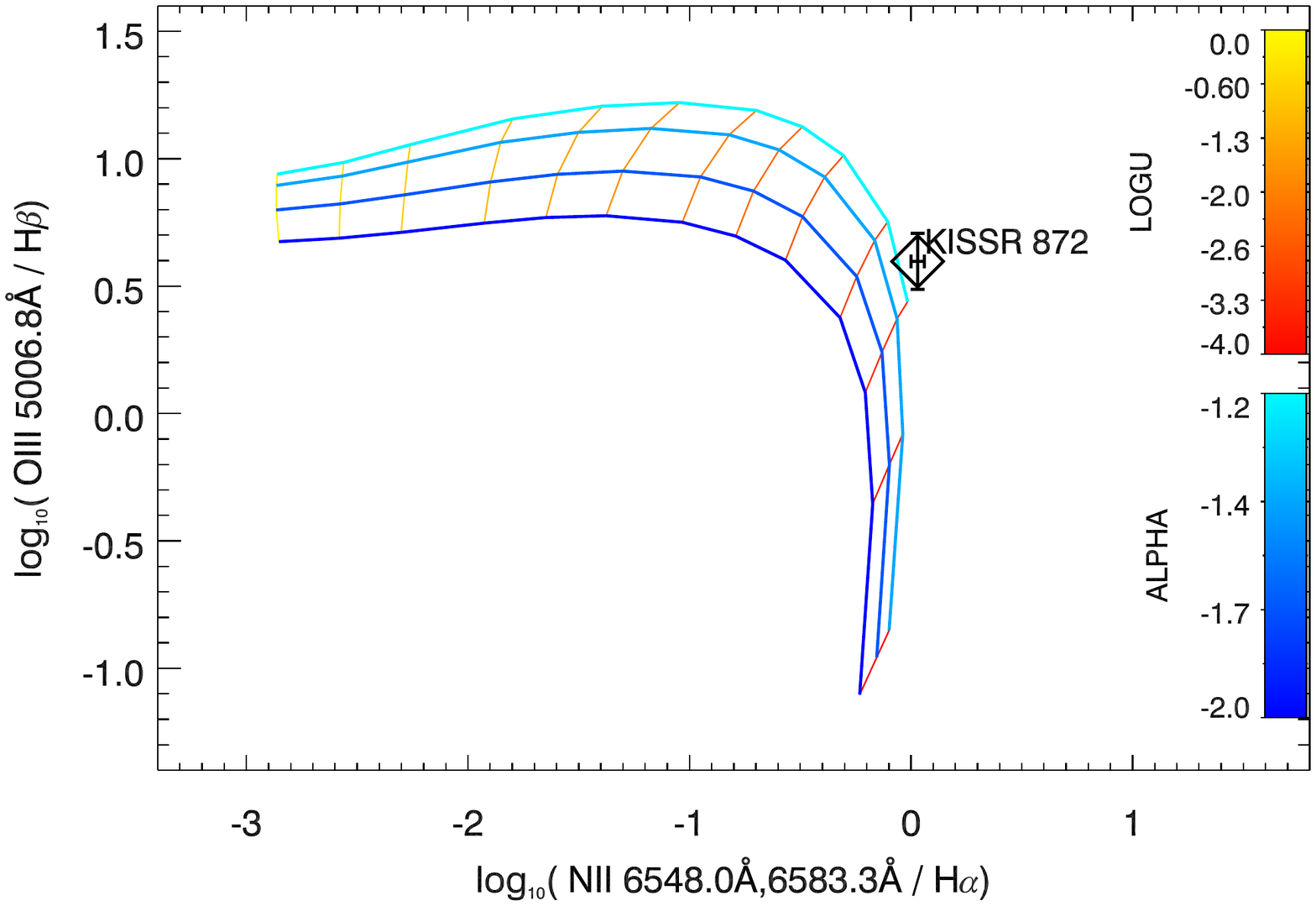}
\includegraphics[width=7cm,trim=0 310 0 200]{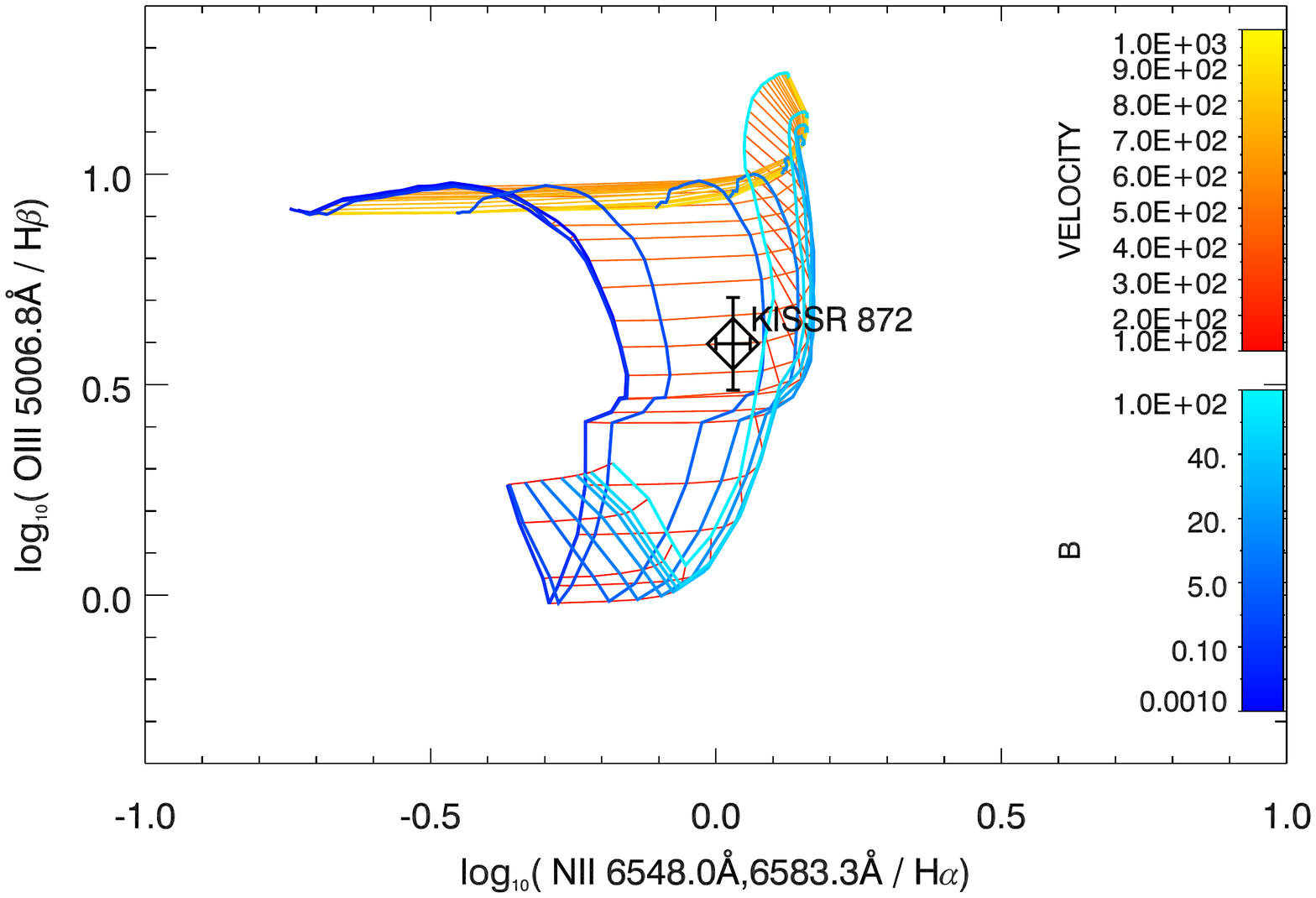}
\includegraphics[width=7cm,trim=0 220 0 100]{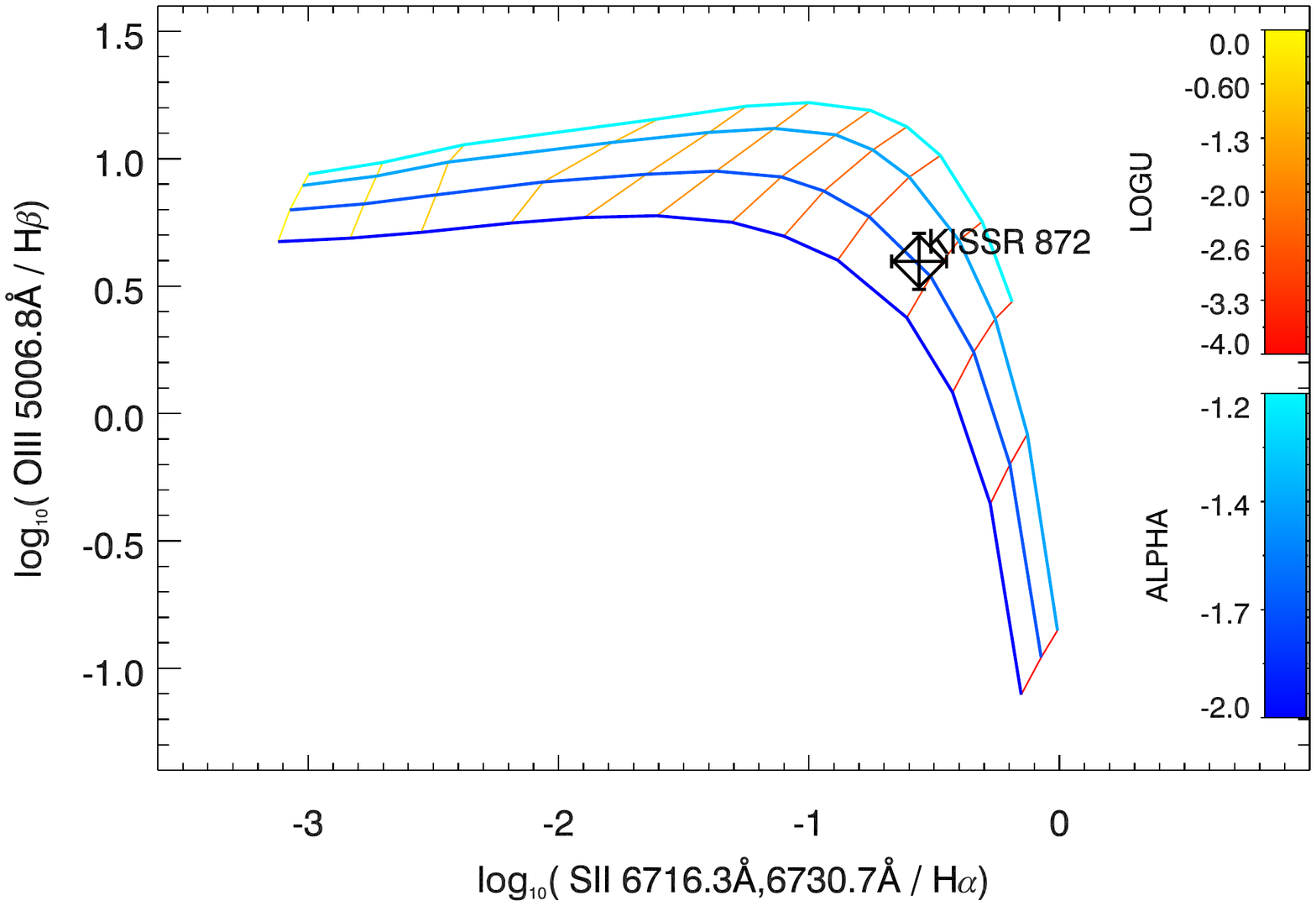}
\includegraphics[width=7cm,trim=0 220 0 100]{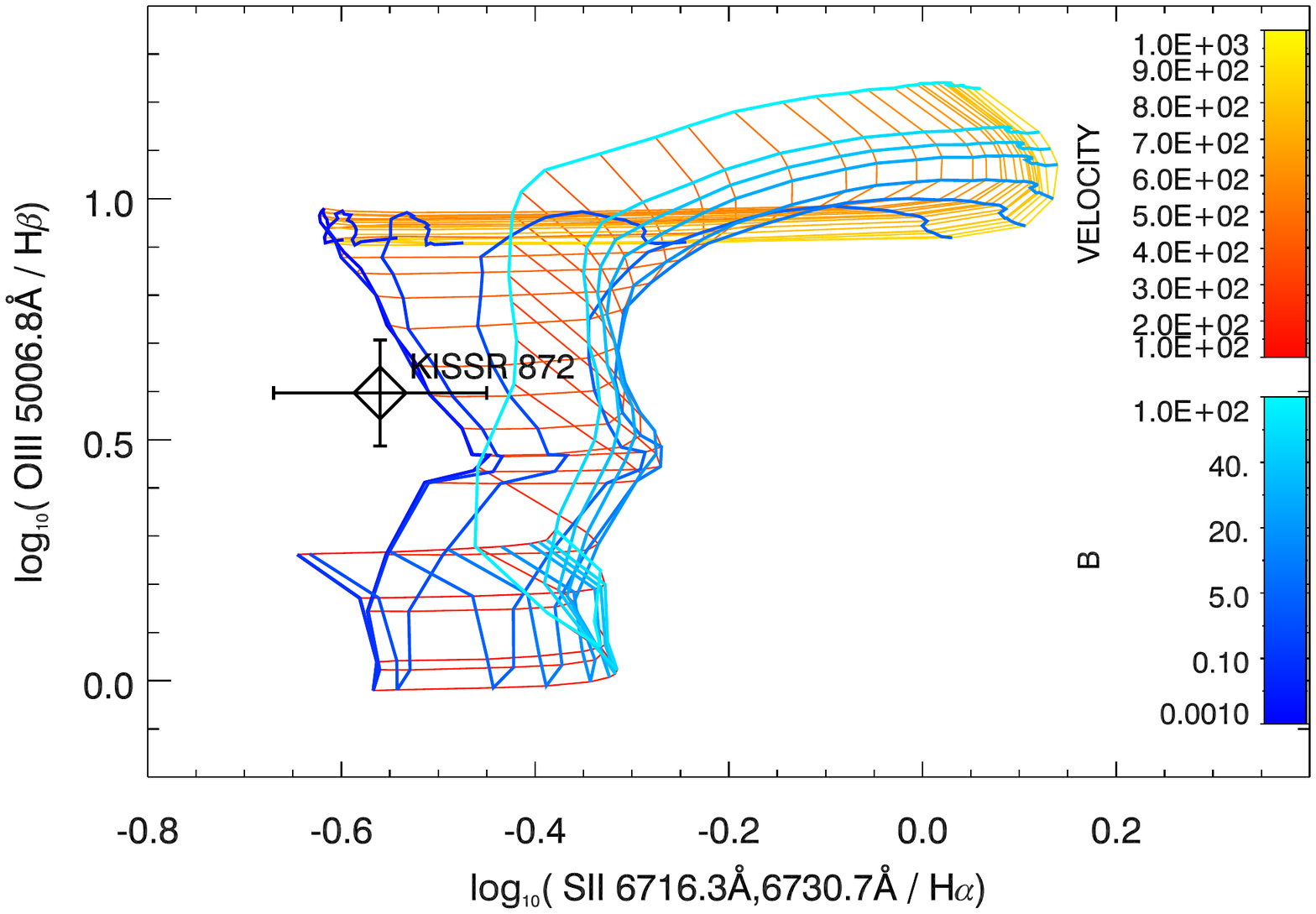}
}
\caption{\small Emission line diagnostic diagrams for KISSR\,872. (Left) AGN dust-free photoionization model grids. (Right) Shock+precursor model grids. Other details same as in Figure~\ref{fig10}.}
\label{fig11}
\end{figure*}

\begin{figure*}
\centering{
\includegraphics[width=7cm,trim=0 310 0 240]{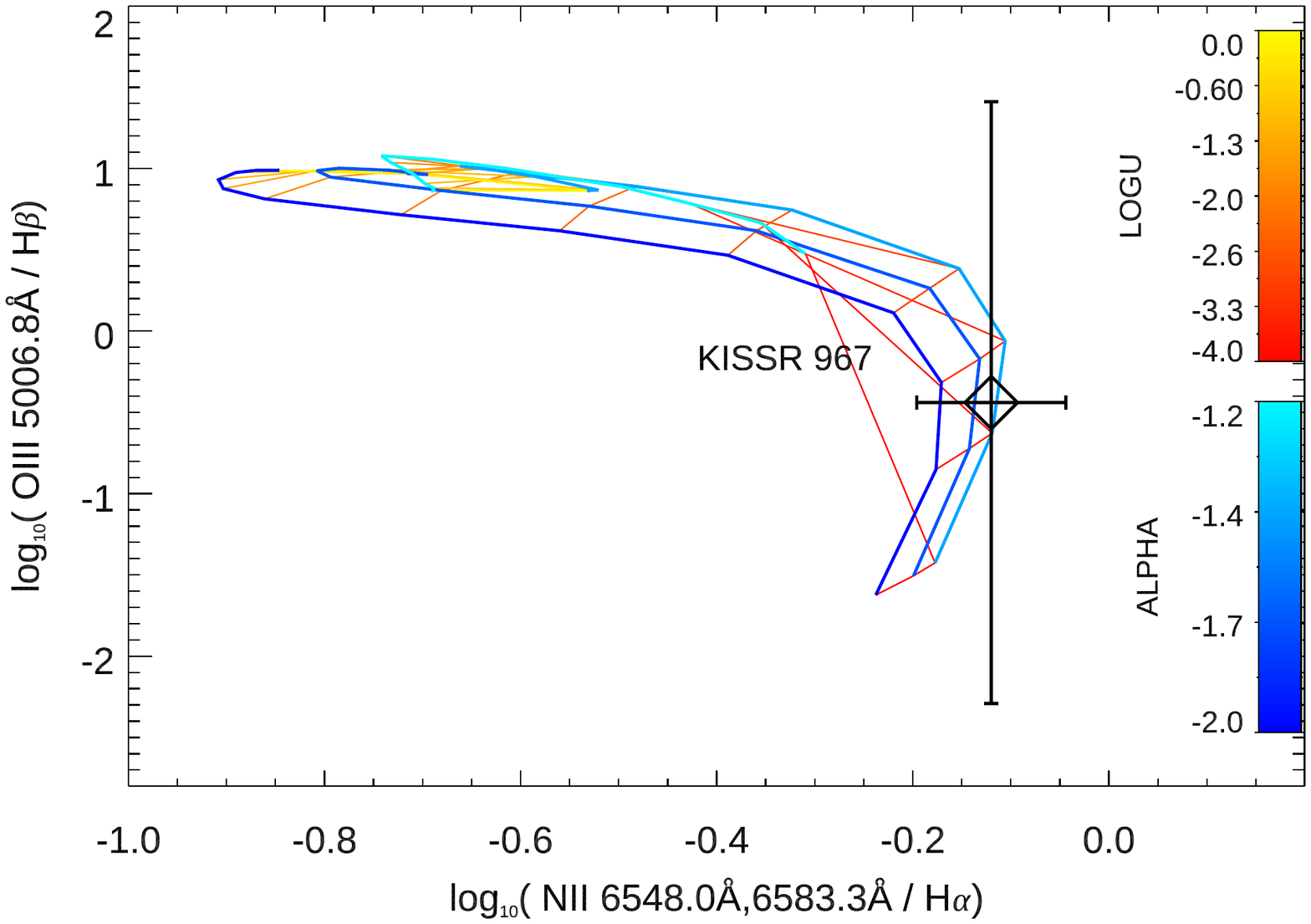}
\includegraphics[width=7cm,trim=0 310 0 240]{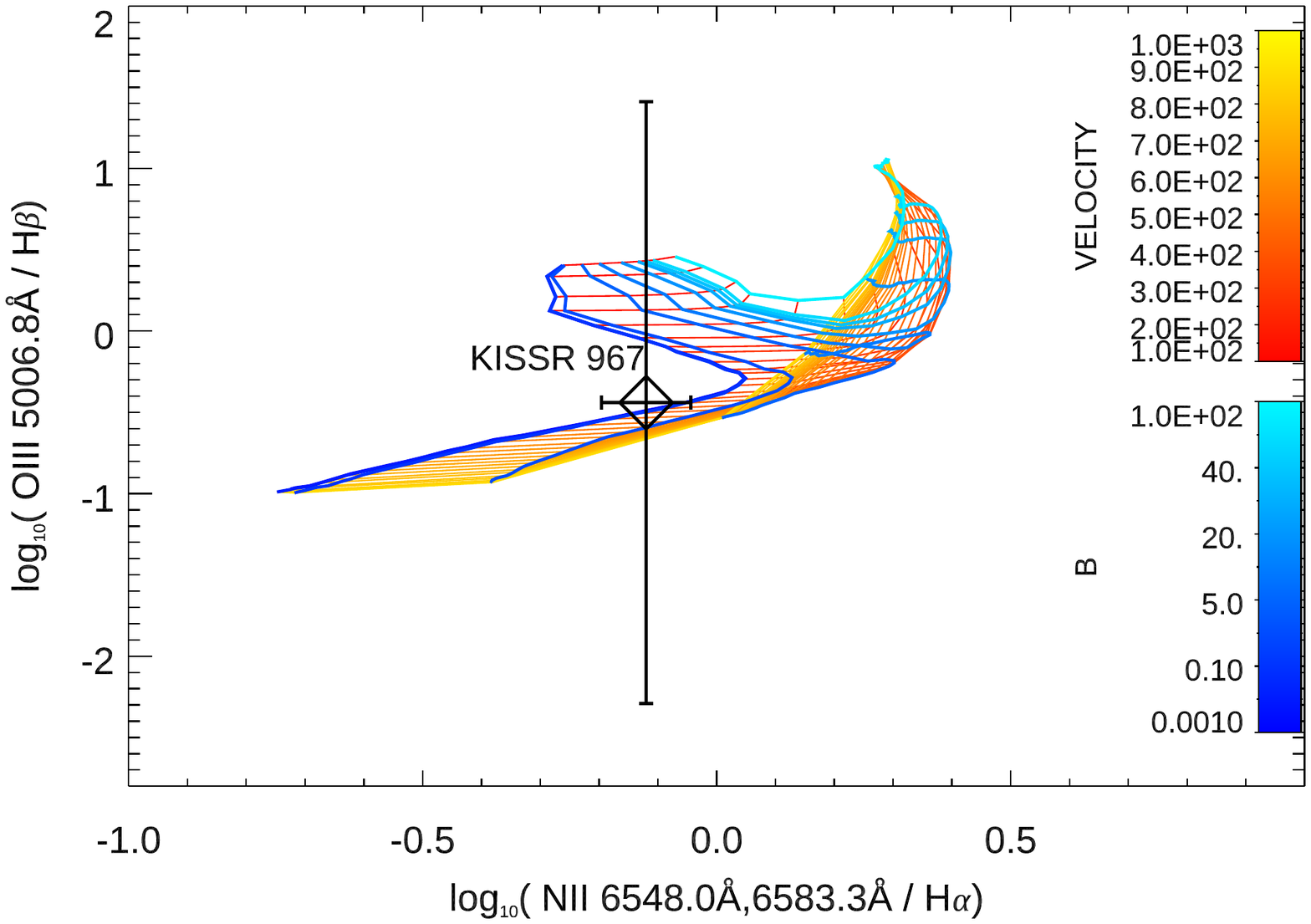}
\includegraphics[width=7cm,trim=0 220 0 120]{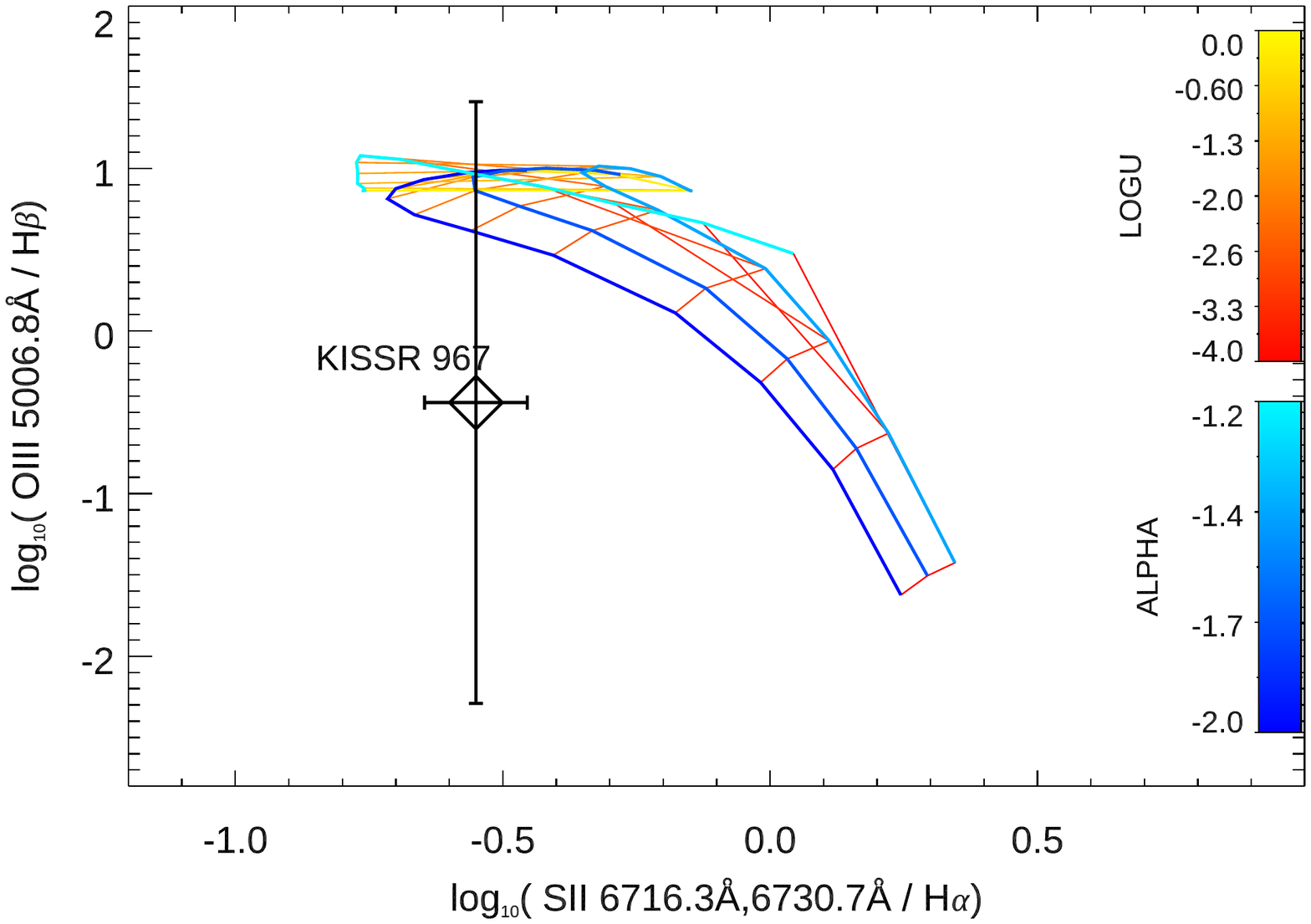}
\includegraphics[width=7cm,trim=0 220 0 120]{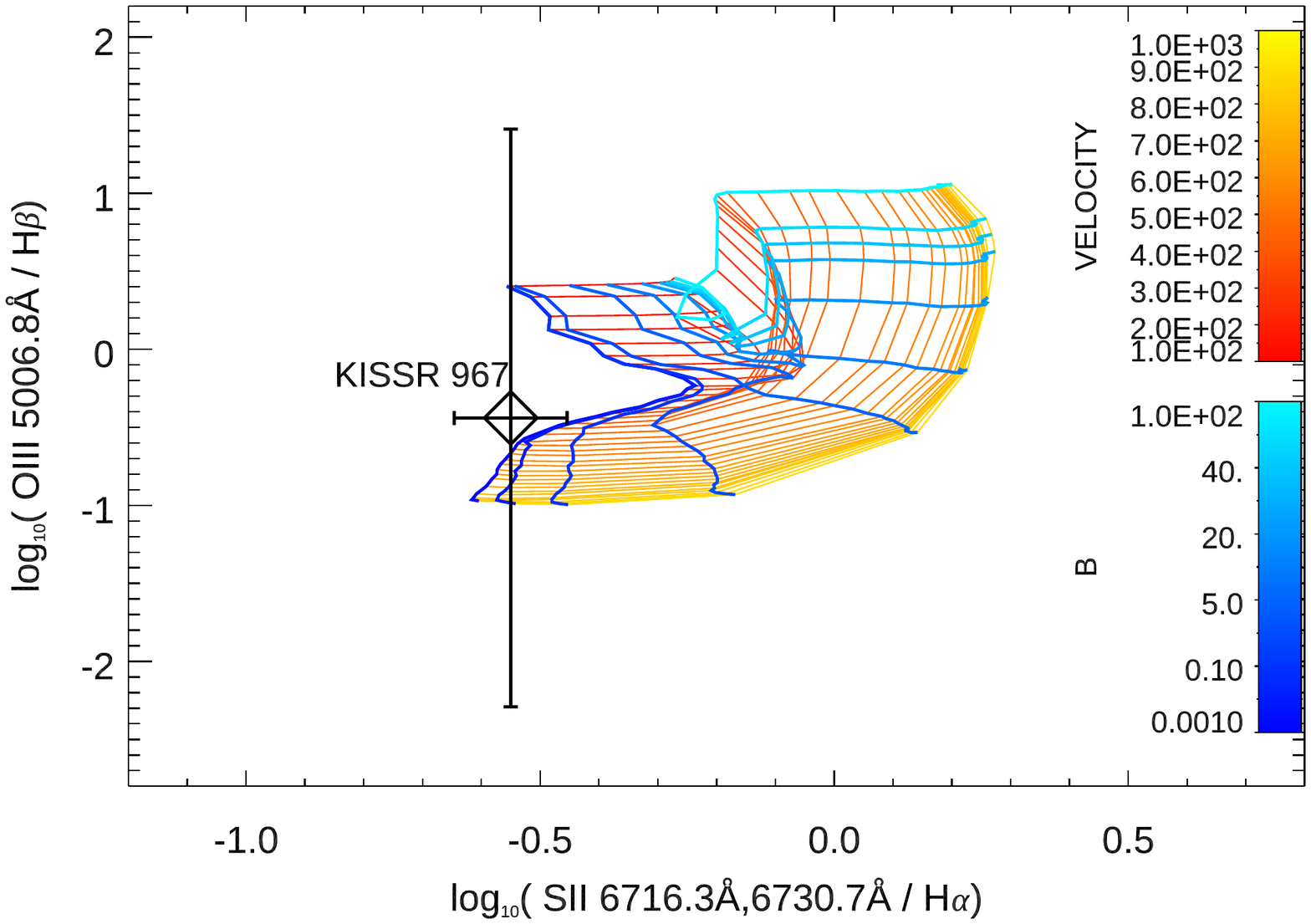}
\includegraphics[width=7cm,trim=0 220 0 180]{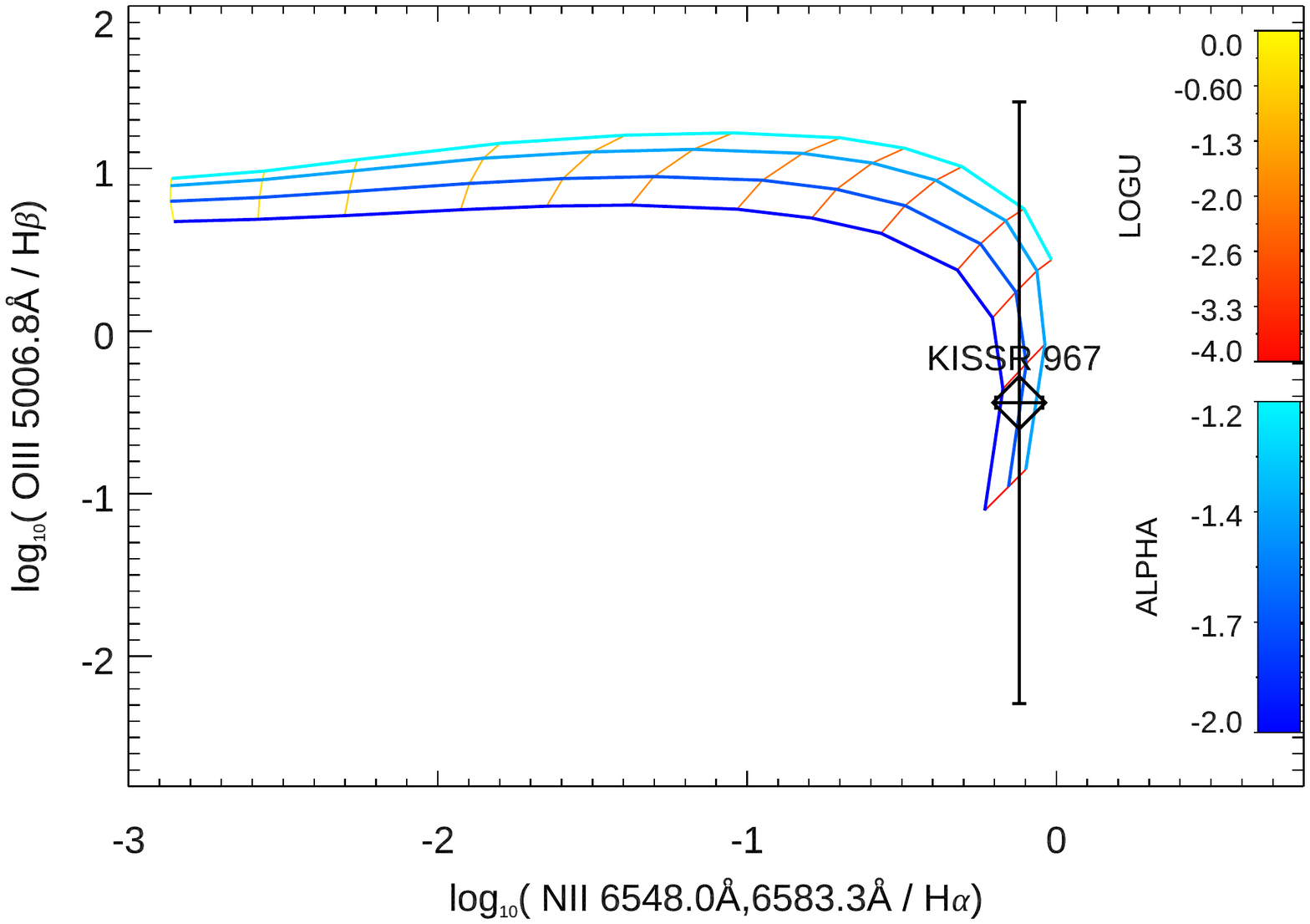}
\includegraphics[width=7cm,trim=0 220 0 180]{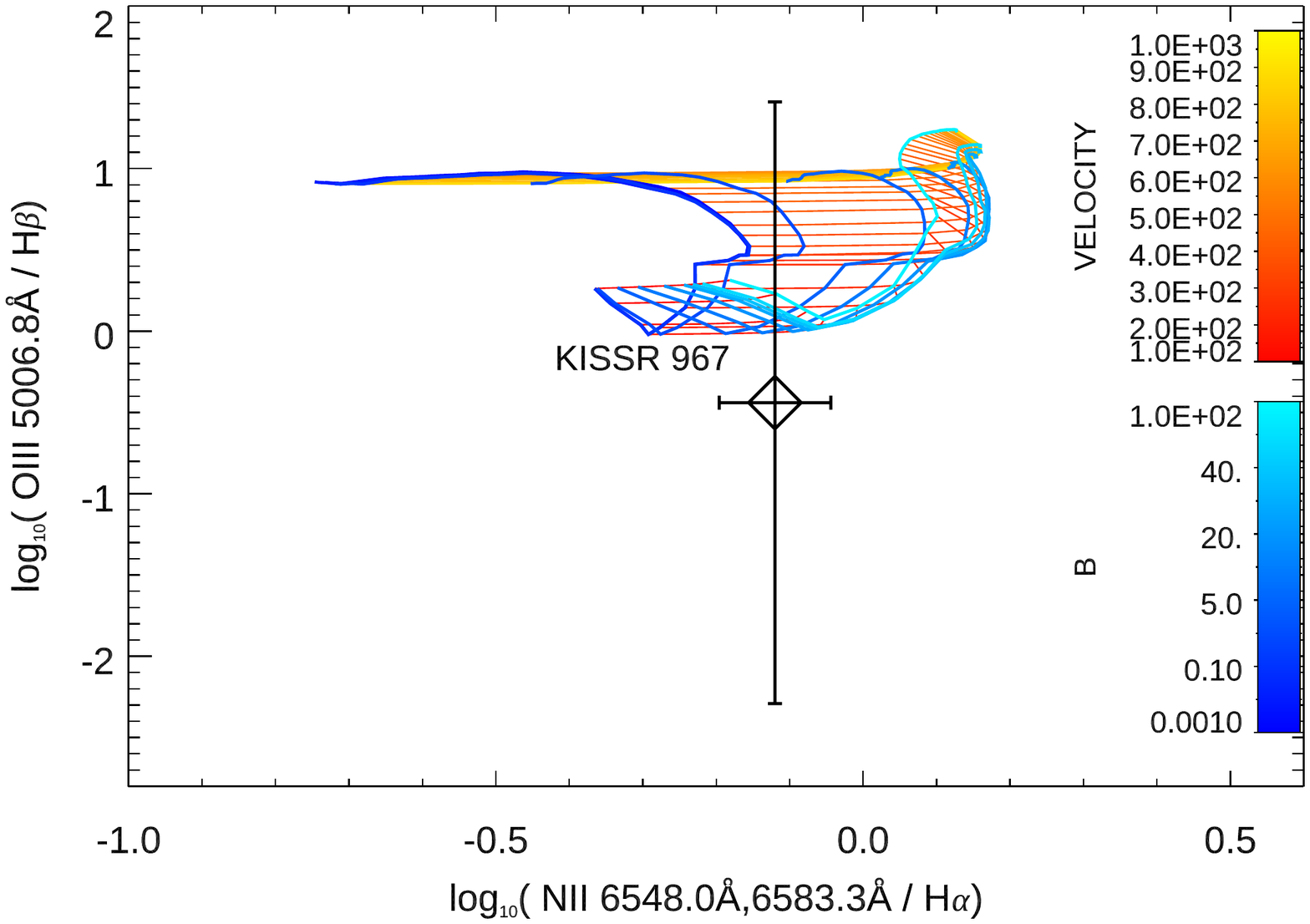}
\includegraphics[width=7cm,trim=0 190 0 150]{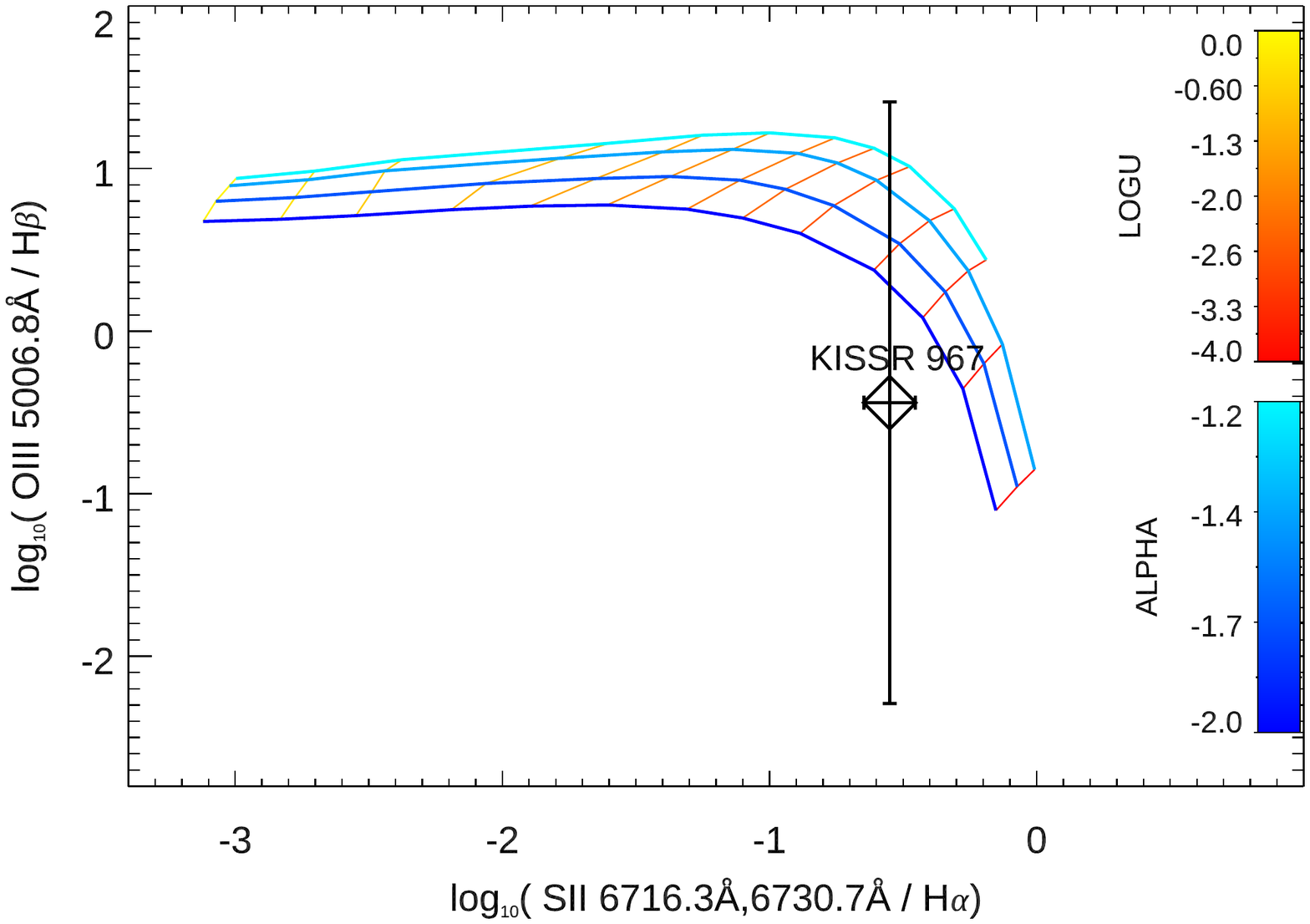}
\includegraphics[width=7cm,trim=0 190 0 150]{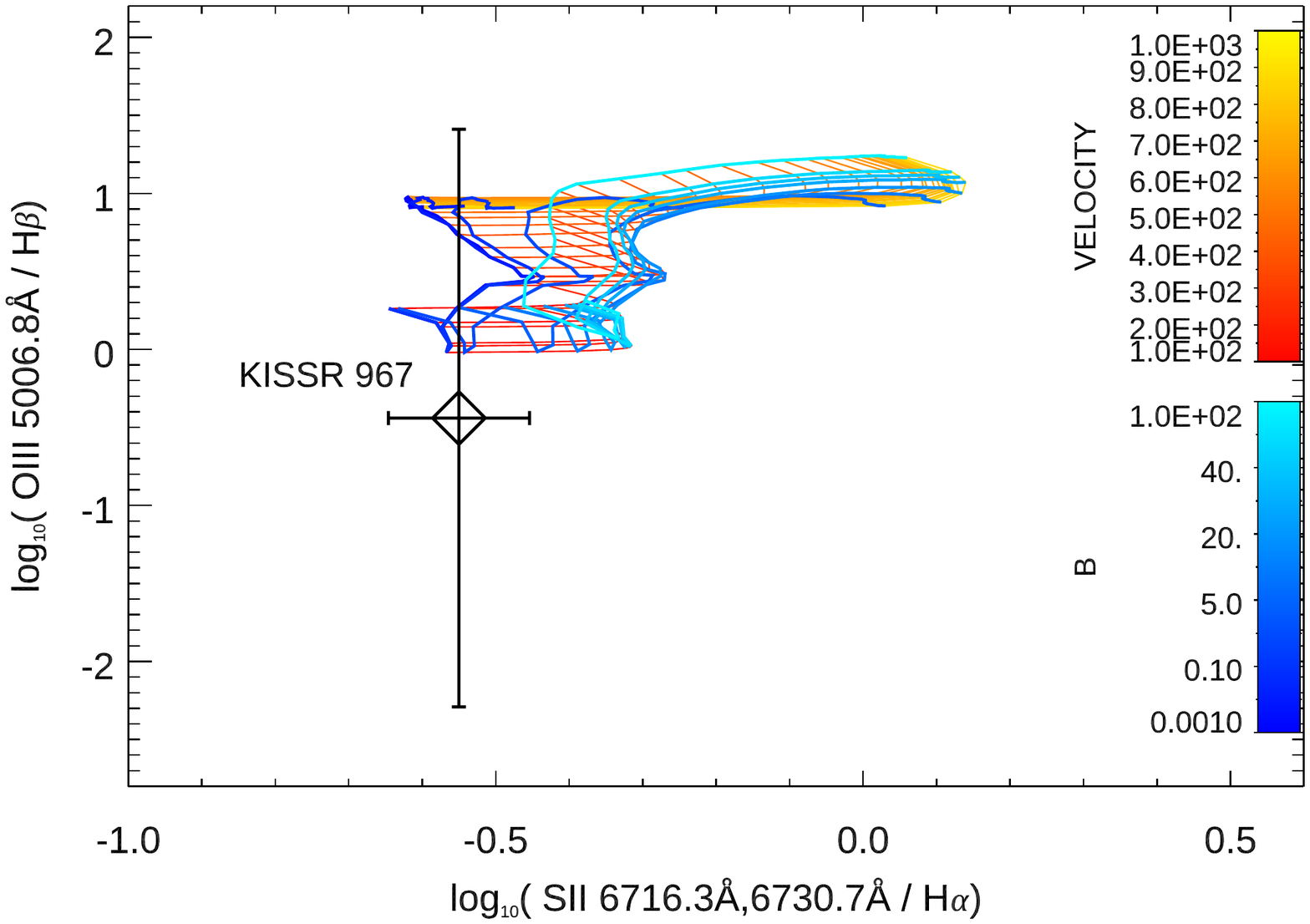}
}
\caption{\small Emission line diagnostic diagrams for KISSR\,967. (Top 2 rows) Left: AGN dusty photoionization model grids, and Right: Shock-only model grids. (Bottom 2 rows) Left: AGN dust-free photoionization model grids. (Right) Shock+precursor model grids. Other details same as in Figure~\ref{fig10}. No meaningful conclusions could be drawn about the best-fit model in KISSR\,967 due to the large errors in line fluxes. 
}
\label{fig12}
\end{figure*}

\begin{figure*}
\centering{
\includegraphics[width=7cm,trim=0 310 0 200]{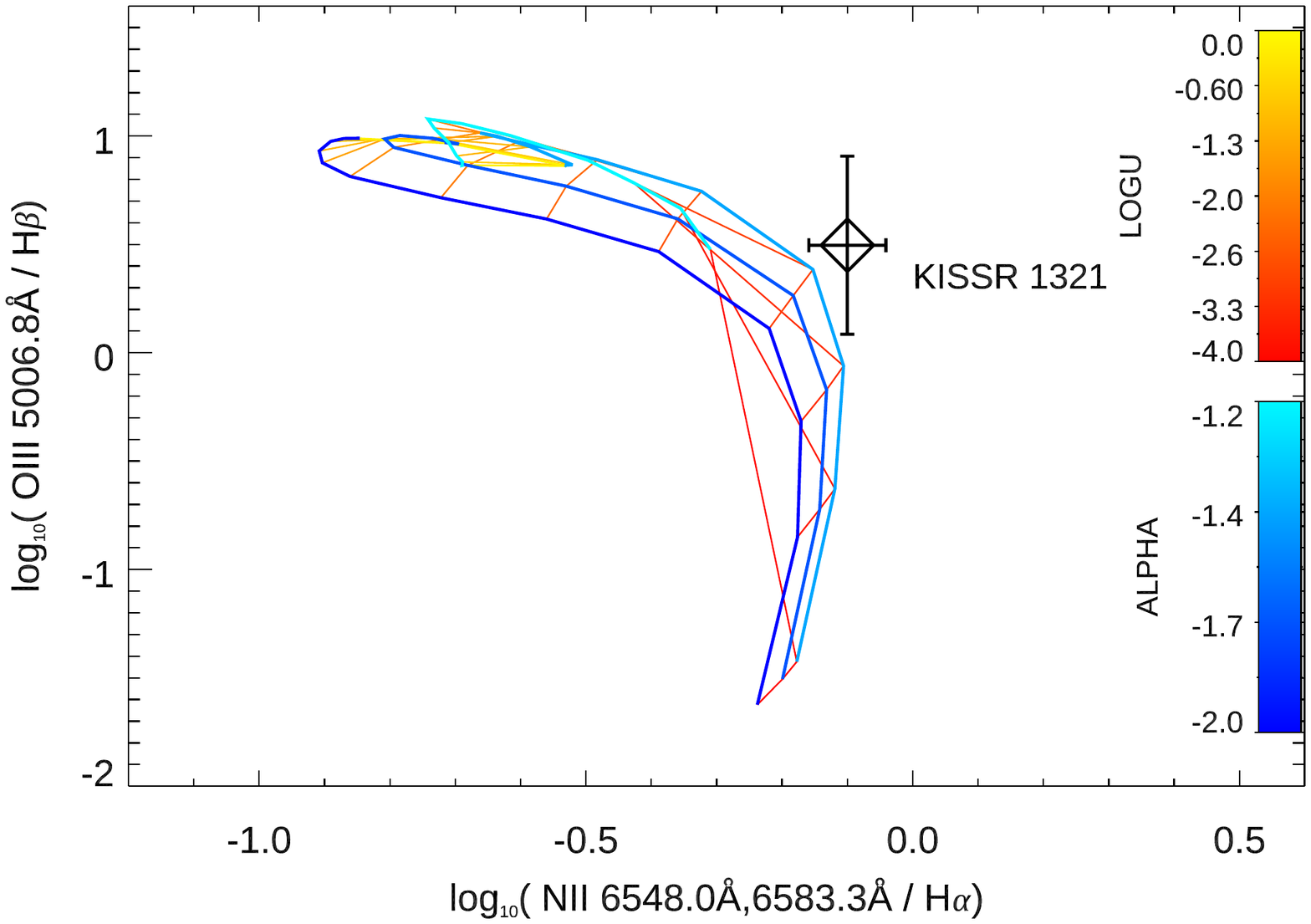}
\includegraphics[width=7cm,trim=0 310 0 200]{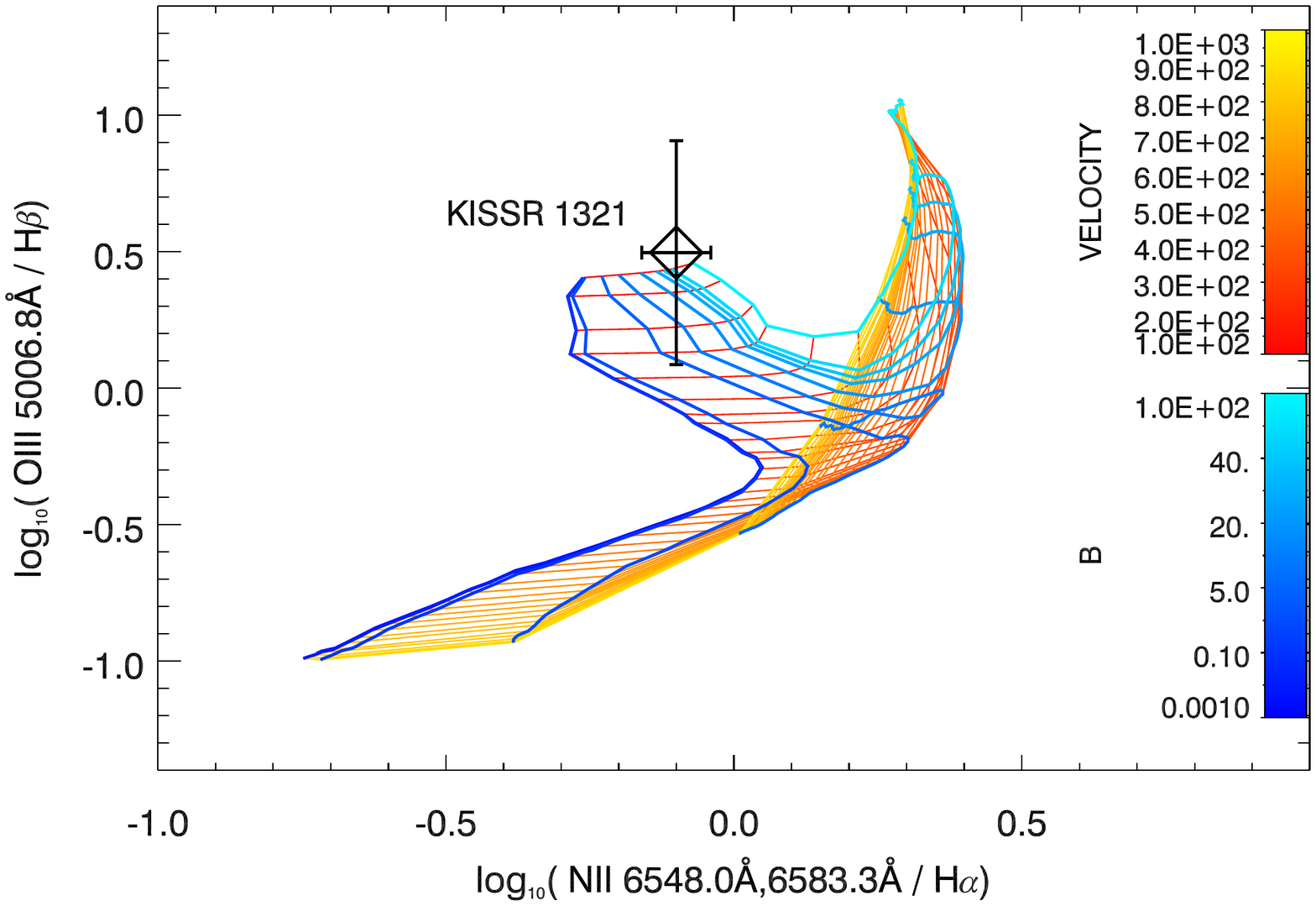}
\includegraphics[width=7cm,trim=0 220 0 100]{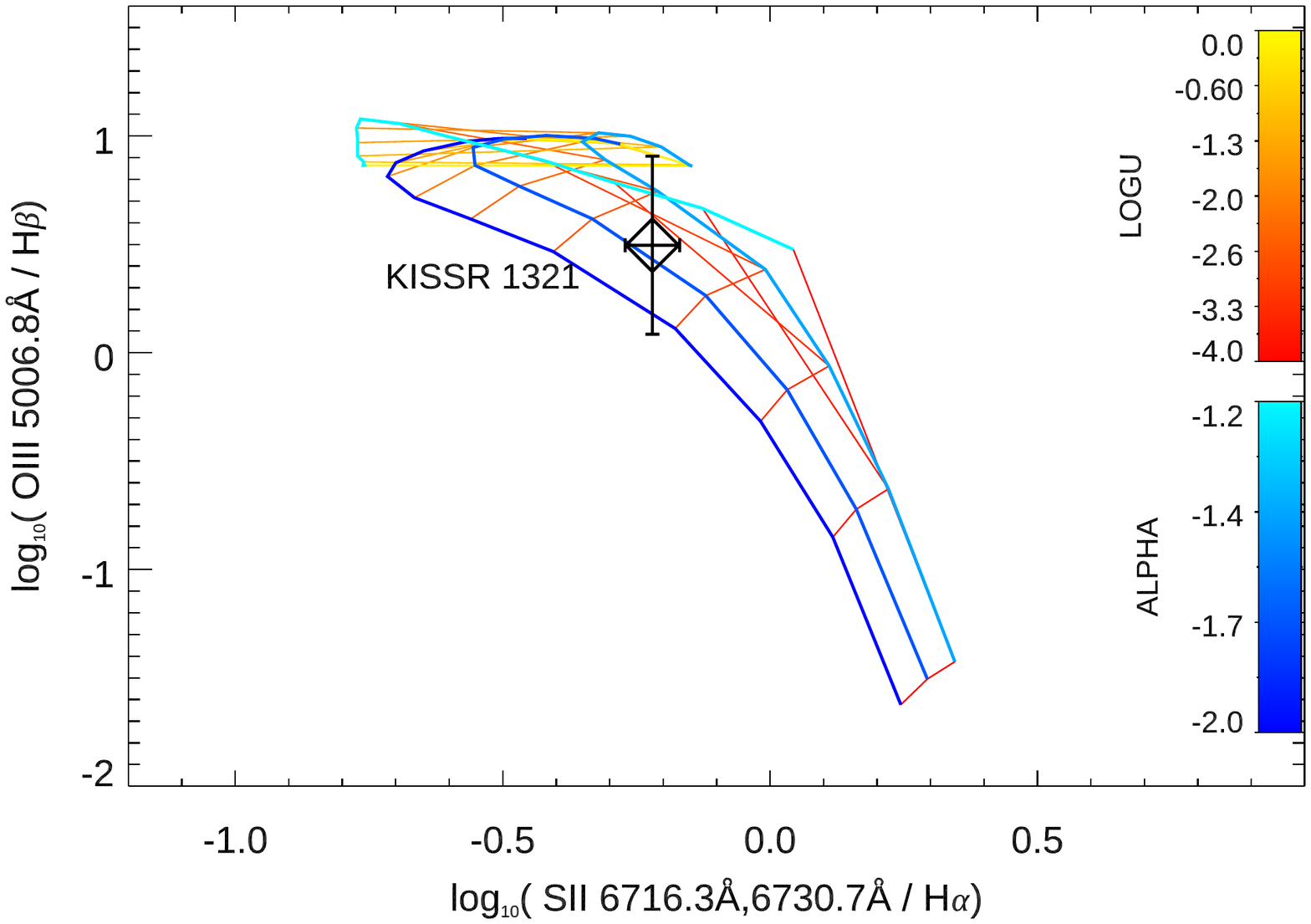}
\includegraphics[width=7cm,trim=0 220 0 100]{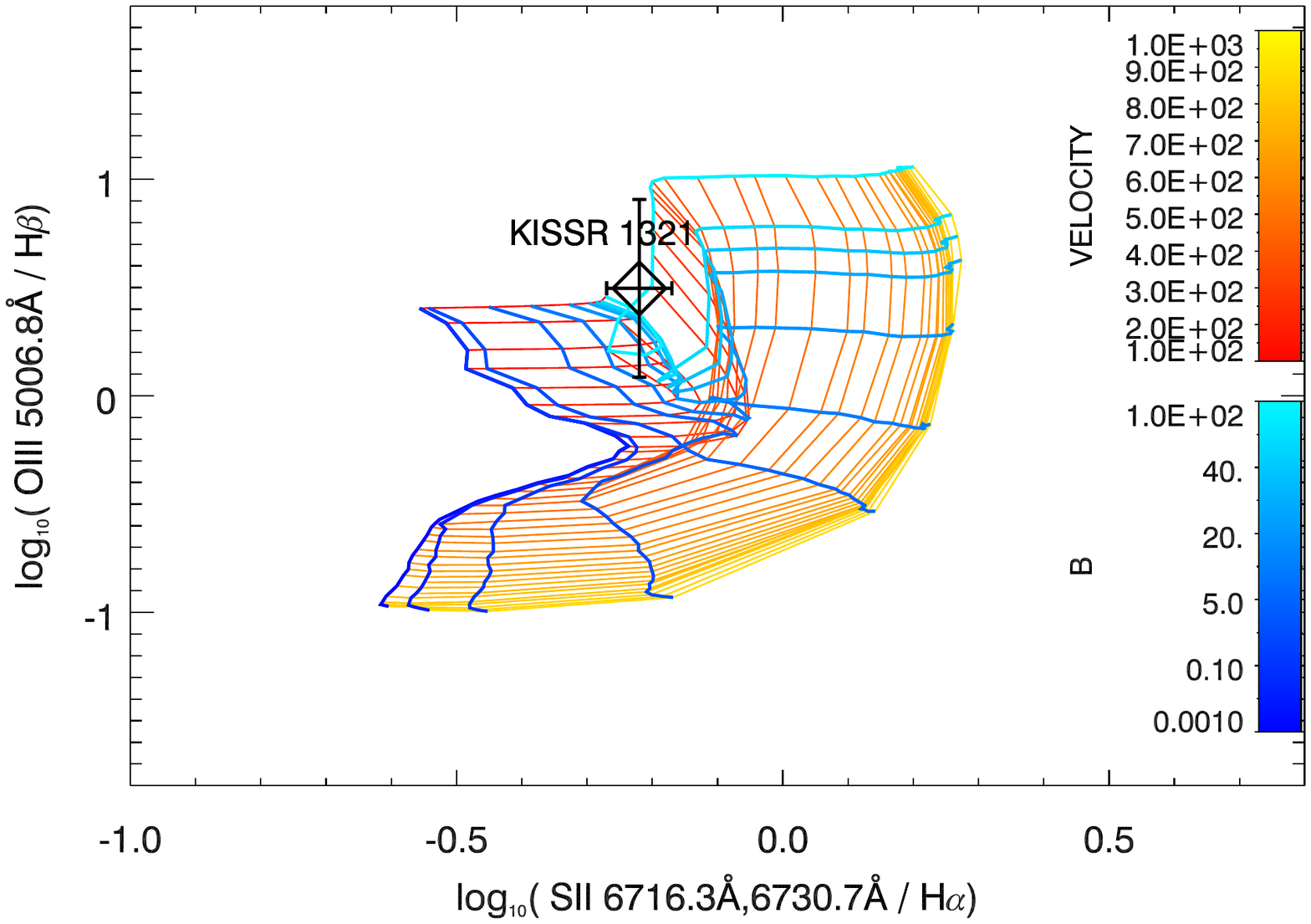}
}
\caption{\small Emission line diagnostic diagrams for KISSR\,1321. (Left) AGN dusty photoionization model grids. (Right) Shock-only model grids. Other details same as in Figure~\ref{fig10}.}
\label{fig13}
\end{figure*}

\begin{figure*}
\centering{
\includegraphics[width=7cm,trim=0 310 0 200]{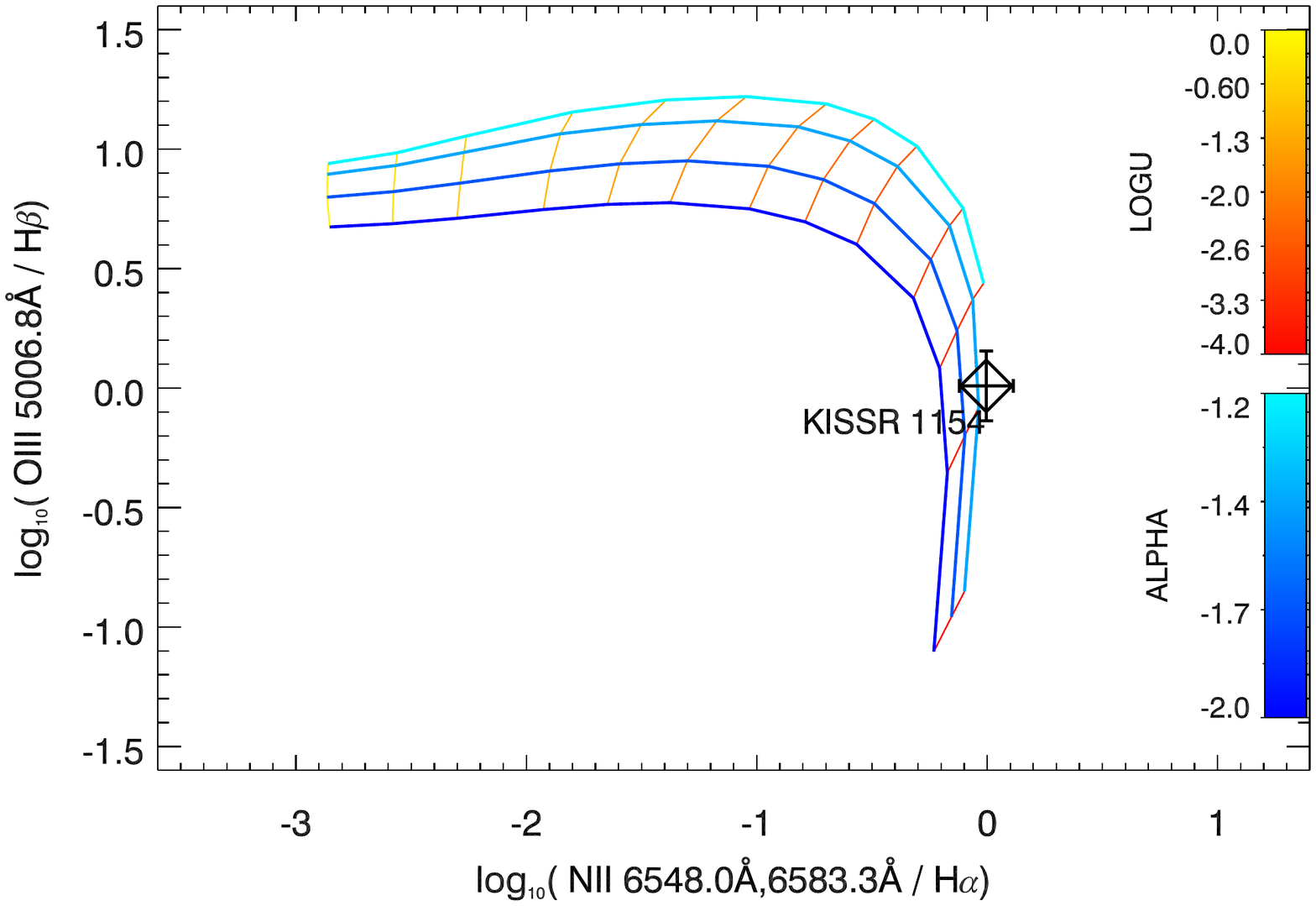}
\includegraphics[width=7cm,trim=0 310 0 200]{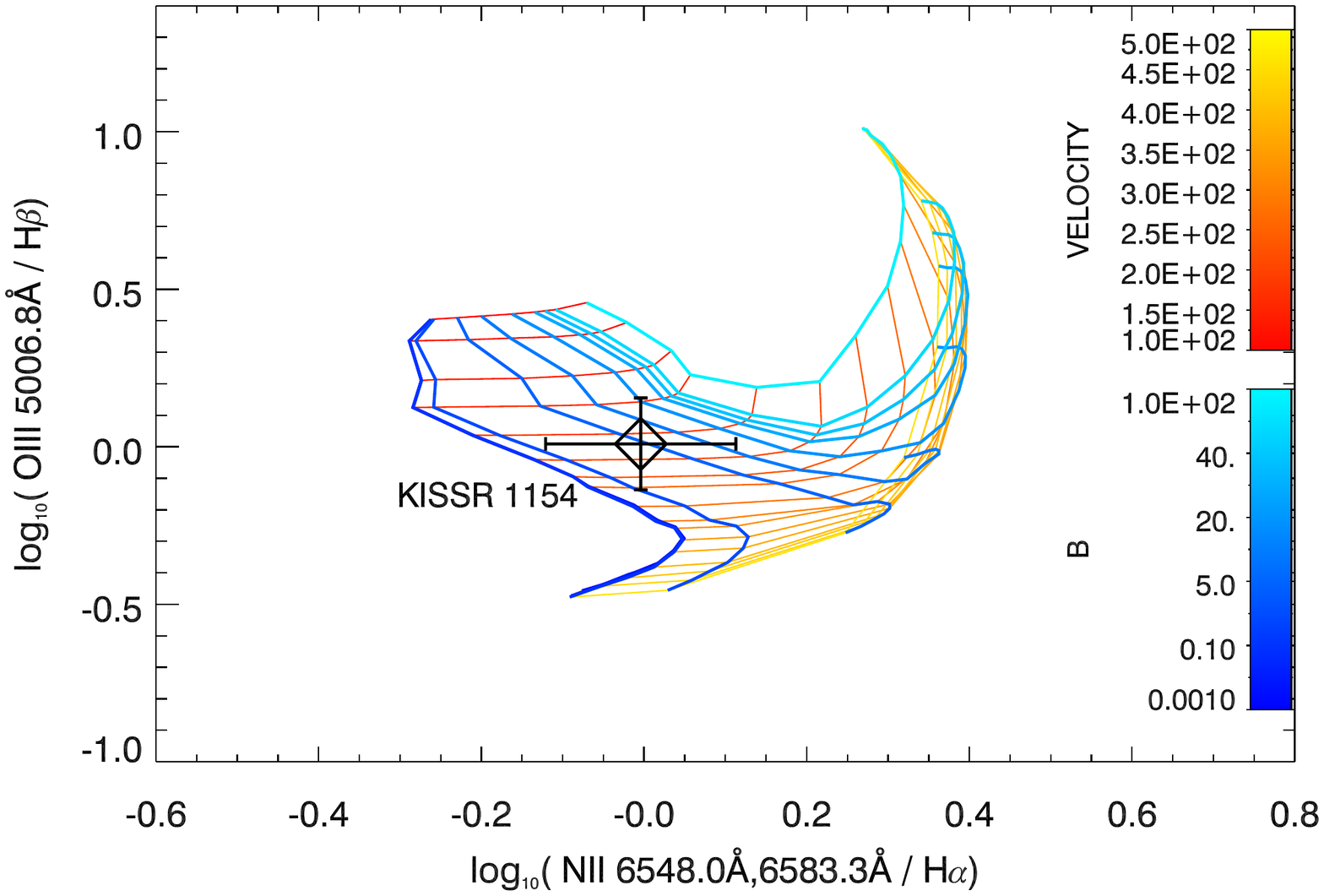}
\includegraphics[width=7cm,trim=0 220 0 100]{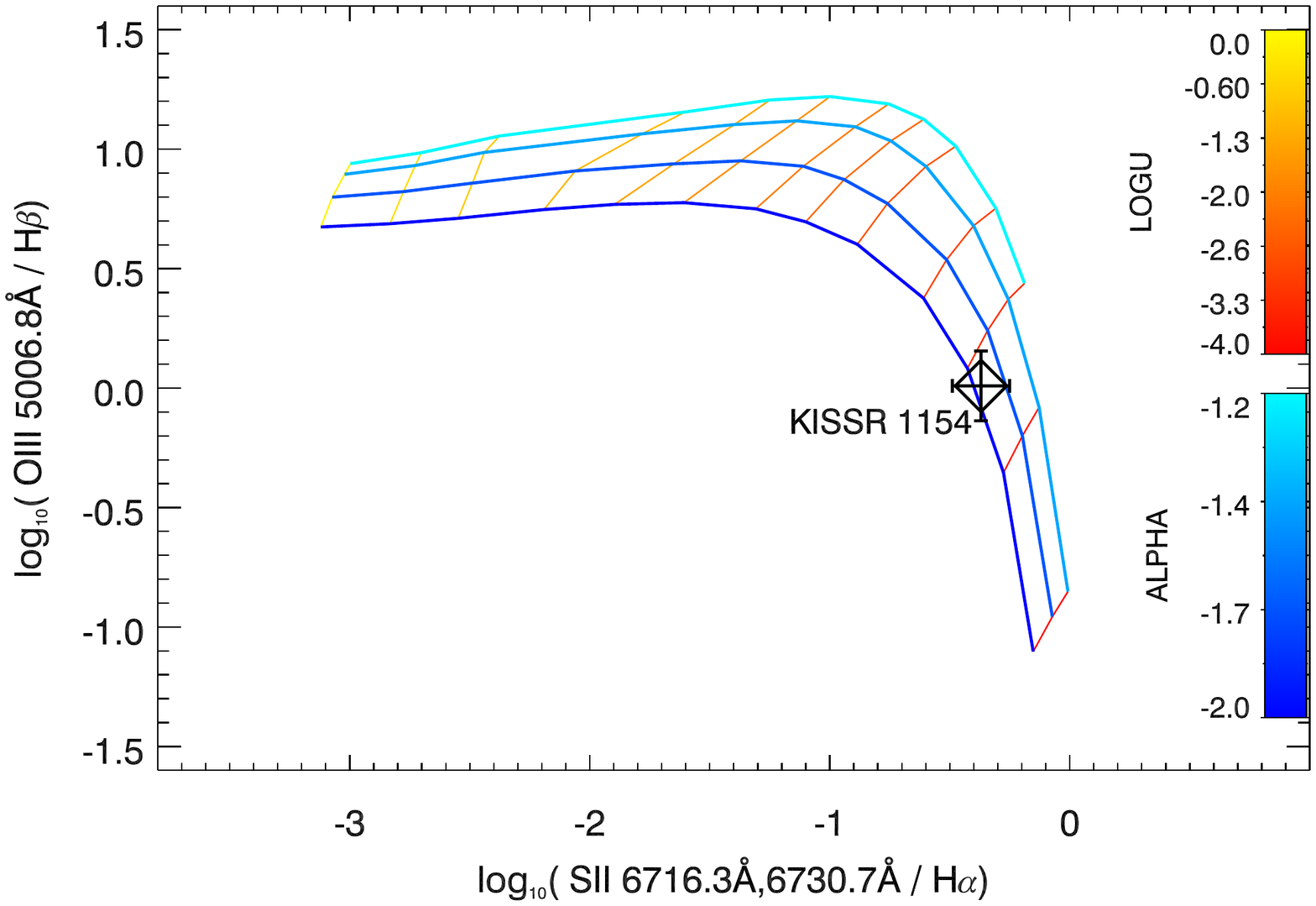}
\includegraphics[width=7cm,trim=0 220 0 100]{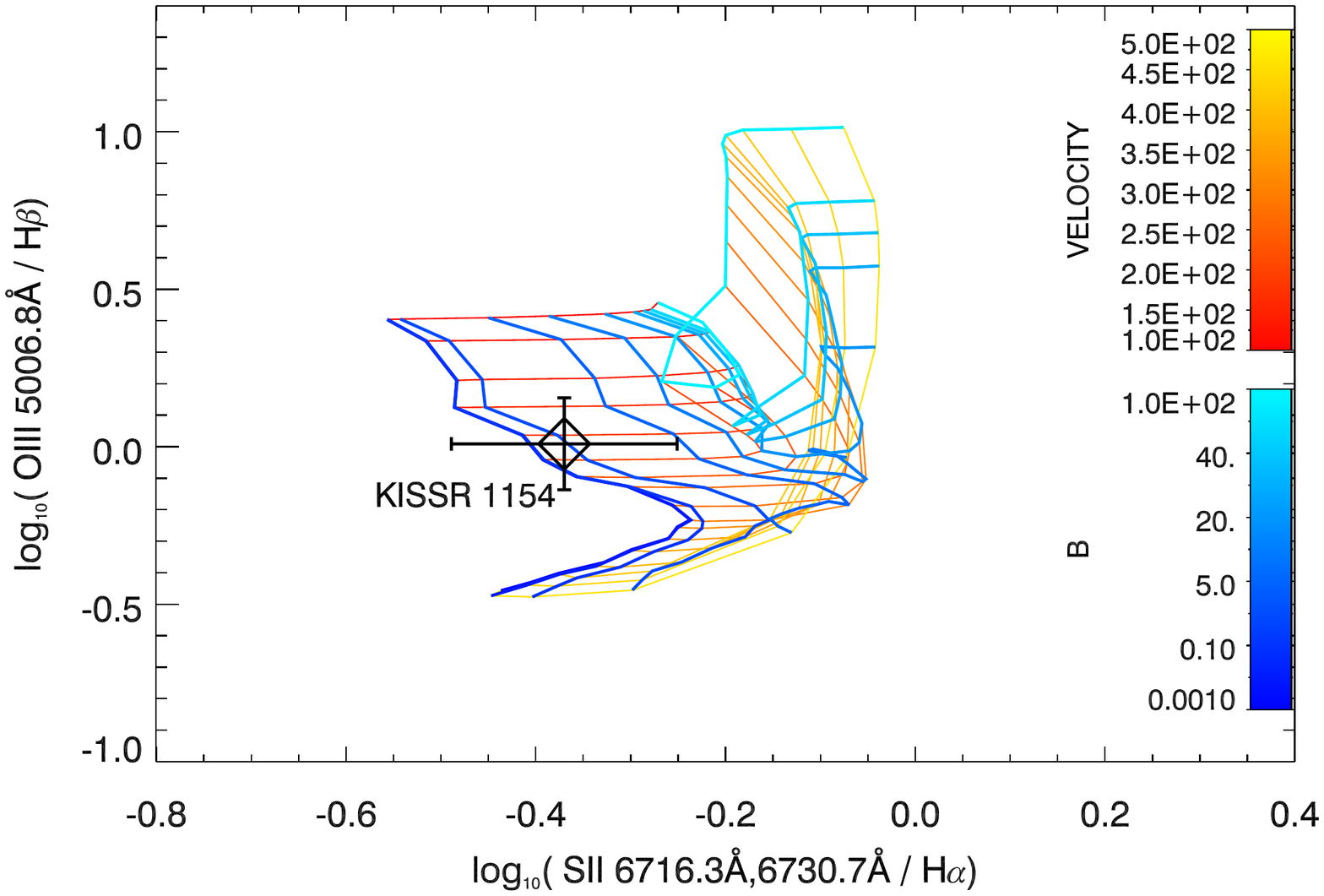}
}
\caption{\small Emission line diagnostic diagrams for KISSR\,1154. (Left) AGN dust-free photoionization model grids. (Right) Shock-only model grids. Other details same as in Figure~\ref{fig10}.}
\label{fig14}
\end{figure*}

\begin{table*}[b]
\scriptsize
\caption{Fitted Line Parameters for KISSR\,618}
\begin{center}
\begin{tabular}{lclcllc}
\hline\hline
{Line} & {$\lambda_{0}$} & {$\lambda_{c}\pm$error} & {$\Delta\lambda\pm$error} & {$f_{p}\pm$error} & {$F\pm$error}& {$L\pm$error}\\
{(1)}   & {(2)}  & {(3)}                 & {(4)} & {(5)} & {(6)} & {(7)}\\ \hline
$[\mathrm {S~II}]$ & 6718.3  &6717.35$\pm$    1.12  &  2.72$\pm$    0.06 &   9.05$\pm$    2.83 &  61.61$\pm$   19.31  &  0.60$\pm$    0.19\\
                                              &&6721.64$\pm$    0.82  &  2.52$\pm$    0.35 &   5.17$\pm$    2.62 &  32.71$\pm$   17.17  &  0.32$\pm$    0.17\\
$[\mathrm {S~II}]$ & 6732.7  &6731.35$\pm$    1.12  &  2.72$\pm$    0.06 &   7.21$\pm$    2.33 &  49.22$\pm$   15.96  &  0.48$\pm$    0.16\\
                                              &&6735.65$\pm$    0.82  &  2.53$\pm$    0.35 &   5.26$\pm$    2.12 &  33.36$\pm$   14.20  &  0.33$\pm$    0.14\\
$[\mathrm {N~II}]$ & 6549.9  &6548.36$\pm$    0.48  &  2.64$\pm$    0.06 &  11.42$\pm$    1.62 &  75.49$\pm$   10.84  &  0.74$\pm$    0.11\\
                                             &&6553.21$\pm$    0.50  &  2.45$\pm$    0.34 &   7.44$\pm$    1.15 &  45.71$\pm$    9.49  &  0.45$\pm$    0.09\\
$[\mathrm {N~II}]$ & 6585.3  &6583.37$\pm$    0.48  &  2.65$\pm$    0.06 &  33.69$\pm$    4.77 & 224.10$\pm$   32.19  &  2.19$\pm$    0.31\\
                                            &&6588.23$\pm$    0.48  &  2.47$\pm$    0.34 &  21.96$\pm$    3.40 & 135.70$\pm$   28.16  &  1.32$\pm$    0.27\\
H$\alpha$         & 6564.6  &6562.88$\pm$    0.21  &  2.64$\pm$    0.06 &  52.84$\pm$    2.27 & 350.21$\pm$   17.07  &  3.42$\pm$    0.17\\
                                            &&6569.64$\pm$    0.45  &  2.46$\pm$    0.06 &  18.42$\pm$    1.12 & 113.44$\pm$    7.41  &  1.11$\pm$    0.07\\
H$\beta$          & 4862.7  &4860.95$\pm$    0.44  &  1.81$\pm$    0.25 &  11.32$\pm$    1.47 &  51.49$\pm$    9.75  &  0.50$\pm$    0.10\\
                                         &&4865.34$\pm$    0.58  &  1.66$\pm$    0.04 &   5.85$\pm$    1.27 &  24.42$\pm$    5.31  &  0.24$\pm$    0.05\\
$[\mathrm {O~III}]$& 4960.3  &4959.97$\pm$    0.49  &  2.15$\pm$    0.12 &  14.77$\pm$    3.45 &  79.70$\pm$   19.15  &  0.78$\pm$    0.19\\
                                          &&4962.95$\pm$    0.48  &  1.21$\pm$    0.77 &   8.95$\pm$    1.88 &  27.18$\pm$   18.11  &  0.27$\pm$    0.18\\
$[\mathrm {O~III}]$& 5008.2  &5007.91$\pm$    0.49  &  2.18$\pm$    0.12 &  43.58$\pm$    3.45 & 237.96$\pm$   22.95  &  2.32$\pm$    0.22\\
                                          &&5010.92$\pm$    0.48  &  1.23$\pm$    0.77 &  26.40$\pm$    1.88 &  81.57$\pm$   51.03  &  0.80$\pm$    0.50\\
\hline
\end{tabular}
\end{center}
{Column~1: Emission lines that were fitted with Gaussian components. Column~2: Rest wavelength in vacuum in $\AA$. Columns~3, 4: Central wavelength and line width ($\sigma$) in $\AA$ along with respective errors. Column~5: Peak line flux in units of $10^{-17}$~erg~cm$^{-2}$~s$^{-1}~\AA^{-1}$ with error. (All line flux densities have been corrected for galactic or foreground reddening.) Column~6: {Total line flux in $10^{-17}$~erg~cm$^{-2}$~s$^{-1}$.} Column~7: Line luminosity in units of $10^{40}$~erg~s$^{-1}$. }
\label{tabline1}
\end{table*}

\begin{table*}
\scriptsize
\caption{Fitted Line Parameters for KISSR\,872}
\begin{center}
\begin{tabular}{lclcllc}
\hline\hline
{Line} & {$\lambda_{0}$} & {$\lambda_{c}\pm$error} & {$\Delta\lambda\pm$error} & {$f_{p}\pm$error} & {$F\pm$error}& {$L\pm$error}\\
{(1)}   & {(2)}  & {(3)}                 & {(4)} & {(5)} & {(6)} & {(7)}\\ \hline
$[\mathrm {S~II}]$ & 6718.3  &6716.37$\pm$    0.98  &  1.95$\pm$    0.10 &  17.14$\pm$    7.62 &  83.74$\pm$   37.49  &  1.05$\pm$    0.47\\
                 &&6720.90$\pm$    1.65  &  2.13$\pm$    0.12 &  10.37$\pm$    1.61 &  55.48$\pm$    9.18  &  0.70$\pm$    0.12\\
$[\mathrm {S~II}]$ & 6732.7  &6730.37$\pm$    0.98  &  1.95$\pm$    0.10 &  15.21$\pm$    6.22 &  74.50$\pm$   30.74  &  0.94$\pm$    0.39\\
                 &&6734.91$\pm$    1.65  &  2.14$\pm$    0.12 &   8.34$\pm$    1.53 &  44.75$\pm$    8.62  &  0.56$\pm$    0.11\\
$[\mathrm {N~II}]$ & 6549.9  &6548.43$\pm$    0.19  &  1.89$\pm$    0.10 &  35.27$\pm$    1.22 & 166.68$\pm$   10.48  &  2.10$\pm$    0.13\\
                 &&6553.74$\pm$    0.20  &  2.07$\pm$    0.12 &  16.34$\pm$    0.48 &  84.69$\pm$    5.45  &  1.06$\pm$    0.07\\
$[\mathrm {N~II}]$ & 6585.3  &6583.43$\pm$    0.10  &  1.90$\pm$    0.10 & 104.03$\pm$    3.60 & 495.19$\pm$   31.12  &  6.23$\pm$    0.39\\
                 &&6588.77$\pm$    0.09  &  2.08$\pm$    0.12 &  48.19$\pm$    0.41 & 251.52$\pm$   14.58  &  3.16$\pm$    0.18\\
H$\alpha$         & 6564.6  &6563.28$\pm$    0.18  &  1.89$\pm$    0.10 & 144.67$\pm$    2.28 & 685.84$\pm$   37.49  &  8.63$\pm$    0.47\\
                 &&6568.38$\pm$    0.20  &  2.07$\pm$    0.11 &  47.47$\pm$    1.68 & 246.80$\pm$   15.62  &  3.10$\pm$    0.20\\
H$\beta$          & 4862.7  &4861.19$\pm$    0.32  &  1.20$\pm$    0.07 &  26.52$\pm$    4.98 &  79.50$\pm$   15.60  &  1.00$\pm$    0.20\\
                 &&4863.77$\pm$    2.30  &  1.35$\pm$    0.07 &  14.91$\pm$    4.58 &  50.52$\pm$   15.76  &  0.64$\pm$    0.20\\
$[\mathrm {O~III}]$& 4960.3  &4959.32$\pm$    0.07  &  2.35$\pm$    0.12 &  10.43$\pm$    0.40 &  61.34$\pm$    3.98  &  0.77$\pm$    0.05\\
                 &&4963.80$\pm$    0.11  &  9.85$\pm$    0.56 &   4.50$\pm$    0.24 & 111.14$\pm$    8.64  &  1.40$\pm$    0.11\\
$[\mathrm {O~III}]$& 5008.2  &5007.26$\pm$    0.41  &  2.37$\pm$    0.18 &  30.78$\pm$    1.17 & 183.08$\pm$   15.84  &  2.30$\pm$    0.20\\
                 &&5011.77$\pm$    0.13  &  9.94$\pm$    0.52 &  13.28$\pm$    0.70 & 331.08$\pm$   24.58  &  4.16$\pm$    0.31\\
H$\gamma$         & 4341.7  &4340.46$\pm$    0.21  &  0.95$\pm$    0.05 &  10.64$\pm$    1.69 &  25.34$\pm$    4.28  &  0.32$\pm$    0.05\\
                 &&4342.22$\pm$    0.47  &  1.10$\pm$    0.40 &   3.26$\pm$    1.34 &   9.02$\pm$    4.94  &  0.11$\pm$    0.06\\
$[\mathrm {O~I}]$  & 6302.0  &6301.81$\pm$    0.11  &  2.30$\pm$    0.12 &   6.69$\pm$    2.21 &  38.55$\pm$   12.89  &  0.48$\pm$    0.16\\
                 &&6309.00$\pm$    2.90  &  0.37$\pm$    0.21 &   3.81$\pm$    3.30 &   3.57$\pm$    3.69  &  0.04$\pm$    0.05\\
$[\mathrm {O~I}]$  & 6365.5  &6366.10$\pm$    0.15  &  2.33$\pm$    0.90 &   2.77$\pm$    2.20 &  16.17$\pm$   14.28  &  0.20$\pm$    0.18\\
                 &&6373.37$\pm$    0.20  &  0.41$\pm$    2.30 &   2.56$\pm$    1.10 &   2.62$\pm$   14.81  &  0.03$\pm$    0.19\\
\hline
\end{tabular}
\end{center}
{Column~1: Emission lines that were fitted with Gaussian components. Column~2: Rest wavelength in vacuum in $\AA$. Columns~3, 4: Central wavelength and line width ($\sigma$) in $\AA$ along with respective errors. Column~5: Peak line flux in units of $10^{-17}$~erg~cm$^{-2}$~s$^{-1}~\AA^{-1}$ with error. (All line flux densities have been corrected for galactic or foreground reddening.) Column~6: {Total line flux in $10^{-17}$~erg~cm$^{-2}$~s$^{-1}$.} Column~7: Line luminosity in units of $10^{40}$~erg~s$^{-1}$. }
\label{tabline2}
\end{table*}

\begin{table*}
\scriptsize
\caption{Fitted Line Parameters for KISSR\,967}
\begin{center}
\begin{tabular}{lclcllc}
\hline\hline
{Line} & {$\lambda_{0}$} & {$\lambda_{c}\pm$error} & {$\Delta\lambda\pm$error} & {$f_{p}\pm$error} & {$F\pm$error}& {$L\pm$error}\\
{(1)}   & {(2)}  & {(3)}                 & {(4)} & {(5)} & {(6)} & {(7)}\\ \hline
$[\mathrm {S~II}]$ & 6718.3  &    6714.62$\pm$    1.47  &  2.85$\pm$    0.07  &  8.73$\pm$    3.42 &  62.46$\pm$   24.47  &  0.97$\pm$    0.38\\
                                             && 6720.11$\pm$    0.83  &  3.02$\pm$    0.37  & 11.50$\pm$    2.44 &  87.02$\pm$   21.32  &  1.34$\pm$    0.33\\
$[\mathrm {S~II}]$ & 6732.7  &  6728.62$\pm$    1.47  &  2.86$\pm$    0.07  &  6.25$\pm$    2.42 &  44.81$\pm$   17.36  &  0.69$\pm$    0.27\\
                                           && 6734.11$\pm$    0.83  &  3.02$\pm$    0.37  &  9.39$\pm$    1.88 &  71.17$\pm$   16.66  &  1.10$\pm$    0.26\\
$[\mathrm {N~II}]$ & 6549.9  &  6547.02$\pm$    0.45  &  2.77$\pm$    0.07  & 15.37$\pm$    2.03 & 106.76$\pm$   14.33  &  1.65$\pm$    0.22\\
                                           && 6553.19$\pm$    0.49  &  2.93$\pm$    0.36  & 14.37$\pm$    0.84 & 105.63$\pm$   14.32  &  1.63$\pm$    0.22\\
$[\mathrm {N~II}]$ & 6585.3  &  6582.02$\pm$    0.45  &  2.79$\pm$    0.07  & 45.33$\pm$    5.97 & 316.87$\pm$   42.52  &  4.90$\pm$    0.66\\
                                             && 6588.21$\pm$    0.45  &  2.95$\pm$    0.36  & 42.39$\pm$    2.47 & 313.51$\pm$   42.49  &  4.84$\pm$    0.66\\
H$\alpha$         & 6564.6  &  6561.02$\pm$    0.50  &  2.78$\pm$    0.07  & 80.91$\pm$   11.11 & 563.49$\pm$   78.67  &  8.71$\pm$    1.22\\
                                       && 6567.24$\pm$    0.51  &  2.94$\pm$    0.07  & 74.76$\pm$    3.90 & 551.00$\pm$   31.95  &  8.52$\pm$    0.49\\
H$\beta$          & 4862.7  &   4860.47$\pm$    2.72  &  1.93$\pm$    0.24  &  5.48$\pm$    6.30 &  26.46$\pm$   30.61  &  0.41$\pm$    0.47\\
                                         && 4863.72$\pm$    0.68  &  2.05$\pm$    0.05  & 12.94$\pm$    5.34 &  66.63$\pm$   27.55  &  1.03$\pm$    0.43\\
$[\mathrm {O~III}]$& 4960.3  & 4960.40$\pm$   20.86  &  1.08$\pm$    0.54  &  2.97$\pm$   36.87 &   8.01$\pm$   99.72  &  0.12$\pm$    1.54\\
                                             && 4962.07$\pm$    1.46  &  0.90$\pm$    2.74  &  2.81$\pm$   36.47 &   6.36$\pm$   84.80  &  0.10$\pm$    1.31\\
$[\mathrm {O~III}]$& 5008.2  & 5008.34$\pm$   20.86  &  1.10$\pm$    0.54  &  8.75$\pm$   36.87 &  24.10$\pm$  102.23  &  0.37$\pm$    1.58\\
                                             && 5010.03$\pm$    1.46  &  0.92$\pm$    2.74  &  8.29$\pm$   36.47 &  19.21$\pm$  101.91  &  0.30$\pm$    1.57\\
\hline
\end{tabular}
\end{center}
{Column~1: Emission lines that were fitted with Gaussian components. Column~2: Rest wavelength in vacuum in $\AA$. Columns~3, 4: Central wavelength and line width ($\sigma$) in $\AA$ along with respective errors. Column~5: Peak line flux in units of $10^{-17}$~erg~cm$^{-2}$~s$^{-1}~\AA^{-1}$ with error. (All line flux densities have been corrected for galactic or foreground reddening.) Column~6: {Total line flux in $10^{-17}$~erg~cm$^{-2}$~s$^{-1}$.} Column~7: Line luminosity in units of $10^{40}$~erg~s$^{-1}$. }
\label{tabline3}
\end{table*}

\begin{table*}
\scriptsize
\caption{Fitted Line Parameters for KISSR\,1154}
\begin{center}
\begin{tabular}{lclcllc}
\hline\hline
{Line} & {$\lambda_{0}$} & {$\lambda_{c}\pm$error} & {$\Delta\lambda\pm$error} & {$f_{p}\pm$error} & {$F\pm$error}& {$L\pm$error}\\
{(1)}   & {(2)}  & {(3)}                 & {(4)} & {(5)} & {(6)} & {(7)}\\ \hline
$[\mathrm {S~II}]$ & 6718.3  & 6716.64$\pm$1.57  &   3.22$\pm$0.24  &  4.07$\pm$1.14  &   32.87$\pm$9.49   &      0.32$\pm$0.09 \\
                                            && 6721.42$\pm$    0.30    &         1.69 $\pm$   0.22    &        5.98  $\pm$  1.67       &       25.39 $\pm$   7.85     &  0.24  $\pm$  0.08\\
$[\mathrm {S~II}]$ & 6732.7  & 6730.64  $\pm$  1.57  &   3.23$\pm$    0.24   &  3.69 $\pm$   0.98  &  29.82 $\pm$   8.26   &  0.29 $\pm$   0.08\\
                               && 6735.43 $\pm$   0.30   & 1.70 $\pm$   0.22   &  5.67 $\pm$   1.53  &  24.16 $\pm$   7.25   &   0.23 $\pm$   0.07\\
$[\mathrm {N~II}]$ & 6549.9  & 6548.48 $\pm$   1.18  &   3.13  $\pm$  0.23   &  4.86  $\pm$  1.16  & 38.09 $\pm$   9.54   &  0.37 $\pm$   0.09 \\
                          && 6552.54  $\pm$  0.14  &  1.63  $\pm$  0.22   &  7.21  $\pm$  1.64   &  29.53 $\pm$   7.79   &  0.28 $\pm$    0.08 \\
$[\mathrm {N~II}]$ & 6585.3  & 6583.48$\pm$    1.18  &   3.15 $\pm$   0.23  &  14.33  $\pm$  3.43  & 113.03 $\pm$  28.31   &  1.09 $\pm$   0.27\\
                                               &&   6587.57 $\pm$   1.18   &  1.65 $\pm$   0.22  &  21.27 $\pm$   4.85  &  87.80  $\pm$ 23.17   &  0.85  $\pm$  0.22\\
H$\alpha$         & 6564.6  & 6562.66$\pm$    0.88  &    3.14 $\pm$   0.23  &  14.79 $\pm$   2.26  & 116.22 $\pm$  19.74  &   1.12  $\pm$  0.19\\
                                          &&   6567.13 $\pm$   0.19  &  1.64  $\pm$  0.12  &  13.44 $\pm$   3.64   & 55.23 $\pm$  15.51   &  0.53 $\pm$   0.15\\
H$\beta$          & 4862.7  & 4860.28$\pm$    0.69  &   2.20 $\pm$   0.29  &   4.93 $\pm$   0.70  &  27.24 $\pm$   5.30 &    0.26   $\pm$ 0.05\\
                                          &&  4865.21$\pm$    0.38   &  0.97 $\pm$   0.07  &   5.51 $\pm$   1.01  &  13.33 $\pm$   2.65  &   0.13   $\pm$ 0.03\\
$[\mathrm {O~III}]$& 4960.3  & 4957.08 $\pm$   0.76  &   2.94 $\pm$   0.32  &   5.58 $\pm$   1.22  &  41.08 $\pm$  10.03  &   0.40 $\pm$  0.10\\
                                             &&    4961.51$\pm$    0.17  &   1.10 $\pm$   0.30  &  16.38 $\pm$   0.81 &   45.23 $\pm$  12.69  &   0.44 $\pm$   0.12\\
$[\mathrm {O~III}]$& 5008.2  & 5004.99 $\pm$   0.76   &  2.97  $\pm$  0.32  &  16.45  $\pm$  1.22 &  122.53  $\pm$ 15.90   &  1.18  $\pm$  0.15\\
                                               &&   5009.47 $\pm$   0.17  &   1.12  $\pm$ 0.30  &  48.32   $\pm$ 0.81  & 135.98 $\pm$  36.92  &   1.31 $\pm$   0.36\\
\hline
\end{tabular}
\end{center}
{Column~1: Emission lines that were fitted with Gaussian components. Column~2: Rest wavelength in vacuum in $\AA$. Columns~3, 4: Central wavelength and line width ($\sigma$) in $\AA$ along with respective errors. Column~5: Peak line flux in units of $10^{-17}$~erg~cm$^{-2}$~s$^{-1}~\AA^{-1}$ with error. (All line flux densities have been corrected for galactic or foreground reddening.) Column~6: {Total line flux in $10^{-17}$~erg~cm$^{-2}$~s$^{-1}$.} Column~7: Line luminosity in units of $10^{40}$~erg~s$^{-1}$. }
\label{tabline4}
\end{table*}

\begin{table*}
\scriptsize
\caption{Fitted Line Parameters for KISSR\,1321}
\begin{center}
\begin{tabular}{lclcllc}
\hline\hline
{Line} & {$\lambda_{0}$} & {$\lambda_{c}\pm$error} & {$\Delta\lambda\pm$error} & {$f_{p}\pm$error} & {$F\pm$error}& {$L\pm$error}\\
{(1)}   & {(2)}  & {(3)}                 & {(4)} & {(5)} & {(6)} & {(7)}\\ \hline
$[\mathrm {S~II}]$ & 6718.3  &      6716.00 $\pm$    0.46  &   3.22 $\pm$    0.07  &  19.62 $\pm$    2.27 &  158.36 $\pm$   18.67   &  1.21 $\pm$    0.14\\
                                               && 6723.02  $\pm$   0.77  &   3.59 $\pm$    0.49  &  10.65  $\pm$   1.24  &  95.91  $\pm$  17.25    & 0.73 $\pm$    0.13\\
$[\mathrm {S~II}]$ & 6732.7  &      6729.99 $\pm$    0.46  &   3.23  $\pm$   0.07  &  13.31   $\pm$  2.00  & 107.67$\pm$    16.36  &   0.82  $\pm$   0.12\\
                                               && 6737.04 $\pm$    0.77   &  3.60 $\pm$    0.49   &  9.85  $\pm$   0.91 &   88.88  $\pm$  14.67  &   0.68  $\pm$   0.11\\
$[\mathrm {N~II}]$ & 6549.9  &       6547.76  $\pm$   0.52  &   3.13  $\pm$   0.07  & 11.33  $\pm$   1.66  &  88.90 $\pm$   13.15  &   0.68  $\pm$   0.10\\
                                              && 6554.39 $\pm$    0.75  &   3.49 $\pm$    0.48   &  8.28  $\pm$   0.77   & 72.52 $\pm$   12.04  &   0.55  $\pm$   0.09\\
$[\mathrm {N~II}]$ & 6585.3  &      6582.76  $\pm$   0.52   &  3.15  $\pm$   0.07  & 33.43 $\pm$    4.89  & 263.83  $\pm$  39.02  &   2.01 $\pm$     0.30\\
                                             && 6589.42 $\pm$    0.52   &  3.51 $\pm$    0.48 &   24.44 $\pm$    2.28 &  215.19 $\pm$   35.71   &  1.64 $\pm$    0.27\\
H$\alpha$         & 6564.6  &          6563.10 $\pm$    0.31   &  3.14  $\pm$   0.07  &  56.61 $\pm$    3.95 &  445.22  $\pm$  32.72   &  3.40  $\pm$   0.25\\
                                              &&6569.74  $\pm$   0.91   &  3.50 $\pm$    0.08  &  16.83  $\pm$   2.36 &  147.75  $\pm$  20.96  &   1.13  $\pm$   0.16\\
H$\beta$          & 4862.7  &          4860.97 $\pm$    1.65   &   2.20  $\pm$  0.30  &  11.48$\pm$    10.57  &  63.38  $\pm$  59.00   &  0.48  $\pm$   0.45\\
                                             &&4863.73  $\pm$   1.44  &   2.48 $\pm$    0.06   &  8.40 $\pm$    8.62  &   52.29  $\pm$  53.70  &   0.40 $\pm$    0.41\\
$[\mathrm {O~III}]$& 4960.3  &     4958.27 $\pm$     0.22   &  1.73 $\pm$    0.15 &   13.17  $\pm$   1.26   & 57.27 $\pm$    7.48   &  0.44 $\pm$     0.06\\
                                                && 4962.81$\pm$     0.18   &  1.51 $\pm$    0.26  &  17.54 $\pm$    0.84  &  66.39  $\pm$  12.02  &   0.51 $\pm$    0.09\\
$[\mathrm {O~III}]$& 5008.2  &     5006.19 $\pm$    0.22   &  1.76 $\pm$    0.15  &  38.85 $\pm$    1.26  & 171.24 $\pm$    16.05   &  1.31  $\pm$   0.12\\
                                              &&5010.78 $\pm$    0.18  &   1.53  $\pm$   0.26  &  51.75  $\pm$   0.84  & 198.74 $\pm$   34.34  &   1.52  $\pm$  0.26\\
\hline
\end{tabular}
\end{center}
{Column~1: Emission lines that were fitted with Gaussian components. Column~2: Rest wavelength in vacuum in $\AA$. Columns~3, 4: Central wavelength and line width ($\sigma$) in $\AA$ along with respective errors. Column~5: Peak line flux in units of $10^{-17}$~erg~cm$^{-2}$~s$^{-1}~\AA^{-1}$ with error. (All line flux densities have been corrected for galactic or foreground reddening.) Column~6: {Total line flux in $10^{-17}$~erg~cm$^{-2}$~s$^{-1}$.} Column~7: Line luminosity in units of $10^{40}$~erg~s$^{-1}$. }
\label{tabline5}
\end{table*}

\vspace{5mm}
\facilities{VLBA, Sloan}
\software{pPXF \citep{Cappellari04,Cappellari17}, MAPPINGS III \citep{Dopita1996,Allen2008}, 
AIPS \citep{vanMoorsel96}}

\bibliographystyle{aasjournal}
\bibliography{ms}
\end{document}